%% file: main.tex
\documentclass[sigconf,nonacm]{acmart}

\usepackage{subfigure}
\usepackage{siunitx}
\usepackage{enumitem}

\usepackage{amsmath}

\usepackage{algorithm}
\usepackage[noend]{algpseudocode}
\let\ReturnInline\Return
\renewcommand{\Return}{\State\ReturnInline}
\algrenewcommand\algorithmicrequire{$\rhd$}
\algrenewcommand\algorithmicensure{$\square$}

\AtBeginDocument{%
  \providecommand\BibTeX{{%
    \normalfont B\kern-0.5em{\scshape i\kern-0.25em b}\kern-0.8em\TeX}}}

\setcopyright{acmcopyright}
\copyrightyear{2018}
\acmYear{2018}
\acmDOI{XXXXXXX.XXXXXXX}

\acmConference[Conference acronym 'XX]{Make sure to enter the correct
  conference title from your rights confirmation emai}{June 03--05,
  2018}{Woodstock, NY}

\newcommand{\su}[1]{{{\color{green} #1}}}

\newcommand{\ignore}[1]{}

\begin{document}

\title{Efficient GPU Implementation of Static and Incrementally Expanding DF-P PageRank for Dynamic Graphs}


\author{Subhajit Sahu}
\email{subhajit.sahu@research.iiit.ac.in}
\affiliation{%
  \institution{IIIT Hyderabad}
  \streetaddress{Professor CR Rao Rd, Gachibowli}
  \city{Hyderabad}
  \state{Telangana}
  \country{India}
  \postcode{500032}
}


\settopmatter{printfolios=true}

\begin{abstract}
PageRank is a widely used centrality measure that "ranks" vertices in a graph by considering the connections and their importance. In this report, we first introduce one of the most efficient GPU implementations of Static PageRank, which recomputes PageRank scores from scratch. It uses a synchronous pull-based atomics-free PageRank computation, with the low and high in-degree vertices being partitioned and processed by two separate kernels. Next, we present our GPU implementation of incrementally expanding (and contracting) Dynamic Frontier with Pruning (DF-P) PageRank, which processes only a subset of vertices likely to change ranks. It is based on Static PageRank, and uses an additional partitioning between low and high out-degree vertices for incremental expansion of the set of affected vertices with two additional kernels. On a server with an NVIDIA A100 GPU, our Static PageRank outperforms Hornet and Gunrock's PageRank implementations by $31\times$ and $5.9\times$ respectively\ignore{, while being $24\times$ faster than our multicore Static PageRank}. On top of the above, DF-P PageRank outperforms Static PageRank by $2.1\times$ on real-world dynamic graphs, and by $3.1\times$ on large static graphs with random batch updates.
\end{abstract}



\maketitle

\section{Introduction}
\label{sec:introduction}
\input{01-introduction}

\section{Related work}
\label{sec:related}
\input{02-related-work}

\section{Preliminaries}
\label{sec:preliminaries}
\input{03-preliminaries}

\section{Approach}
\label{sec:approach}
\input{04-approach}

\section{Evaluation}
\label{sec:evaluation}
\input{05-evaluation}

\section{Conclusion}
\label{sec:conclusion}
\input{06-conclusion}

\begin{acks}
I would like to thank Prof. Kishore Kothapalli, Prof. Sathya Peri, and Prof. Hemalatha Eedi for their support.\ignore{Note that Britannia Industries Ltd., the owner of the 50-50 biscuit brand, did not sponsor our work.}
\end{acks}

\bibliographystyle{ACM-Reference-Format}
\bibliography{main}

\clearpage
\appendix
\input{aa-appendix}

\end{document}

%% file: 01-introduction.tex
PageRank is an algorithm for assessing the significance of nodes within a network by assigning numerical scores based on link structures \cite{rank-page99}. It operates on the principle that pages with more high-quality inbound links are of greater importance and thus should have higher ranks. While originally devised to rank web pages in search results, this metric finds applications in various domains, such as urban planning \cite{urban-zhang18}, video object tracking \cite{gong2013pagerank}, traffic flow prediction \cite{traffic-kim15}, dynamic asset valuation \cite{sawilla2006abstracting}, protein target identification \cite{banky2013equal}\ignore{, brain region importance assessment \cite{zuo2012network}}, software system characterization \cite{chepelianskii2010towards}, and identification of crucial species for environmental health \cite{allesina2009googling}\ignore{, and quantifying the scientific impact of researchers \cite{rank-senanayake15}}.

With the rise of extensive interconnected data, interest in parallel PageRank computation has surged. This has led to a number of implementations of parallel PageRank for multicore CPUs \cite{rank-garg16, rank-beamer17, rank-lakhotia18, grutzmacher2020acceleration, huang2020accelerating, chen2021hipa}. Unfortunately, multicore CPUs only offer a limited\ignore{parallelism and} memory bandwidth. This makes them unsuitable for graph algorithms, such as PageRank, which have a low computation-to-communication ratio. GPUs, on the other hand, boast extremely high bandwidth memory, connected in close proximity to thousands of lightweight cores with user-managed caches. Further, the GPU hardware is designed to be able to switch between running threads at no cost in order to support memory access latency hiding. When graphs algorithms are suitably designed, they can significantly outperform a parallel CPU-based implementation.

In recent years, significant research effort has focused on developing efficient parallel implementations of PageRank for GPUs \cite{duong2012parallel, rank-nvgraph, wang2016gunrock, busato2018hornet, dathathri2018gluon, nodehi2018tigr, grutzmacher2018high, piccinotti2019solving, grutzmacher2020acceleration, kang2020computing, wang2021grus, chen2022atos, chen2022scalable, yang2022graphblast, concessao2023meerkat}.\ignore{FPGAs \cite{rank-guoqiang20}, SpMV ASICs \cite{rank-sadi18}, CPU-GPU hybrids \cite{rank-giri20}, CPU-FPGA hybrids \cite{usta2020accelerating, rank-li21, rank-hassan21, rank-mughrabi21}, and distributed systems \cite{rank-sarma13, kang2022analyzing, vandromme2022scaling}.} These implementations, referred to as Static PageRank, compute ranks from scratch for a given graph, assuming the graph remains static over time. In this paper, we present our GPU implementation of Static PageRank,\footnote{\label{foo:repository}\url{https://github.com/puzzlef/pagerank-cuda-dynamic}} which utilizes a synchronous pull-based atomics-free computation method. It partitions and processes low and high in-degree vertices separately using two distinct kernels. Additionally, it avoids computation of global teleport rank contribution due to dead ends (vertices with no outgoing edges) --- we ensure the input graph contains no dead ends. Our implementation represents, to the best of our knowledge, the most efficient implementation for parallel PageRank computation on the GPU. Our GPU implementation of Static PageRank is compared with other state-of-the-art implementations in Table \ref{tab:compare-large}. It includes both direct and indirect comparisons, with details given in Sections \ref{sec:static-comparison} and \ref{sec:static-comparison-indirect} respectively.

\input{src/tab-compare-large}

Real-world graphs often exhibit dynamic characteristics, undergoing frequent edge updates \cite{agarwal2012real, barros2021survey}. Recomputing PageRank scores for vertices upon each update, known as Static PageRank, can be resource-intensive. A strategy to reduce computation involves initiating PageRank computation from previous vertex ranks, thereby minimizing iterations required for convergence. We refer to this as the \textit{Naive-dynamic (ND)} approach. To further optimize runtime, recalculating ranks solely for potentially affected vertices is crucial. One common approach, which we refer to as the \textit{Dynamic Traversal (DT)} approach, entails identifying reachable vertices from updated graph regions and processing only those \cite{rank-desikan05, kim2015incremental, rank-giri20, sahu2022dynamic}. However, marking all reachable vertices as affected, even for minor changes, may lead to unnecessary computation, particularly in dense graph regions. Our previous work \cite{sahu2024df} addressed these concerns by introducing incrementally expanding \textit{Dynamic Frontier (DF)} and \textit{Dynamic Frontier with Pruning (DF-P)} approaches. These approaches process only a subset of vertices likely to change ranks, and were implemented as parallel multicore algorithms \cite{sahu2024df}. Here, we present our GPU implementation of DF-P PageRank,\footnote{See Footnote \ref{foo:repository}} based on Static PageRank. It features partitioning between low and high out-degree vertices for incremental expansion of affected vertices using two additional kernels. Table \ref{tab:compare} shows the performance comparison of DF-P PageRank with Static, ND, DT, and DF PageRank.

\input{src/tab-compare}

%% file: src/tab-compare-large.tex
\begin{table}[hbtp]
  \centering
  \caption{Speedup of our GPU implementation of Static PageRank compared to other state-of-the-art implementations. Direct comparisons are based on running the given implementation on our server, while indirect comparisons (denoted with a $*$) involve comparing results obtained by the given implementation relative to a common reference (Hornet/Gunrock).}
  \label{tab:compare-large}
  \begin{tabular}{|c|c||c|}
    \toprule
    \textbf{PageRank implementation} &
    \textbf{Published} &
    \textbf{Our Speedup} \\
    \midrule
    Hornet \cite{busato2018hornet} & 2018 & $31\times$ \\ \hline
    Gunrock \cite{wang2016gunrock} & 2016 & $5.9\times$ \\ \hline
    Galois with Gluon \cite{dathathri2018gluon} & 2018 & $448\times^*$ \\ \hline
    Tigr \cite{nodehi2018tigr} & 2018 & $40\times^*$ \\ \hline
    Grus \cite{wang2021grus} & 2021 & $7.2\times^*$ \\ \hline
    Atos \cite{chen2022atos} & 2022 & $7.6\times^*$ \\ \hline
    Multi-GPU Atos ($1$ GPU) \cite{chen2022scalable} & 2022 & $7.8\times^*$ \\ \hline
    Multi-GPU Atos ($4$ GPUs) \cite{chen2022scalable} & 2022 & $4.8\times^*$ \\ \hline
    GraphBLAST \cite{yang2022graphblast} & 2022 & $11\times^*$ \\ \hline
    Meerkat \cite{concessao2023meerkat} & 2023 & $18\times^*$ \\ \hline
  \bottomrule
  \end{tabular}
\end{table}

%% file: src/tab-compare.tex
\begin{table}[hbtp]
  \centering
  \caption{Speedup of our GPU implementation of Dynamic Frontier with Pruning (DF-P) PageRank compared to other approaches of updating PageRank scores (on GPU), on real-world dynamic graphs (Table \ref{tab:dataset}), and on large static graphs with random batch updates (Table \ref{tab:dataset-large}), respectively.}
  \label{tab:compare}
  \begin{tabular}{|l||c|}
    \toprule
    \textbf{PageRank approach} &
    \textbf{Speedup of DF-P} \\
    \midrule
    Static \cite{rank-page99} & $2.1\times$, $3.1\times$ \\ \hline
    Naive-dynamic (ND) \cite{rank-page99, rank-zhang17} & $1.5\times$, $1.7\times$ \\ \hline
    Dynamic Traversal (DT) \cite{rank-desikan05, kim2015incremental, rank-giri20, sahu2022dynamic} & $1.8\times$, $13.1\times$ \\ \hline
    Dynamic Frontier (DF) \cite{sahu2024df} & $2.1\times$ $1.3\times$ \\ \hline
  \bottomrule
  \end{tabular}
\end{table}

%% file: 02-related-work.tex
\subsection{Static PageRank}

One of the earliest GPU implementations of PageRank is by Wu et al. \cite{rank-wu10}. They implement Static PageRank on AMD GPUs using OpenCL. For this, they develop a Sparse Matrix-Vector Multiplication (SpMV) routine as a primitive. Here, as a pre-processing step, they sort the rows of the PageRank sparse matrix based on the number of non-zero elements, while keeping track of the original row number. Subsequently, they employ three distinct kernels to process the rows: One thread per row (1T1R), one quarter wavefront per row (16T1R), and one wavefront per row (1W1R), depending on the number of non-zero elements in each row. Cevahir et al. \cite{cevahir2010efficient} conduct PageRank computation on an NVIDIA GPU cluster (TSUBAME 1.2), with each node housing two Tesla GPUs. They partition the graph into chunks, assigning one chunk to each node. Each chunk is split into two portions, each of which can be loaded into the device memory of each GPU. They allocate one MPI process for each GPU, and utilize an SpMV kernel on the GPU for computation. Rungsawang and Manaskasemsak \cite{rank-rungsawang12} improve upon the work of Cevahir et al. \cite{cevahir2010efficient} by devising an algorithm that does not require large graphs to fit within the limited device memory. They achieve this by partitioning the graph between nodes and then further dividing each partition into chunks using the CPU at each node. These chunks are sized to fit into the device memory of the Tesla GPU at each node, which then processes them one after the other. Duong et al. \cite{rank-duong12} introduce a push-based GPU implementation of Static PageRank, employing atomic operations. They utilize a single thread per vertex for rank computation and handle the global teleport rank contribution (dangling value) with atomic operations. Additionally, they propose a multi-GPU implementation where the input graph is partitioned among GPUs, allowing independent computation of PageRank scores on each GPU. After each iteration, the computed ranks are synchronized on the CPU, where they are combined and redistributed among the GPUs\ignore{for subsequent iterations}.

However, Wu et al. \cite{rank-wu10}, Cevahir et al. \cite{cevahir2010efficient}, and Rungsawang and Manaskasemsak \cite{rank-rungsawang12} must first build the PageRank matrix before performing PageRank computation using an SpMV kernel. Our GPU implementation of PageRank does not require the building of any sparse matrix --- we directly utilize the adjacency list (stored in CSR format) of the graph for PageRank computation. Rungsawang and Manaskasemsak \cite{rank-rungsawang12} simply assign a thread for processing each vertex in the graph, which fails to achieve good load balancing and suffers from thread divergence, while Wu et al. \cite{rank-wu10} sort the rows by the number of non-zero elements, for load balancing using three different kernels, which can be an expensive operation. In contrast, we partition the vertices in the graph into low and high degree vertex sets and process them with two kernels.
Duong et al. \cite{rank-duong12} use a push-based PageRank computation, relying on atomic operations per edge, and also compute the global teleport contribution due to dead ends using atomic add operations. This can result in substantial memory contention among GPU threads. In contrast, our pull-based approach requires only one write per vertex, and by eliminating dead ends during graph loading we avoid the need for finding the global teleport contribution. Duong et al. \cite{rank-duong12} also do not partition vertices into low and high-degree sets.

We now discuss a number of graph processing frameworks, which include a GPU implementation of PageRank. Wang et al. \cite{wang2016gunrock} present the Gunrock multi-GPU CUDA library, which provides a high-level vertex/edge frontier-based bulk-synchronous / asynchronous API, and a number of high performance primitives. The PageRank implementation of Gunrock adopts a push-based approach, executing atomic adds per edge, employs a parallel for loop (using thrust) over the range of vertex IDs, utilizing either a thread- or a block-per-vertex approach. Further, its computes the global teleport contribution to each vertex attributable to dead ends. Busato et al. \cite{busato2018hornet} introduce Hornet, a platform-independent data structure optimized for dynamic sparse graphs and matrices. Hornet accommodates large-scale data growth without necessitating any data reallocation or re-initialization throughout the dynamic evolution process. The GPU-based graph processing framework based on the Hornet data structure is known as cuHornet. In its PageRank implementation, Hornet follows a push-based approach, akin to Gunrock. It calculates the rank contribution of each vertex separately, stored in a distinct vector, and utilizes an additional kernel to compute ranks from these contributions. Similar to Gunrock, Hornet employs a parallel for loop (using a custom kernel) over all vertices, adopting a thread-per-vertex approach.

We now discuss issues with the PageRank implementation of Gunrock \cite{wang2016gunrock} and Hornet \cite{busato2018hornet}. Both utilize push-based PageRank computation involving atomic add operations per edge, leading to significant memory contention among GPU threads. However, our approach is pull-based, and necessitates only a single write per vertex. Hornet computes vertex rank contributions separately and employs an additional kernel for rank computation, whereas we calculate rank contributions on the fly and utilize only a kernel pair for rank computation. Gunrock performs a parallel for (using thrust) over the range of vertex IDs, while Hornet performs a parallel for (using custom kernel) over all vertices using a thread per vertex. Unlike Gunrock and Hornet, we partition vertices into low and high in-degree sets and employ both thread- and block-per-vertex kernels. Gunrock uses a kernel to compute the global teleport contribution due to dead ends, while we preemptively eliminate dead ends during graph loading. Finally, Hornet uses a naive rank vector vector norm computation with atomic operations, while we implement a more efficient parallel reduce operation.

Wang et al. \cite{wang2021grus} introduce Grus, a GPU-based graph processing framework optimized for Unified Memory (UM) efficiency. Their focus lies in minimizing data migration, reducing page faults, and mitigating page migration overhead. Their PageRank implementation is push-based, and utilizes an adaptive UM policy, prioritizing frontier and rank data with high priority, Compressed Sparse Row (CSR) index array with medium priority, and CSR edges array with low priority. They use a bitmap-directed frontier, based on an 8-bit integer array alongside a queue, which eliminates the need for atomic operations. Load balancing is achieved with a warp-centric approach. Chen et al. \cite{chen2022atos} critique existing frameworks like Gunrock for launching each graph frontier as a separate GPU kernel in the Bulk Synchronous Parallel (BSP) model, leading to potential issues with parallelism, finish times, and kernel launch overhead, especially for small frontiers. To address this, they propose Atos, a persistent task scheduler designed to minimize kernel launch overhead and support asynchronous execution. For PageRank, they present a push-based asynchronous implementation, employing a queue-based frontier to track vertices for the next iteration, based on Adpative PageRank by Kamvar et al. \cite{kamvar2004adaptive}. In another study, Chen et al. \cite{chen2022scalable} broaden their Atos dynamic scheduling framework to multi-node GPU systems, accommodating Partitioned Global Address Space (PGAS) style lightweight one-sided memory operations within and between nodes. Yang et al. \cite{yang2022graphblast} introduce GraphBLAST, a high-performance linear algebra-based graph framework on the GPU. They address the lack of efficient GraphBLAS implementations for GPUs, highlighting the performance gap compared to state-of-the-art GPU graph frameworks, like the GraphBLAS Template Library (GBTL). They focus on exploiting input and output sparsity to simplify algorithm development and improve performance. They employ edge-balanced load balancing approach (merge-based) with segmented scan for PageRank, alongside a heuristic to switch between push- and pull-based approaches, favoring an early switch akin to Ligra \cite{shun2013ligra}. Concessao et al. \cite{concessao2023meerkat} introduce Meerkat, a library-based framework for dynamic graph algorithms optimized for GPU architectures. Leveraging a GPU-tailored graph representation and the warp-cooperative execution model, Meerkat enhances performance by exploiting common iteration patterns such as iterating over vertex neighborhoods efficiently. The framework supports dynamic edge additions and deletions, including batched versions. In their implementation of PageRank, Meerkat first computes the rank contribution of each vertex before calculating the rank itself\ignore{(similar to Hornet \cite{busato2018hornet})}. It employs a pull-based warp-per-vertex approach for parallel computation, where threads collaborate within a warp to compute vertex ranks, and achieves coalesced writes for rank updates.

We now note some concerns regarding GPU-based implementations of PageRank algorithms in existing frameworks. Grus \cite{wang2021grus} and Atos \cite{chen2022atos} adopt a push-based PageRank computation method. This involves atomic add operations per edge, and can introduce needless memory contention among GPU threads. Our approach is pull-based, and requires only a single write per vertex. Atos additionally employs a queue-based frontier for tracking vertices to be processed in the next iteration. This requires atomic operations, and is not conducive to parallelism. While Grus and Meerkat \cite{concessao2023meerkat} utilize warp-centric load balancing, they do not partition vertices into low and high in-degree sets. We partition vertices into such sets for load balancing, and to minimize thread divergence. Finally, Meerkat computes the rank contribution of each vertex, followed by a separate kernel for computing the ranks. It the uses another kernel to compute the common teleport contribution with atomics. However, as mentioned earlier, we calculate rank contributions on the fly, and employ only a pair of kernels for rank computation.

\subsection{Dynamic PageRank}

Early work in dynamic graph algorithms in the sequential setting includes the sparsification method proposed by Eppstein et al. \cite{graph-eppstein97} and Ramalingam's bounded incremental computation approach \cite{incr-ramalingam96}.\ignore{The latter advocates measuring the work done as part of the update in proportion to the effect the update has on the computation.} Several approaches have been suggested for incremental computation of approximate PageRank values in a dynamic or evolving graph. Chien et al. \cite{rank-chien01} identify a small region near updated vertices in the graph and represent the rest of the graph as a single vertex in a smaller graph. PageRanks are computed for this reduced graph and then transferred back to the original graph. Chen et al. \cite{chen2004local} propose various methods to estimate the PageRank score of a webpage using a small subgraph of the entire web, by expanding backwards from the target node along reverse hyperlinks. Bahmani et al. \cite{bahmani2010fast} analyze the efficiency of Monte Carlo methods for incremental PageRank computation. Zhan et al. \cite{zhan2019fast} introduce a Monte Carlo-based algorithm for PageRank tracking on dynamic networks, maintaining $R$ random walks starting from each node. Pashikanti et al. \cite{rank-pashikanti22} also employ a similar Monte Carlo-based approach for updating PageRank scores upon vertex and edge insertions/deletions.

A few approaches have been devised to update exact PageRank scores on dynamic graphs. Zhang \cite{rank-zhang17} uses a simple incremental PageRank computation system for dynamic graphs, which we refer to as the \textit{Naive-dynamic (ND)} approach, on hybrid CPU and GPU platforms\ignore{ --- employing the Update-Gather-Apply-Scatter (UGAS) computation model}. Additionally, Ohsaka et al. \cite{ohsaka2015efficient} propose a method for locally updating PageRank using the Gauss-Southwell method, prioritizing the vertex with the greatest residual for initial updating; however, their algorithm is inherently sequential. A widely adopted approach for updating PageRank \cite{rank-desikan05, kim2015incremental, rank-giri20, sahu2022dynamic} is based on the observation that changes in the out-degree of a node do not influence its PageRank score, adhering to the first-order Markov property. The portion of the graph undergoing updates, involving edge insertions or deletions, is used to identify the affected region of the graph in a preprocessing step. This is typically accomplished through Breadth-First Search (BFS) or Depth-First Search (DFS) traversal from vertices connected to the inserted or deleted edges. Subsequently, PageRanks are computed solely for this region. Desikan et al. \cite{rank-desikan05} originally proposed this, which we term as the \textit{Dynamic Traversal (DT)} approach in this report. Kim and Choi \cite{kim2015incremental} apply this approach with an asynchronous PageRank implementation, while Giri et al. \cite{rank-giri20} utilize it with collaborative executions on multi-core CPUs and massively parallel GPUs. Sahu et al. \cite{sahu2022dynamic} employ this strategy on a Strongly Connected Component (SCC)-based graph decomposition to limit computation to reachable SCCs from updated vertices, on multi-core CPUs and GPUs. Sallinen et al. \cite{sallinen2023real} recently proposed an event-based approach for asynchronously updating PageRank scores of vertices in dynamic graphs.\ignore{In the paper, each vertex stores its present rank, and computes its total rank contribution to its out-going neighbors, and thus the value it has previously distributed to its out-neighbors. For each edge addition, the vertex computes the previous value it sent, then sends a negative value representing the difference of the prior value to the new. For the new edge, it simply sends the adjusted flow as a single delta. For edge deletion, the algorithm is similar; negative flow is passed across the removed edge, redirecting positive flow to the remaining edges.} However, their per-thread event processing and termination detection methods are relatively complicated.

In our prior research \cite{sahu2024df}, we introduced two parallel approaches, Dynamic Frontier (DF) and Dynamic Frontier with Pruning (DF-P), for updating PageRank on dynamic graphs while using the power-iteration method \cite{rank-page99}\ignore{, achieving notable performance on multicore processors}.\ignore{Our findings indicated that DF/DF-P PageRank outperformed Static and DT PageRank by $5.2\times$/$15.2\times$ and $1.3\times$/$3.5\times$ respectively on real-world dynamic graphs, and by $7.2\times$/$9.6\times$ and $4.0\times$/$5.6\times$ on large static graphs with random batch updates.} In the report, we recommended DF-P PageRank for real-world dynamic graphs, suggesting a switch to DF PageRank if higher error rates were observed. For large graphs with random batch updates, we recommended DF-P PageRank, except for web graphs, where we suggested choosing DF PageRank.

%% file: 03-preliminaries.tex
\subsection{PageRank algorithm}
\label{sec:pagerank}

The PageRank, represented as $R[v]$, of a vertex $v \in V$ in the graph $G(V, E)$, measures its importance based on incoming links. Equation \ref{eq:pr} describes the PageRank computation for vertex $v$ in graph $G$, where $V$ denotes the set of vertices, $E$ represents the set of edges, $G.in(v)$ and $G.out(v)$ denote incoming and outgoing neighbors of $v$, respectively, and $\alpha$ is the damping factor. Initially, each vertex has a PageRank of $1/|V|$. The \textit{power-iteration} method iteratively updates these values until they converge within a specified iteration tolerance $\tau$. This is often measured using the $L_1$-norm \cite{ohsaka2015efficient}, though $L_2$ and $L_\infty$-norm are also occasionally used.

Core to the PageRank algorithm is the random surfer model. It conceives a surfer navigating the web by following links on each page. The damping factor $\alpha$, defaulting to $0.85$, indicates the likelihood of the surfer continuing along a link rather than jumping randomly. PageRank for each page reflects the long-term probability of the surfer visiting that page, originating from a random page and following links. PageRank values essentially constitute the eigenvector of a transition matrix, encoding probabilities of page transitions in a Markov Chain.

Dead ends, also referred to as dangling vertices, present a significant challenge in PageRank computation due to their lack of out-links, which compels the surfer to transition to a random web page. As a consequence, dead ends evenly distribute their rank across all vertices in the graph, necessitating computation in each iteration, thereby incurring overhead. We tackle this problem by introducing self-loops to all vertices in the graph \cite{kolda2009generalized, rank-andersen07, rank-langville06}, a strategy observed to be particularly effective in streaming environments and spam-link applications \cite{kolda2009generalized}.

\begin{equation}
\label{eq:pr}
    R[v] = \alpha \times \sum_{u \in G.in(v)} \frac{R[u]}{|G.out(u)|} + \frac{1 - \alpha}{|V|}
\end{equation}

\subsection{Fundamentals of a GPU}

The fundamental building block of NVIDIA GPUs is the Streaming Multiprocessor (SM). Each SM consists of multiple CUDA cores, which are responsible for executing parallel threads. SMs also include shared memory, registers, and special function units. The number of SMs varies across different GPU models, and each SM operates independently, executing multiple threads in parallel \cite{cuda-sanders10, gpu-nickolls10}.

The memory hierarchy of NVIDIA GPUs includes global memory, shared memory, and local memory. Global memory is the largest and slowest memory type, providing high-capacity storage for data. Shared memory is a fast, low-latency memory that is shared among threads within an SM, facilitating efficient data sharing and communication. It is used to store data that is frequently accessed by multiple threads, which can improve performance by reducing the number of memory accesses. Local memory is per-thread private memory used to store variables that do not fit in registers \cite{cuda-sanders10, gpu-nickolls10}.

\ignore{\paragraph{CUDA Cores}}

\ignore{CUDA cores are the primary execution units within an SM. These cores are responsible for executing individual threads in parallel. Each CUDA core is capable of performing arithmetic and logic operations, as well as handling memory access and control flow. The number of CUDA cores per SM varies across different GPU architectures, with newer generations featuring an increased number of cores \cite{cuda-sanders10, gpu-nickolls10}.}

\ignore{\paragraph{Behavioral fundamentals}}

Threads on a GPU are structured differently from those on a multicore CPU, organized into a hierarchy of warps, thread blocks, and grids. A warp comprises threads executing instructions in synchronous manner, i.e., in lockstep. NVIDIA GPUs typically have a warp size of 32 threads. Thread blocks, on the other hand, are collections of threads executing on the same SM. Within a thread block, warps execute in lockstep, with the SM scheduling alternate warps if any threads stall, such as during memory requests. All threads within a block can communicate through a user-managed cache, known as shared memory, which is private to each SM. Lastly, a grid encompasses multiple thread blocks, each functioning independently. Grids offer a higher-level structure for managing parallelism, optimizing GPU resource utilization. Thread blocks within a grid communicate solely through global memory\ignore{, which, although slower than shared memory, facilitates data exchange between different blocks}.

\subsection{Dynamic Graphs}
\label{sec:about-dynamic}

A dynamic graph can be construed as a succession of graphs, where $G^t(V^t, E^t)$ denotes the graph at time step $t$. The transitions between successive time steps $t-1$ and $t$, from $G^{t-1}(V^{t-1}, E^{t-1})$ to $G^t(V^t, E^t)$, can be denoted as a batch update $\Delta^t$ at time step $t$. This update encompasses a collection of edge deletions $\Delta^{t-}$, defined as $\{(u, v)\ |\ u, v \in V\} = E^{t-1} \setminus E^t$, and a set of edge insertions $\Delta^{t+}$, defined as $\{(u, v)\ |\ u, v \in V\} = E^t \setminus E^{t-1}$.

\ignore{\paragraph{Interleaving graph updates with computation:}}

\ignore{We consider graph updated to be batched, with changes to the graph structure and the algorithm execution being interleaved, permitting only one writer on the graph structure concurrently at any given time. If parallel updating of the graph structure is necessary during computation, a graph snapshot is needed to be obtained, upon which the computation may be performed (isolating it from the graphs updates). See for instance, the Aspen graph processing framework which reduces snapshot acquisition costs \cite{graph-dhulipala19}.}

\subsection{Existing approaches for updating PageRank on Dynamic Graphs}

\subsubsection{Naive-dynamic (ND) approach}
\label{sec:about-naive}

This approach simply updates vertex ranks in dynamic networks by initializing them with ranks from the previous graph snapshot and executing the PageRank algorithm on all vertices. Rankings obtained with this method are at least as accurate as those from the static algorithm.\ignore{Zhang et al. \cite{rank-zhang17} have explored the \textit{Naive-dynamic} approach in the hybrid CPU-GPU setting.}

\subsubsection{Dynamic Traversal (DT) approach}
\label{sec:about-traversal}

The Dynamic Traversal (DT) approach was initially proposed by Desikan et al. \cite{rank-desikan05}. It entails skipping the processing of vertices unaffected by the given batch update. For each edge deletion or insertion $(u, v)$ in the batch update, all the vertices reachable from vertex $u$ in either graph $G^{t-1}$ or $G^t$ are flagged as affected, typically using DFS or BFS.\ignore{Giri et al. \cite{rank-giri20} have explored the \textit{Dynamic Traversal} approach in the hybrid CPU-GPU setting. On the other hand, Banerjee et al. \cite{rank-sahu22} have explored this approach in the CPU and GPU settings separately where they compute the ranks of vertices in topological order of strongly connected components (SCCs) to minimize unnecessary computation. They borrow this ordered processing of SCCs from the original static algorithm proposed by Garg et al. \cite{rank-garg16}.}

%% file: 04-approach.tex
\subsection{Our GPU-based Static PageRank}
\label{sec:static}

We first discuss our GPU implementation of Static PageRank. Its psuedocode is given in Algorithm \ref{alg:static}. The explanation of the psuedocode is given in Section \ref{sec:static-impl}.

\paragraph{Copying data to the device:}

We begin by copying the CSR representation of the transpose of the current graph $G^{t'}$ to the GPU.

\paragraph{Partitioning vertices into low-degree and high-degree sets:}

We employ a pair of kernels, namely a thread-per-vertex kernel and a block-per-vertex kernel, for rank computation on the GPU. For this, we partition the vertex IDs for the two kernels by their in-degree. The thread-per-vertex kernel handles vertices with low in-degrees, while the block-per-vertex kernel computes ranks of high in-degree vertices. Using two different kernels allows us to improve processing performance for high-degree vertices, while minimizing thread divergence for low-degree vertices. The psuedocode for partitioning\ignore{vertex IDs} is given in Algorithm \ref{alg:partition}, and is explained in Section \ref{sec:partition}.

\paragraph{Rank computation of each vertex:}

We find that, unlike on multicore CPUs, employing a synchronous implementation of the PageRank algorithm, utilizing two separate rank vectors, yields superior performance. Consequently, we opt for a \textit{synchronous} implementation of the PageRank algorithm. During the rank computation phase in each iteration, we employ two distinct kernels. The \textit{thread-per-vertex} kernel manages vertices with low in-degree, assigning one thread per vertex to independently update each vertex's rank in a sequential manner. Conversely, the \textit{block-per-vertex} kernel allocates a thread block per vertex. Within each thread block, threads compute the rank contribution of in-neighbors to the vertex in a strided manner. These partial rank contributions are stored in shared memory, and a block reduction (summation) operation is performed to obtain the net rank contribution. Subsequently, the first thread in the thread block updates the rank of the\ignore{respective} vertex.

\paragraph{Convergence detection:}

To determine if the ranks of vertices have converged, we calculate the $L_\infty$-norm of the difference between the current and previous ranks. If this value is below the specified iteration tolerance $\tau$, the algorithm terminates, indicating convergence. If not, we swap the current and previous ranks and proceed to the next iteration. The calculation of the $L_\infty$-norm involves two kernels. The first kernel computes the $L_\infty$-norm of the rank differences for each thread block in a grid and stores the results in a temporary buffer. The second kernel computes the net $L_\infty$-norm of the results in the buffer, which is then transferred to the CPU.

\subsection{Our Static PageRank implementation}
\label{sec:static-impl}

Algorithm \ref{alg:static} outlines the psuedocode of our GPU-based Static PageRank. It takes as input the transpose of the current graph snapshot $G^{t'}$, and returns the computed rank vector $R$.

The algorithm begins by initializing the rank vectors $R$ and $R_{new}$, where each vertex's initial rank is set to $1/|V^{t'}|$ (lines \ref{alg:static--initialize-begin}-\ref{alg:static--initialize-end}). We then partition the vertex IDs in $P'$ based on their in-degree (line \ref{alg:static--partition}). This partitioning step is crucial for efficient computation on the GPU. Next, in lines \ref{alg:static--compute-begin}-\ref{alg:static--compute-end}, PageRank iterations are performed. Within each iteration, the ranks are updated using the \texttt{updateRanks()} function (line \ref{alg:static--update}), which computes the updated PageRank scores $R_{new}$ for each vertex based on the previous iteration's ranks $R$ (synchronous). When invoking \texttt{updateRanks()}, we disable the use of vertex and neighbor affected flags ($\delta_V$ and $\delta_N$) in order the utilize the same function for Static PageRank (\texttt{updateRanks()} is also used with DF-P PageRank). After updating the ranks, we calculate the $L_\infty$-norm difference $\Delta R$ between the current $R_{new}$ and previous ranks $R$ to check for convergence (line \ref{alg:static--error}). If the change in ranks falls below the specified tolerance $\tau$, the algorithm terminates (line \ref{alg:static--converged}). Finally, the algorithm returns the converged rank vector $R$ (line \ref{alg:static--return}).

Algorithm \ref{alg:static} uses a pull-based approach for PageRank computation, where each vertex's rank is updated through a single write by a thread. This is in contrast to a push-based approach, where each thread calculates and sums the outgoing PageRank contribution of its vertex to its neighbors, necessitating atomic updates \cite{verstraaten2015quantifying}. We find this to be more efficient and employ it for all implementations \cite{sahu2024df}. We also employ an synchronous implementation of Static PageRank, using two separate rank vectors, which we observe to perform better than an asynchronous implementation.

\input{src/alg-static}

\subsection{Our GPU-based Dynamic Frontier with Pruning (DF-P) PageRank}
\label{sec:frontier}

When dealing with a batch update comprising edge deletions $(u, v) \in \Delta^{t-}$ and insertions $(u, v) \in \Delta^{t+}$, if the total update size $|\Delta^{t-} \cup \Delta^{t+}|$ is relatively small compared to the total edge count $|E|$, only a small fraction of vertices are expected to undergo rank changes. To handle this, our earlier work proposed \textbf{Dynamic Frontier (DF)} and \textbf{Dynamic Frontier with Pruning (DF-P)} approaches, which employ an incremental process to identify affected vertices and update their ranks. Our research showed that DF/DF-P PageRank outperformed Static and Dynamic Traversal (DT) PageRank by $5.2\times$/$15.2\times$ and $1.3\times$/$3.5\times$\ignore{respectively} on real-world dynamic graphs, and by $7.2\times$/$9.6\times$ and $4.0\times$/$5.6\times$ on large graphs with random batch updates \cite{sahu2024df}.

Unfortunately, multicore CPUs have limited memory bandwidth and parallelism, making them unsuitable for graph algorithms like PageRank, which have a low computation-to-communication ratio. In contrast, GPUs offer high-bandwidth memory, connected in close proximity to thousands of lightweight cores with user-managed caches. Moreover, GPU hardware is designed to switch running threads at no cost to support memory access latency hiding. When appropriately designed, graph algorithms can significantly outperform CPU-based implementations on GPUs. This makes a GPU implementation of DF and DF-P PageRank attractive. In this section, we describe the design of DF and DF-P PageRank for GPUs.

\paragraph{Copying data to the device:}

\ignore{A GPU has its own memory that supports a high volume of data transfers.}We\ignore{first} copy the CSR representation of the current graph $G^t$, its transpose $G^{t'}$, the previous ranks of each vertex $R^{t-1}$, and the batch update consisting of edge deletions $\Delta^{t-}$ (both source and target vertex IDs in separate arrays) and edge insertions $\Delta^{t+}$ (source vertex IDs only) to the global memory of the GPU. The CSR of $G^{t'}$ is used for computing PageRank scores, while that of $G^t$ is used for the initial and incremental marking of affected vertices. Note that we only need the source vertex IDs of edge insertions $(u, v) \in \Delta^{t+}$ as we only need to mark the outgoing neighbors of $u$ as affected. On the other hand, for edge deletions $(u, v) \in \Delta^{t-}$, we need both the source and target vertex IDs as we need to mark both the outgoing neighbors of $u$ and the target vertices of the edge deletions, i.e., $v$, as affected.

\paragraph{Partitioning vertices into low-degree and high-degree sets:}

The work to be performed by each thread is proportional to the in-degree of each vertex during the rank computation, and to the out-degree of each vertex during incremental marking of affected vertices. Thus, similar to Static PageRank, we use a pair of kernels, i.e., a \textit{thread-per-vertex} kernel and a \textit{block-per-vertex} kernel, for both the rank computation and the incremental marking phases, by partitioning the vertex IDs for the two kernels by their in-degree for the rank computation phase, and by their out-degree for the incremental marking phase. This also helps minimize thread divergence with the thread-per-vertex kernel. The psuedocode for partitioning the vertices (by in- or out-degree of vertices) is given in Algorithm \ref{alg:partition}, with its explanation given in Section \ref{sec:partition}.

\paragraph{Marking the initial set of affected vertices:}

Upon each edge deletion $(u, v) \in \Delta^{t-}$ and insertion $(u, v) \in \Delta^{t+}$ in the batch update, with the DF and DF-P approaches, we need to mark the outgoing neighbors of vertex $u$ in both the previous $G^{t-1}$ and current graph $G^t$ as affected. This is equivalent to marking the outgoing neighbors of the source vertex, $u$, for each edge deletion and insertion $(u, v) \in \Delta^{t-} \cup \Delta^{t+}$, and marking the target vertex, $v$, of each edge deletion $(u, v) \in \Delta^{t-}$ as affected. We use one kernel to mark the target vertex, $v$, for each edge deletion $(u, v) \in \Delta^{t-}$ as affected, and use a temporary array to indicate that the outgoing neighbors of the source vertex, $u$, for each edge deletion and insertion $(u, v) \in \Delta^{t-} \cup \Delta^{t+}$ need to be marked as affected later. A thread-per-vertex kernel and block-per-vertex kernel are then used to actually mark the outgoing neighbors of such vertices as affected. The vertex IDs between the two kernels are partitioned by out-degree, as mentioned above.

\paragraph{Rank computation and incremental expansion / contraction of the set of affected vertices:}

Similar to our Static PageRank, we adopt a synchronous implementation of the PageRank algorithm, employing a kernel pair during each iteration's rank computation phase. In addition, with the DF and DF-P approaches, if the relative change in the rank of a vertex $u$ is greater than the frontier tolerance $\tau_f$, we need to incrementally mark the outgoing neighbors $v \in G^t.out(u)$ of $u$ as affected \cite{sahu2024df}. However, performing this incremental marking during the rank computation phase may introduce significant thread divergence (the work is proportional to the out-degree of $u$, and not its in-degree). To avoid this, we use a temporary array to indicate that the neighbors of $u$ need to be incrementally marked as affected. This can later be used with a pair of kernels to actually mark the outgoing neighbors of such vertices as affected.

With the DF-P approach, if the relative change in the rank of a vertex $u$ falls within the prune tolerance $\tau_p$, the vertex $u$ needs to be marked as not affected \cite{sahu2024df}. This is achieved by directly unflagging it in the set of affected vertices, which entails $O(1)$ work and does not introduce significant thread divergence. Additionally, considering the potential pruning of vertices and the incorporation of a self-loop for each vertex in the graph (as discussed in Sections \ref{sec:dataset} and \ref{sec:batch-generation}), we employ a closed-loop formula for computing the rank of each vertex (Equation \ref{eq:pr-prune}). This formula is specifically designed to accommodate the presence of the self-loop, thereby mitigating the need for recursive rank calculations stemming from it \cite{sahu2024df}. We refer the reader to our prior work \cite{sahu2024df} for an example of the DF and DF-P approaches.

\paragraph{Convergence detection:}

This is similar to that of Static PageRank.

\begin{flalign}
\label{eq:pr-prune}
  R[v] & = \frac{1}{1 - \alpha / |G.out(v)|} \left(\alpha K + \frac{1 - \alpha}{|V|}\right) && \\
    \text{where, } K & = \left(\sum_{u \in G.in(v)} \frac{R[u]}{|G.out(u)|}\right) - \frac{R[v]}{|G.out(v)|}
\end{flalign}

\ignore{\subsubsection{A simple example}}

\ignore{Figure \ref{fig:about-frontier} provides an illustration of DF and DF-P PageRank. Initially, as demonstrated in Figures \ref{fig:about-frontier-df1} and \ref{fig:about-frontier-dfp1}, the graph consists of $16$ vertices and $23$ edges. Subsequent to this, Figures \ref{fig:about-frontier-df2} and \ref{fig:about-frontier-dfp2} display a batch update executed on the original graph, entailing an edge insertion from vertex $4$ to $12$ and an edge deletion from vertex $2$ to $1$. Post-batch update, we proceed with the initial phase of DF/DF-P PageRank, wherein we identify and mark the outgoing neighbors of vertices $2$ and $4$ as affected, specifically vertices $1$, $8$, $12$, and $14$. These affected vertices are distinguished with a yellow fill. It is noteworthy that vertices $2$ and $4$ remain unmarked as affected. This is due to the fact that changes in a vertex's out-degree do not influence its PageRank score (refer to Equation \ref{eq:pr}). Subsequently, we commence the first iteration of the PageRank algorithm.}

\ignore{In the first iteration (depicted in Figures \ref{fig:about-frontier-df3} and \ref{fig:about-frontier-dfp3}), the ranks of affected vertices undergo updating. Now, say the relative change in rank of vertices $1$, $8$, $12$, and $14$ surpasses the frontier tolerance $\tau_f$ --- these vertices are highlighted with a red border in the figures. Consequently, in both DF and DF-P PageRank, we proceed to incrementally mark the outgoing neighbors of vertices $1$, $8$, $12$, and $14$ as affected. Specifically, vertices $3$, $5$, $9$, $10$, $14$, and $15$ are identified as such.}

\ignore{\input{src/fig-about-frontier}}

\ignore{In the second iteration, as illustrated in Figures \ref{fig:about-frontier-df4} and \ref{fig:about-frontier-dfp4}, another round of updates is applied to the ranks of the impacted vertices. Notably, the ranks of vertices $3$, $5$, $9$, $14$, and $15$ exhibit a relative change exceeding the designated frontier tolerance $\tau_f$. Consequently, employing DF/DF-P PageRank, we identify the outgoing neighbors of these vertices, namely vertices $4$, $6$, $10$, $15$, and $16$, as affected. Conversely, the relative change in rank of vertices $1$, $8$, and $12$ remains below the prune tolerance threshold $\tau_p$. Consequently, utilizing DF-P PageRank, these vertices are no longer classified as affected, indicating a probable convergence of their ranks. This action effectively contracts the frontier of affected vertices. However, if a vertex's rank has not yet converged, it might be re-designated as affected by one of its in-neighbors. Subsequently, in the ensuing iteration, the ranks of affected vertices undergo further updates. Should the change in rank for each vertex fall within the defined iteration tolerance $\tau$\ignore{(we use $L\infty$-norm for convergence detection)}, it signifies convergence of the ranks, and the algorithm halts.}

\ignore{\paragraph{Contrasting with Dynamic Traversal (DT) PageRank:}}

\ignore{We now compare DF and DF-P PageRank with DT PageRank (see Figures \ref{fig:about-frontier-dt1}-\ref{fig:about-frontier-dt4}). In Figure \ref{fig:about-frontier-dt2}, the identical batch update applied to the original graph is depicted, akin to Figures \ref{fig:about-frontier-df2} and \ref{fig:about-frontier-dfp2}. In response to this update, DT PageRank designates all vertices reachable from $2$ and $4$ as affected, i.e., all vertices except $7$, $11$, and $13$. Subsequently, the ranks of this subset of affected vertices undergo updates in each iteration\ignore{(while the ranks of unaffected vertices remain unchanged)}, continuing until convergence is achieved.}

\subsection{Determining suitable Partitioning approach}
\label{sec:parition-determine}

In order to optimize the performance of DF and DF-P PageRank for the GPU, we attempt three different partitioning techniques for work distribution between the thread-per-vertex and block-per-vertex kernels --- for updating ranks of vertices in the graph and incrementally marking affected vertices.

With the first technique, which we refer to as \textit{Don't Partition}, we do not partition the graph, and instead selectively execute the thread/block-per-vertex kernels on each vertex, depending on the in/out-degree of the vertex --- for both the rank computation phase and the incremental marking of affected vertices. With the second technique, which we refer to as \textit{Partition $G'$} ($G'$ stands for transpose of $G$, the current graph), we partition the graph into low in-degree and high in-degree vertices, and run the kernels on respective partitions for updating ranks --- the incremental marking of affected vertices is still done selectively. With the third technique, which we refer to as \textit{Partition $G$, $G'$}, we partition the graph by both in-degrees and out-degrees, and run the kernels of respective partitions (i.e., thread-per-vertex kernel on low degree vertices, and block-per-vertex kernel on high degree vertices) for both rank computation and incremental marking of affected vertices.

Figure \ref{fig:adjust-partition} illustrates the relative runtime with each partitioning technique. Here, the measured runtime includes the overall runtime of the DF/DF-P PageRank, and not just the time needed for partitioning the vertices, or performing rank computation. Results indicate that the \textit{Partition $G$, $G'$} technique performs the best, as shown in Figure \ref{fig:adjust-partition}. Note that partitioning the vertex IDs by out-degree, i.e., \textit{Partition $G$}, is useful for incremental marking of affected vertices, but comes with added runtime cost --- hence the small improvement in performance when moving from \textit{Partition $G'$} to \textit{Partition $G$, $G'$}.

\input{src/fig-adjust-partition}

\subsection{Our DF* PageRank implementation}

Algorithm \ref{alg:frontier} presents the psuedocode of GPU-based DF and DF-P PageRank, which aims to efficiently compute PageRank on large-scale graphs with dynamic updates. The algorithm takes several inputs, including the current graph snapshot $G^t$ and its transpose $G^{t'}$, edge deletions $\Delta^{t-}$ and insertions $\Delta^{t+}$, and the previous rank vector $R^{t-1}$. It returns the updated rank vector $R$.

We start by initializing the rank vectors $R$ and $R_{new}$ with the previous rank vector $R^{t-1}$ (line \ref{alg:frontier--initialize}). Next, we partition the vertex IDs based on their out- and in-degree, aiming for efficient computation on the GPU (lines \ref{alg:frontier--partition-begin}-\ref{alg:frontier--partition-end}). Then, we mark the initial set of affected vertices using the provided edge deletions and insertions, and expand the affected set to include relevant neighbor vertices (lines \ref{alg:frontier--mark-begin}-\ref{alg:frontier--mark-end}). Subsequently, we start with PageRank iterations (lines \ref{alg:frontier--compute-begin}-\ref{alg:frontier--compute-end}), continuing until either the maximum number of iterations $MAX\_ITERATIONS$ is reached or the change in ranks falls below the specified tolerance $\tau$ (line \ref{alg:frontier--converged}). Within each iteration, we update the rank vector $R_{new}$ based on the affected vertices (line \ref{alg:frontier--update}), while marking the vertices whose neighbors must be incrementally marked as affected (with relative change in rank greater than the frontier tolerance $\tau_f$) or contracting the set of affected vertices if the change in rank of a vertex is small (with relative change in rank below the prune tolerance $\tau_p$, only with DF-P PageRank). The $L_\infty$-norm difference between the current $R_{new}$ and previous ranks $R$ is then computed to check for convergence, and the rank vectors are swapped for the next iteration (line \ref{alg:frontier--error}). In line \ref{alg:frontier--converged}, we perform a convergence check. If convergence has not yet been achieved, we incrementally expand the set of affected vertices (line \ref{alg:frontier--remark}) from the vertices identified during rank computation (line \ref{alg:frontier--update}). Finally, we return the updated rank vector $R$ (line \ref{alg:frontier--return})\ignore{, providing the computed PageRank scores for each vertex in the graph after dynamic updates}. The details for \texttt{updateRanks()}, \texttt{partition()}, and \texttt{initialAffected()}/\texttt{expandAffected()} functions is given in Sections \ref{sec:update}, \ref{sec:partition}, and \ref{sec:affected}, respectively.

Algorithm \ref{alg:frontier} also uses a pull-based synchronous implementation of PageRank, similar to Algorithm \ref{alg:static}. This is in contrast to our multicore CPU implementation of DF and DF-P PageRank, where we observe that an asynchronous implementation offers better performance \cite{sahu2024df, sahu2024incrementally}. We also utilize synchronous implementations for Naive-dynamic (ND) and Dynamic Traversal (DT) PageRank.

\input{src/alg-frontier}

%% file: src/alg-static.tex
\begin{algorithm}[!hbt]
\caption{Our GPU-based Static PageRank.}
\label{alg:static}
\begin{algorithmic}[1]
\Require{$G^{t'}(V^{t'}, E^{t'})$: Transpose of current input graph}
\Require{$R, R_{new}$: Rank vector in the previous, current iteration}
\Ensure{$P'$: Partitioned vertex IDs, low in-degree first }
\Ensure{$N'_P$: Number of vertices with low in-degree}
\Ensure{$\Delta R$: $L\infty$-norm between previous and current ranks}
\Ensure{$\tau$: Iteration tolerance}

\Statex

\Function{static}{$G^{t'}$}
  \State $\rhd$ Initialize ranks
  \State $R \gets R_{new} \gets \{\}$ \label{alg:static--initialize-begin}
  \ForAll{$v \in V^{t'}$ \textbf{in parallel}}
    \State $R[v] \gets R_{new}[v] \gets 1/|V^{t'}|$
  \EndFor \label{alg:static--initialize-end}
  \State $\rhd$ Partition vertex IDs by in-degree 
  \State $\{P', N'_P\} \gets partition(G^{t'})$ \label{alg:static--partition} \Comment{Alg. \ref{alg:partition}}
  \State $\rhd$ Perform PageRank iterations
  \ForAll{$i \in [0 .. MAX\_ITERATIONS)$} \label{alg:static--compute-begin}
    \State $updateRanks(\cdots, \cdots, R_{new}, R, G^{t'}, P', N'_P)$ \Comment{Alg. \ref{alg:update}} \label{alg:static--update}
    \State $\Delta R \gets l_{\infty}NormDelta(R_{new}, R)$ \textbf{;} $swap(R_{new}, R)$ \label{alg:static--error}
    \If{$\Delta R \leq \tau$} \textbf{break} \label{alg:static--converged}
    \EndIf
  \EndFor \label{alg:static--compute-end}
  \State \ReturnInline{$R$} \label{alg:static--return}
\EndFunction
\end{algorithmic}
\end{algorithm}

%% file: src/fig-about-frontier.tex
\begin{figure*}[hbtp]
  \centering
  \subfigure[Initial graph]{
    \label{fig:about-frontier-df1}
    \includegraphics[width=0.23\linewidth]{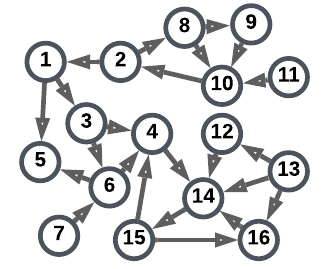}
  }
  \subfigure[Marking initial affected vertices (DF)]{
    \label{fig:about-frontier-df2}
    \includegraphics[width=0.23\linewidth]{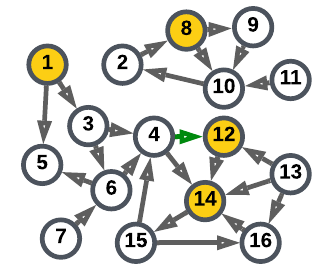}
  }
  \subfigure[After first iteration (DF)]{
    \label{fig:about-frontier-df3}
    \includegraphics[width=0.23\linewidth]{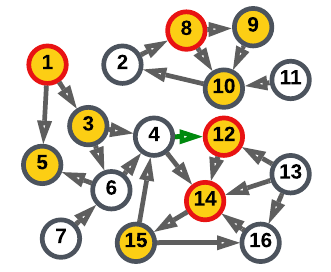}
  }
  \subfigure[After second iteration (DF)]{
    \label{fig:about-frontier-df4}
    \includegraphics[width=0.23\linewidth]{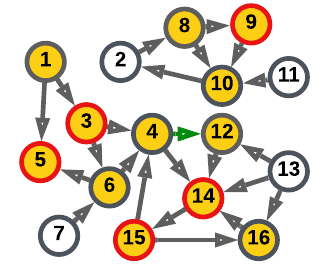}
  } \\[2ex]
  \subfigure[Initial graph]{
    \label{fig:about-frontier-dfp1}
    \includegraphics[width=0.23\linewidth]{out/about-frontier-11.pdf}
  }
  \subfigure[Marking initial affected vertices (DF-P)]{
    \label{fig:about-frontier-dfp2}
    \includegraphics[width=0.23\linewidth]{out/about-frontier-32.pdf}
  }
  \subfigure[After first iteration (DF-P)]{
    \label{fig:about-frontier-dfp3}
    \includegraphics[width=0.23\linewidth]{out/about-frontier-33.pdf}
  }
  \subfigure[After second iteration (DF-P)]{
    \label{fig:about-frontier-dfp4}
    \includegraphics[width=0.23\linewidth]{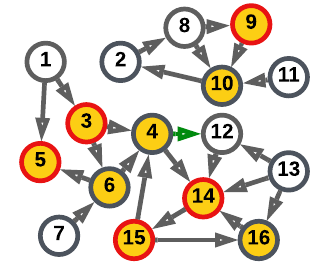}
  } \\[2ex]
  \subfigure[Initial graph]{
    \label{fig:about-frontier-dt1}
    \includegraphics[width=0.23\linewidth]{out/about-frontier-11.pdf}
  }
  \subfigure[Marking affected vertices (DT)]{
    \label{fig:about-frontier-dt2}
    \includegraphics[width=0.23\linewidth]{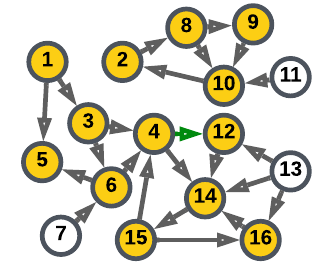}
  }
  \subfigure[After first iteration (DT)]{
    \label{fig:about-frontier-dt3}
    \includegraphics[width=0.23\linewidth]{out/about-frontier-22.pdf}
  }
  \subfigure[After second iteration (DT)]{
    \label{fig:about-frontier-dt4}
    \includegraphics[width=0.23\linewidth]{out/about-frontier-22.pdf}
  } \\[-2ex]
  \caption{An illustration of our \textit{Dynamic Frontier (DF)} and \textit{Dynamic Frontier with Pruning (DF-P)} approaches, in subfigures (a)-(d) and (e)-(h) respectively, compared against the \textit{Dynamic Traversal (DT)} approach, shown in subfigures (i)-(l) \cite{sahu2024df}.}
  \label{fig:about-frontier}
\end{figure*}

%% file: src/fig-adjust-partition.tex
\begin{figure}[!hbt]
  \centering
  \subfigure{
    \label{fig:adjust-partition--all}
    \includegraphics[width=0.98\linewidth]{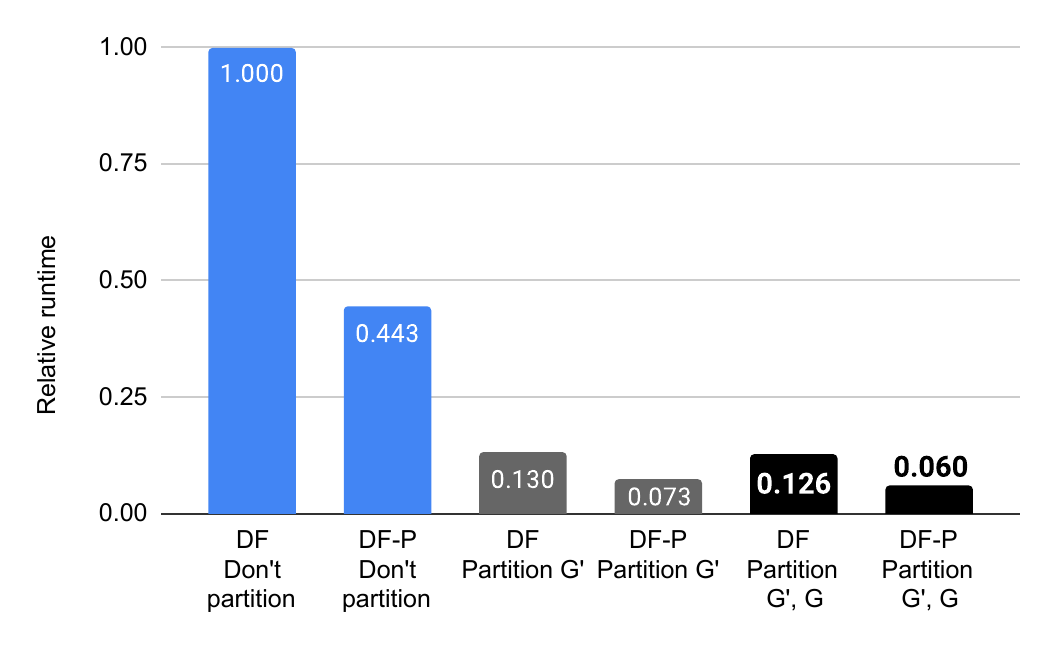}
  } \\[-2ex]
  \caption{Mean relative runtime with our \textit{Dynamic Frontier (DF)} and \textit{Dynamic Frontier with Pruning (DF-P)} approaches across three different levels of work-partitioning for GPU computation. Here, \textit{Partition $G$} denotes partitioning the vertices of the current graph $G$ by their out-degree, while \textit{Partition $G'$} signifies partitioning the vertices by their in-degree. Note that $G'$ stands for the transpose of the current graph $G$.}
  \label{fig:adjust-partition}
\end{figure}

%% file: src/alg-frontier.tex
\begin{algorithm}[!hbt]
\caption{Our GPU-based Dynamic Frontier (DF*) PageRank.}
\label{alg:frontier}
\begin{algorithmic}[1]
\Require{$G^t(V^t, E^t), G^{t'}$: Current input graph, and its transpose}
\Require{$\Delta^{t-}, \Delta^{t+}$: Edge deletions and insertions (input)}
\Require{$R^{t-1}$: Previous rank vector}
\Require{$R, R_{new}$: Rank vector in the previous, current iteration}
\Ensure{$\delta_V, \delta_N$: Is a vertex, or neighbors of a vertex affected}
\Ensure{$P, P'$: Partitioned vertex IDs --- low out-, in-degree first }
\Ensure{$N_P, N'_P$: Number of vertices with low out-, in-degree}
\Ensure{$\Delta R$: $L\infty$-norm between previous and current ranks}
\Ensure{$\tau$: Iteration tolerance}

\Statex

\Function{dynamicFrontier}{$G^t, G^{t'}, \Delta^{t-}, \Delta^{t+}, R^{t-1}$}
  \State $\rhd$ Initialize ranks
  \State $R \gets R_{new} \gets R^{t-1}$ \label{alg:frontier--initialize}
  \State $\rhd$ Partition vertex IDs by out- and in-degree 
  \State $\{P, N_P\} \gets partition(G^t)$ \label{alg:frontier--partition-begin} \Comment{Alg. \ref{alg:partition}}
  \State $\{P', N'_P\} \gets partition(G^{t'})$ \label{alg:frontier--partition-end} \Comment{Alg. \ref{alg:partition}}
  \State $\rhd$ Mark initial set of affected vertices \label{alg:frontier--mark-begin}
  \State $\{\delta_V, \delta_N\} \gets initialAffected(G^t, \Delta^{t-}, \Delta^{t+})$ \Comment{Alg. \ref{alg:affected}}
  \State $expandAffected(\delta_V, \delta_N, G^t, P, N_P)$ \label{alg:frontier--mark-end} \Comment{Alg. \ref{alg:affected}}
  \State $\rhd$ Perform PageRank iterations
  \ForAll{$i \in [0 .. MAX\_ITERATIONS)$} \label{alg:frontier--compute-begin}
    \State $\delta_N \gets \{\}$
    \State $updateRanks(\delta_V, \delta_N, R_{new}, R, G^t, P', N'_P)$ \Comment{Alg. \ref{alg:update}} \label{alg:frontier--update}
    \State $\Delta R \gets l_{\infty}NormDelta(R_{new}, R)$ \textbf{;} $swap(R_{new}, R)$ \label{alg:frontier--error}
    \If{$\Delta R \leq \tau$} \textbf{break} \label{alg:frontier--converged}
    \EndIf
    \State $expandAffected(\delta_V, \delta_N, G^t, P, N_P)$ \Comment{Alg. \ref{alg:affected}} \label{alg:frontier--remark}
  \EndFor \label{alg:frontier--compute-end}
  \State \ReturnInline{$R$} \label{alg:frontier--return}
\EndFunction
\end{algorithmic}
\end{algorithm}

%% file: 05-evaluation.tex
\subsection{Experimental Setup}
\label{sec:setup}

\subsubsection{System used}

Experiments are performed on a system featuring an NVIDIA A100 GPU with $108$ SMs and $64$ CUDA cores per SM. The GPU has $80$ GB of global memory, with a bandwidth of $1935$ GB/s. Each SM has a shared memory capacity of $164$ KB. The system also has an AMD EPYC-7742 processor with $64$ cores, running at a frequency of $2.25$ GHz.\ignore{Each core has $4$ MB L1 cache, a $32$ MB L2 cache, and shares a $256$ MB L3 cache.} The server is set up with $512$ GB of DDR4 system memory and runs Ubuntu $20.04$.

\subsubsection{Configuration}

We utilize 32-bit integers to represent vertex IDs and 64-bit floating-point numbers for vertex ranks. Affected vertices are denoted by an 8-bit integer vector. Our configuration sets the damping factor to $\alpha = 0.85$ \cite{rank-langville06}, the frontier tolerance $\tau_f$ to $10^{-6}$ \cite{sahu2024df}, the prune tolerance $\tau_p$ to $10^{-6}$ \cite{sahu2024df}, and the iteration tolerance $\tau$ to $10^{-10}$ using the $L_\infty$-norm \cite{rank-dubey22, rank-plimpton11}. We cap the maximum number of iterations $MAX\_ITERATIONS$ at $500$ \cite{nvgraph}. Compilation is performed using NVCC, a part of CUDA toolkit $11.4$, and OpenMP $5.0$ (for CPU code). OpenMP's \textit{dynamic schedule} with a chunk size of $2048$ is used for CPU-based computations to facilitate dynamic workload balancing among threads.

\subsubsection{Dataset}
\label{sec:dataset}

To experiment with real-world dynamic graphs, we employ five temporal networks sourced from the Stanford Large Network Dataset Collection \cite{snapnets}. Details of these graphs are summarized in Table \ref{tab:dataset}. Here, the number of vertices range from $24.8$ thousand to $2.60$ million, temporal edges from $507$ thousand to $63.4$ million, and static edges from $240$ thousand to $36.2$ million. For investigations involving large static graphs with random batch updates, we utilize $12$ graphs as listed in Table \ref{tab:dataset}, obtained from the SuiteSparse Matrix Collection \cite{suite19}. Here, the number of vertices range from $3.07$ to $214$ million, and edges from $25.4$ million to $3.80$ billion. To mitigate dead ends (vertices without out-links) and the associated overhead of global teleport rank computation in each iteration, we augment all vertices with self-loops\ignore{within the graph} \cite{kolda2009generalized, rank-andersen07, rank-langville06}.

\input{src/tab-dataset}
\input{src/tab-dataset-large}

\subsubsection{Batch Generation}
\label{sec:batch-generation}

For experiments involving real-world dynamic graphs, we first load $90\%$ of each graph from Table \ref{tab:dataset}, and add self-loops to all vertices. Subsequently, we load $B$ edges in $100$ batch updates. Here, $B$ denotes the desired batch size, specified as a fraction of the total number of temporal edges $|E_T|$ in the graph.\ignore{Additionally, self-loops are added to all vertices with each batch update.} For experiments on large graphs with random batch updates, we select each base (static) graph from Table \ref{tab:dataset-large} and generate a random batch update comprising an $80\% : 20\%$ mix of edge insertions and deletions to emulate realistic batch updates. To prepare the set of edges for insertion, we select vertex pairs with equal probability. For edge deletions, we uniformly delete each existing edge. To simplify, we ensure no new vertices are added or removed from the graph. The batch size is measured as a fraction of edges in the original graph, ranging from $10^{-7}$ to $0.1$ (i.e., $10^{-7}|E|$ to $0.1|E|$), with multiple batches generated for each size for averaging. Self-loops are added to all vertices alongside each batch update, as\ignore{mentioned} earlier.

\subsubsection{Measurement}
\label{sec:measurement}

We assess the runtime of each approach on the entire updated graph, including partitioning, initial marking of affected vertices, and convergence detection time. However, we exclude the time taken for memory transfers (to and from the GPU) as well as allocation/deallocation. The mean time and error for a specific method at a given batch size are computed as the geometric mean across input graphs. Additionally, we evaluate the error/accuracy of each approach by measuring the $L1$-norm \cite{ohsaka2015efficient} of the returned ranks compared to ranks obtained from a reference Static PageRank run on the updated graph with an extremely low iteration tolerance of $\tau = 10^{-100}$, limited to $500$ iterations.

\subsection{Performance of Our Static PageRank}
\label{sec:static-comparison}

\input{src/fig-compare}

We now evaluate the performance of our GPU implementation of Static PageRank, and compare it with the performance of Static PageRank in the Hornet \cite{busato2018hornet} and Gunrock \cite{wang2016gunrock} graph processing frameworks on large (static) graphs from Table \ref{tab:dataset-large}. For Hornet, we use a CUDA C++ program to read each input graph with \texttt{GraphStd::re} \texttt{ad()}, perform \texttt{HornetInit}, create a \texttt{HornetGraph}, and set up \texttt{Stati} \texttt{cPageRank} with a damping factor $\alpha$ of $0.85$, and iteration tolerance $\tau$ of $10^{-10}$, and limit the maximum number of iterations to $500$. We also define a new PageRank operator called \texttt{Max}, to compute $L_\infty$-norm of the absolute difference between the previous and current rank vectors (since we use $L_\infty$-norm for convergence detection), and use this for convergence detection (with \texttt{forAllnumV()}) instead of $L1$-norm (used by default). To perform the PageRank computation, we then use \texttt{StaticPageRank::run()}, and measure its runtime with \texttt{Timer<DEVICE>}. For Gunrock, we use a CUDA C++ program to read each input graph in Table \ref{tab:dataset-large} with \texttt{io::matrix\_market\_t::load()}, convert it to a CSR representation with \texttt{format::csr\_t::from\_coo()}, and build a graph with \texttt{graph::build::from\_csr()}. We then perform PageRank computation with \texttt{gunrock::pr::run()} upon the loaded graph with a damping factor $\alpha$ of $0.85$, an iteration tolerance of $10^{-10}$, and limit the number of iterations to $500$ by modifying the \texttt{gunrock::pr::en} \texttt{actor\_t::is\_} \texttt{converged()} function (note that Gunrock uses $L_\infty$-norm for convergence detection by default), and record the runtime reported by \texttt{gunrock::pr::run()}.\ignore{Neither Hornet nor Gunrock offer GPU implementation of dynamic PageRank algorithms.}

Figure \ref{fig:compare--runtime} illustrates the runtime of Hornet, Gunrock, and our Static PageRank on the GPU, for each graph in the dataset. On the \textit{sk-2005} graph, our Static PageRank computes the ranks of vertices with an iteration tolerance $\tau$ of $10^{-10}$ in $4.2$ seconds, achieving a processing rate of $471$ million edges/s. Figure \ref{fig:compare--speedup} shows the speedup of Our Static PageRank with respect to Hornet and Gunrock. Our Static PageRank is on average $31\times$ faster than Hornet, and $5.9\times$ times faster than Gunrock. This speedup is particularly high on the \textit{webbase-2001} graph and road networks with Hornet, and on the \textit{indochina-2004} graph with Gunrock. Further, our GPU implementation of Static PageRank is on average $24\times$ times faster than our parallel multicore implementation of Static PageRank.

\subsection{Performance of Our DF-P PageRank}

\subsubsection{Results on real-world dynamic graphs}

We now compare the performance of our GPU implementation of Dynamic Frontier (DF) and Dynamic Frontier with Pruning (DF-P) PageRank with Static, Naive-dynamic (ND), and Dynamic Traversal (DT) PageRank on real-world dynamic graphs from Table \ref{tab:dataset}. This is done on batch updates of size $10^{-5}|E_T|$ to $10^{-3}|E_T|$ in multiples of $10$. For each batch size, as mentioned in Section \ref{sec:batch-generation}, we load $90\%$ of the graph, add self-loops to all vertices in the graph, and then load $B$ edges (where $B$ is the batch size) consecutively in $100$ batch updates. Figure \ref{fig:temporal-summary--runtime-overall} displays the overall runtime of each approach across all graphs for each batch size, while Figure \ref{fig:temporal-summary--error-overall} illustrates the overall rank error compared to a reference Static PageRank run (as described in Section \ref{sec:measurement}). Additionally, Figures \ref{fig:temporal-summary--runtime-graph} and \ref{fig:temporal-summary--error-graph} present the mean runtime and rank error of the approaches on each dynamic graph in the dataset. Figure \ref{fig:temporal-compare} presents a comparison of the overall runtime and error between the GPU implementation of each approach and its respective CPU counterpart. Finally, Figures \ref{fig:temporal-sx-mathoverflow}, \ref{fig:temporal-sx-askubuntu}, \ref{fig:temporal-sx-superuser}, \ref{fig:temporal-wiki-talk-temporal}, and \ref{fig:temporal-sx-stackoverflow} show the runtime and rank error of the approaches on each dynamic graph in Table \ref{tab:dataset}, upon each consecutive batch update.

\input{src/fig-temporal-summary}
\input{src/fig-8020-runtime}
\input{src/fig-8020-error}

Figure \ref{fig:temporal-summary--runtime-overall} shows that DF PageRank is, on average, $1.4\times$ faster than Static PageRank for batch updates of size $10^{-5}|E_T|$. In contrast, DF-P PageRank demonstrates average speedups of $3.6\times$, $2.0\times$, and $1.3\times$ over Static PageRank for batch update sizes of $10^{-5}|E_T|$, $10^{-4}|E_T|$, and $10^{-3}|E_T|$, respectively. Furthermore, DF-P PageRank achieves average speedups of $4.2\times$, $2.8\times$, and $3.6\times$ compared to DT PageRank for the same batch updates. This speedup is particularly pronounced on the \textit{sx-mathoverflow} graph, as indicated by Figure \ref{fig:temporal-summary--runtime-graph}. Regarding rank error, Figures \ref{fig:temporal-summary--error-overall} and \ref{fig:temporal-summary--error-graph} illustrate that DF and DF-P PageRank generally exhibit higher error on average compared to ND and DT PageRank but lower error than Static PageRank. This makes the ranks obtained with DF and DF-P PageRank acceptable. DF-P PageRank can thus be the default choice for updating PageRank scores on dynamic graphs. However, if elevated error levels are observed (during intermediate empirical tests), transitioning to ND PageRank is advisable.

\subsubsection{Results on large graphs with random updates}

We also assess the performance of our GPU implementation of DF and DF-P PageRank alongside Static, ND, and DT PageRank on large (static) graphs listed in Table \ref{tab:dataset-large}, with randomly generated batch updates. As elaborated in Section \ref{sec:batch-generation}, the batch updates vary in size from $10^{-7}|E|$ to $0.1|E|$ (in multiples of $10$), comprising $80\%$ edge insertions and $20\%$ edge deletions to mimic realistic scenarios. Self-loops are added to all vertices with each batch update. Figure \ref{fig:8020-runtime} illustrates the runtime of Static, ND, DT, DF, and DF-P PageRank, while Figure \ref{fig:8020-error} displays the error in ranks obtained with each approach. Figures \ref{fig:8020-runtime-compare} and \ref{fig:8020-error-compare} depict a comparison of the overall runtime and error between the GPU and CPU implementation of each approach.

Figure \ref{fig:8020-runtime--mean} illustrates that for batch updates ranging from $10^{-7}|E|$ to $10^{-4}|E|$, DF PageRank achieves an average speedup of $2.5\times$, $1.3\times$, and $10.4\times$ compared to Static, ND, and DT PageRank, respectively. Further, DF-P PageRank achieves average speedups of $3.1\times$, $1.7\times$, and $13.1\times$ over Static, ND, and DT PageRank, respectively. This acceleration is particularly pronounced on road networks and protein k-mer graphs, characterized by a low average degree (as depicted in Figure \ref{fig:8020-runtime--all}). It may be noted that DT PageRank exhibits slower performance than ND PageRank on large graphs with uniformly random batch updates. This is attributed to DT PageRank marking a significant number of vertices as affected, as updates are scattered randomly across the graph, rendering most of the graph reachable from the updated regions \cite{sahu2024incrementally}. Particularly on road networks and protein k-mer graphs, characterized by a low average degree and a large diameter, DT PageRank's performance is further hindered due to limited parallelism exploitable by the GPU. Figures \ref{fig:8020-error--mean} and \ref{fig:8020-error--all} indicate that DF-P PageRank generally exhibits higher error compared to ND, DT, and DF PageRank, but lower error than Static PageRank (up to a batch size of $10^{-4}|E|$). Hence, we recommend utilizing DF-P PageRank for large random batch updates up to a batch size of $10^{-4}|E|$. For larger batch updates, we advise the reader to switch to ND PageRank instead.

\ignore{\subsubsection{Comparison of vertices marked as affected}}

\ignore{Figure \ref{fig:measure-affected} displays the (mean) percentage of vertices marked as affected by Dynamic Traversal (DT), our improved Dynamic Frontier (DF), and Dynamic Frontier with Pruning (DF-P) PageRank on real-world dynamic graphs from Table \ref{tab:dataset}. This analysis is conducted on batch updates of size $10^{-5}|E_T|$ to $10^{-3}|E_T|$ in multiples of $10$ (see Section \ref{sec:batch-generation} for details). For DF and DF-P PageRank, affected vertices are marked incrementally --- therefore, we count all vertices that were ever flagged as affected.}

\ignore{As Figure \ref{fig:measure-affected} indicates, the proportion of vertices marked as affected by DF and DF-P PageRank is lower than DT PageRank for batch updates of size $10^{-5}|E_T|$, but comparable for larger batch updates. Therefore, the performance improvement with DF and DF-P PageRank is primarily attributed to the incremental marking of affected vertices. Additionally, it's worth noting that the percentage of vertices marked as affected is generally low across all approaches. This is likely because updates in real-world dynamic graphs tend to be concentrated in specific regions of the graph rather than being scattered throughout.}

\ignore{\input{src/fig-measure-affected}}

%% file: src/tab-dataset.tex
\begin{table}[hbtp]
  \centering
  \caption{List of $5$ real-world dynamic graphs\ignore{, i.e., temporal networks}, sourced from the Stanford Large Network Dataset Collection \cite{snapnets}. Here, $|V|$ denotes the number of vertices, $|E_T|$ represents the count of temporal edges (inclusive of duplicates), and $|E|$ indicates the number of static edges (without duplicates).}
  \label{tab:dataset}
  \begin{tabular}{|c||c|c|c|c|}
    \toprule
    \textbf{Graph} &
    \textbf{\textbf{$|V|$}} &
    \textbf{\textbf{$|E_T|$}} &
    \textbf{\textbf{$|E|$}} \\
    \midrule
    sx-mathoverflow & 24.8K & 507K & 240K \\ \hline
    sx-askubuntu & 159K & 964K & 597K \\ \hline
    sx-superuser & 194K & 1.44M & 925K \\ \hline
    wiki-talk-temporal & 1.14M & 7.83M & 3.31M \\ \hline
    sx-stackoverflow & 2.60M & 63.4M & 36.2M \\ \hline
  \bottomrule
  \end{tabular}
\end{table}

%% file: src/tab-dataset-large.tex
\begin{table}[hbtp]
  \centering
  \caption{List of $12$ graphs obtained from the SuiteSparse Matrix Collection \cite{suite19}, with directed graphs marked by $*$. Here, $|V|$ denotes the number of vertices, $|E|$ denotes the total number of edges (including self-loops), and $D_{avg}$ denotes the average degree of a vertex in each graph.}
  \label{tab:dataset-large}
  \begin{tabular}{|c||c|c|c|c|}
    \toprule
    \textbf{Graph} &
    \textbf{\textbf{$|V|$}} &
    \textbf{\textbf{$|E|$}} &
    \textbf{\textbf{$D_{avg}$}} \\
    \midrule
    \multicolumn{4}{|c|}{\textbf{Web Graphs (LAW)}} \\ \hline
    indochina-2004$^*$ & 7.41M & 199M & 26.8 \\ \hline  
    arabic-2005$^*$ & 22.7M & 654M & 28.8 \\ \hline  
    uk-2005$^*$ & 39.5M & 961M & 24.3 \\ \hline  
    webbase-2001$^*$ & 118M & 1.11B & 9.4 \\ \hline  
    it-2004$^*$ & 41.3M & 1.18B & 28.5 \\ \hline  
    sk-2005$^*$ & 50.6M & 1.98B & 39.1 \\ \hline  
    \multicolumn{4}{|c|}{\textbf{Social Networks (SNAP)}} \\ \hline
    com-LiveJournal & 4.00M & 73.4M & 18.3 \\ \hline  
    com-Orkut & 3.07M & 237M & 77.3 \\ \hline  
    \multicolumn{4}{|c|}{\textbf{Road Networks (DIMACS10)}} \\ \hline
    asia\_osm & 12.0M & 37.4M & 3.1 \\ \hline  
    europe\_osm & 50.9M & 159M & 3.1 \\ \hline  
    \multicolumn{4}{|c|}{\textbf{Protein k-mer Graphs (GenBank)}} \\ \hline
    kmer\_A2a & 171M & 531M & 3.1 \\ \hline  
    kmer\_V1r & 214M & 679M & 3.2 \\ \hline  
  \bottomrule
  \end{tabular}
\end{table}

%% file: src/fig-compare.tex
\begin{figure*}[hbtp]
  \centering
  \subfigure[Runtime in seconds (logarithmic scale) with \textit{Hornet}, \textit{Gunrock}, \textit{Our} Static PageRank.]{
    \label{fig:compare--runtime}
    \includegraphics[width=0.98\linewidth]{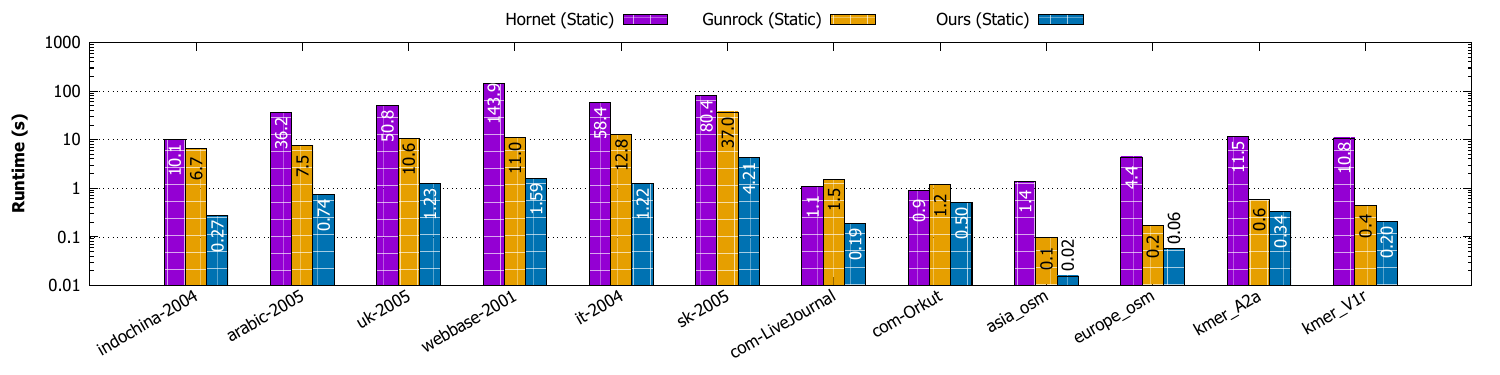}
  } \\[-0ex]
  \subfigure[Speedup of \textit{Our} Static PageRank (logarithmic scale) with respect to \textit{Hornet} and \textit{Gunrock}.]{
    \label{fig:compare--speedup}
    \includegraphics[width=0.98\linewidth]{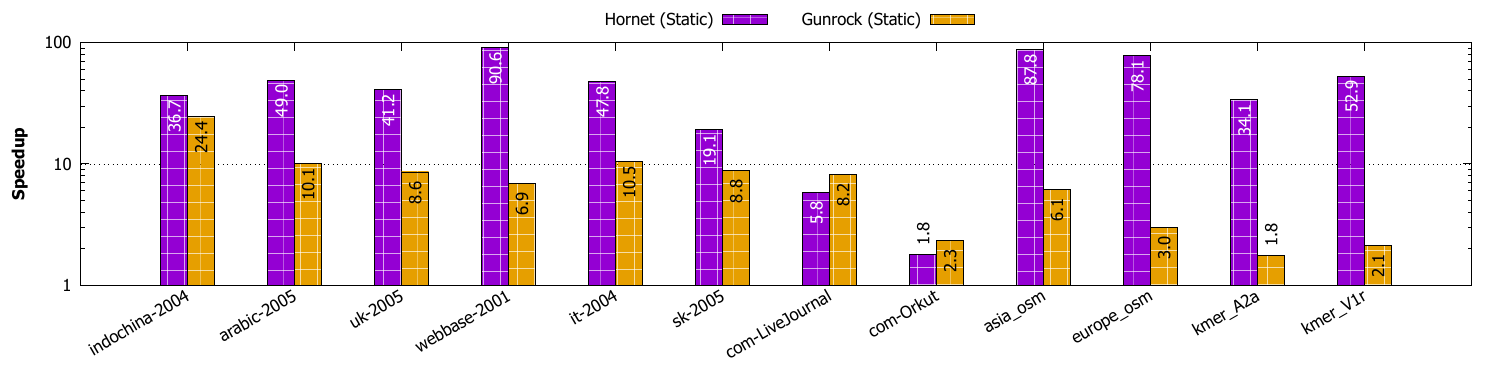}
  } \\[-2ex]
  \caption{Runtime in seconds and speedup (log-scale) with \textit{Hornet}, \textit{Gunrock}, \textit{Our} Static PageRank for each graph in the dataset.}
  \label{fig:compare}
\end{figure*}

%% file: src/fig-temporal-summary.tex
\begin{figure*}[!hbt]
  \centering
  \subfigure[Overall Runtime]{
    \label{fig:temporal-summary--runtime-overall}
    \includegraphics[width=0.48\linewidth]{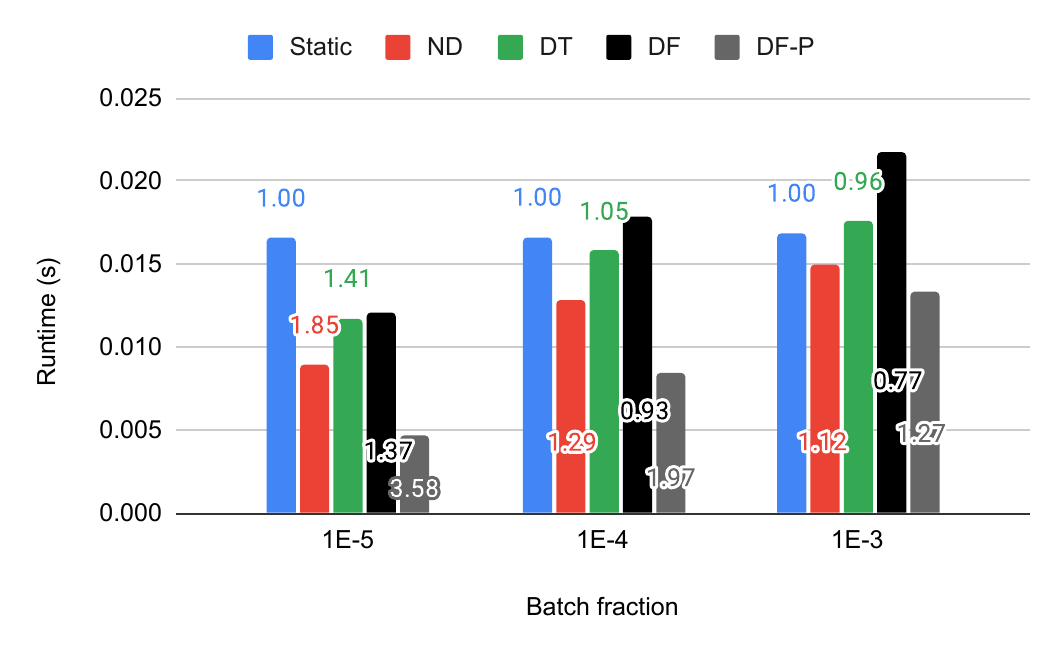}
  }
  \subfigure[Overall Error in ranks obtained]{
    \label{fig:temporal-summary--error-overall}
    \includegraphics[width=0.48\linewidth]{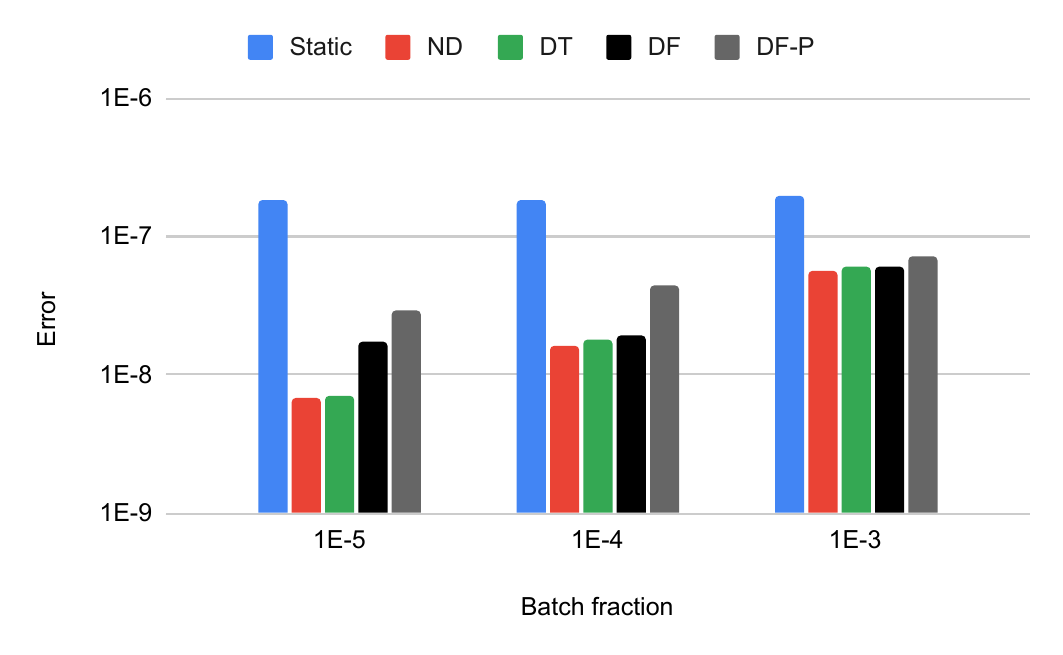}
  } \\[2ex]
  \includegraphics[width=0.48\linewidth]{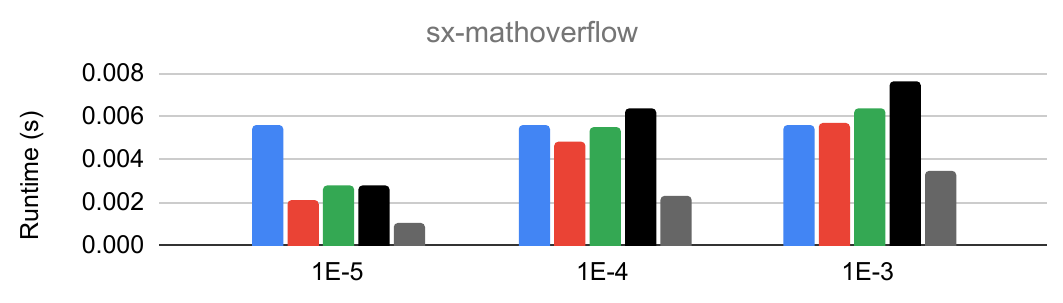}
  \includegraphics[width=0.48\linewidth]{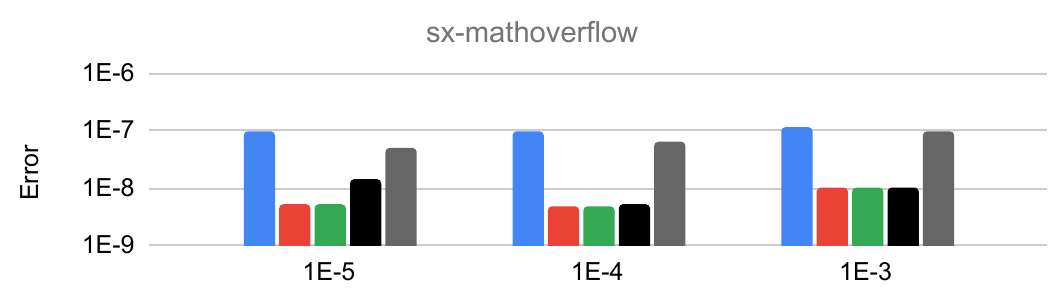}
  \includegraphics[width=0.48\linewidth]{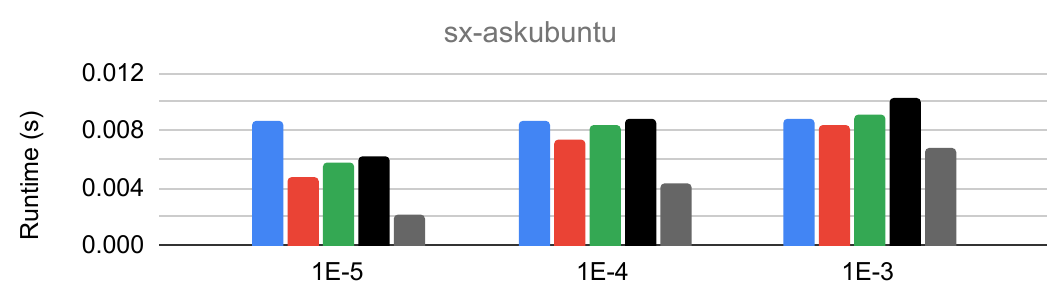}
  \includegraphics[width=0.48\linewidth]{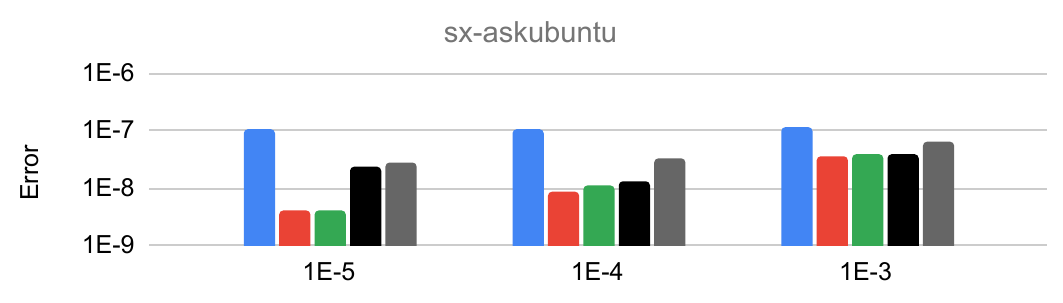}
  \includegraphics[width=0.48\linewidth]{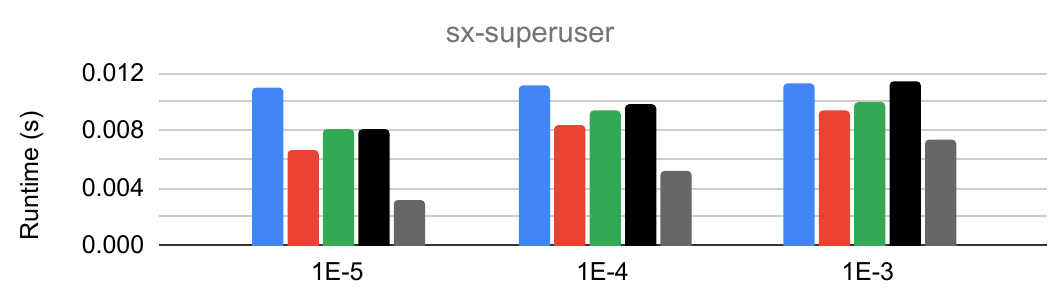}
  \includegraphics[width=0.48\linewidth]{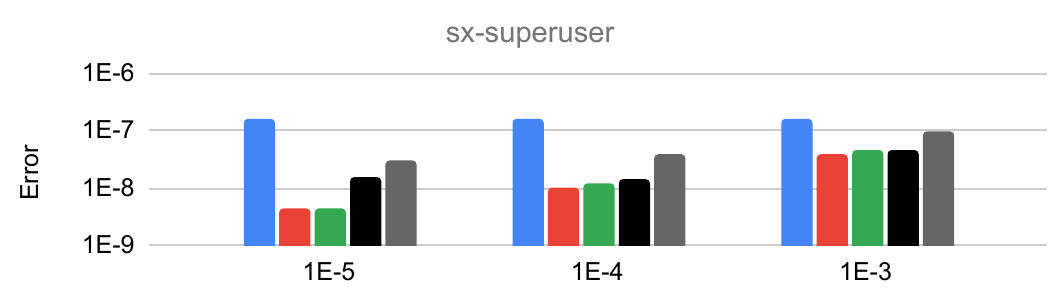}
  \includegraphics[width=0.48\linewidth]{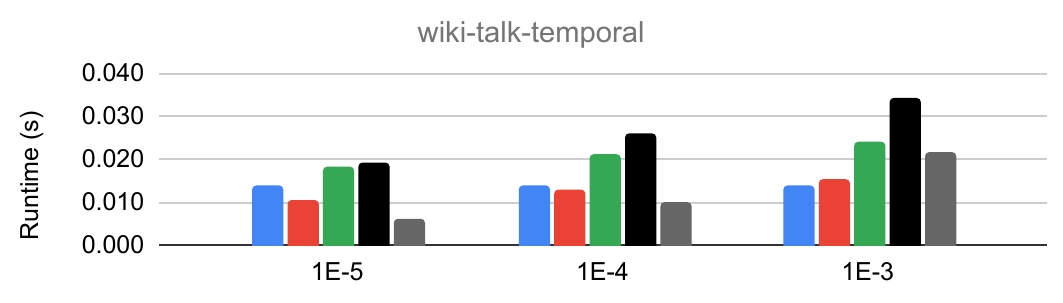}
  \includegraphics[width=0.48\linewidth]{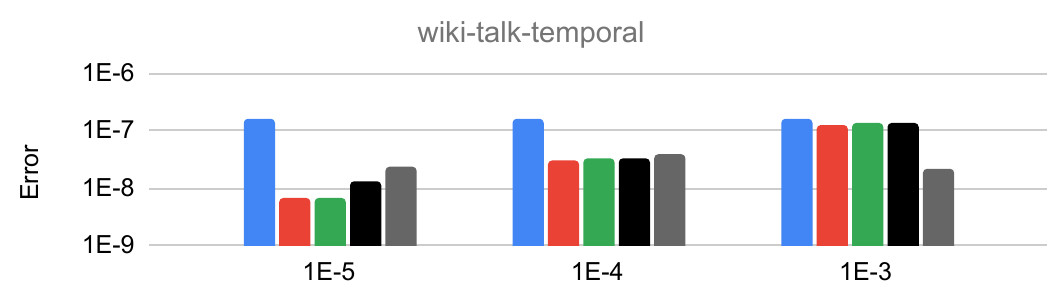}
  \subfigure[Runtime on each dynamic graph]{
    \label{fig:temporal-summary--runtime-graph}
    \includegraphics[width=0.48\linewidth]{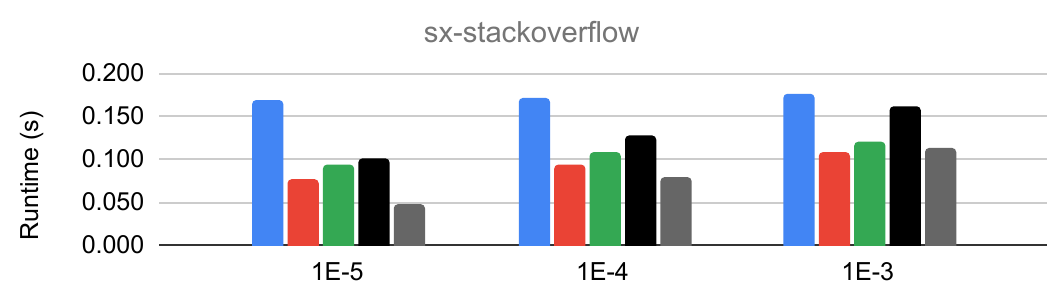}
  }
  \subfigure[Error in ranks obtained on each dynamic graph]{
    \label{fig:temporal-summary--error-graph}
    \includegraphics[width=0.48\linewidth]{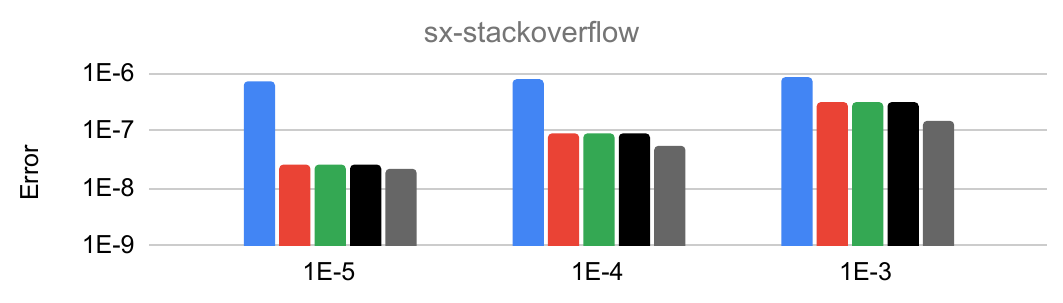}
  } \\[-2ex]
  \caption{Mean Runtime and Error in ranks obtained with our GPU implementation of \textit{Static}, \textit{Naive-dynamic (ND)}, \textit{Dynamic Traversal (DT)}, \textit{Dynamic Frontier (DF)}, and \textit{Dynamic Frontier with Pruning (DF-P)} PageRank on real-world dynamic graphs, with batch updates of size $10^{-5}|E_T|$ to $10^{-3}|E_T|$. Here, (a) and (b) show the overall runtime and error across all temporal graphs, while (c) and (d) show the runtime and rank error for each approach (relative to reference Static PageRank, see Section \ref{sec:measurement}). In (a), the speedup of each approach with respect to Static PageRank is labeled.}
  \label{fig:temporal-summary}
\end{figure*}

%% file: src/fig-8020-runtime.tex
\begin{figure*}[hbtp]
  \centering
  \subfigure[Overall result]{
    \label{fig:8020-runtime--mean}
    \includegraphics[width=0.38\linewidth]{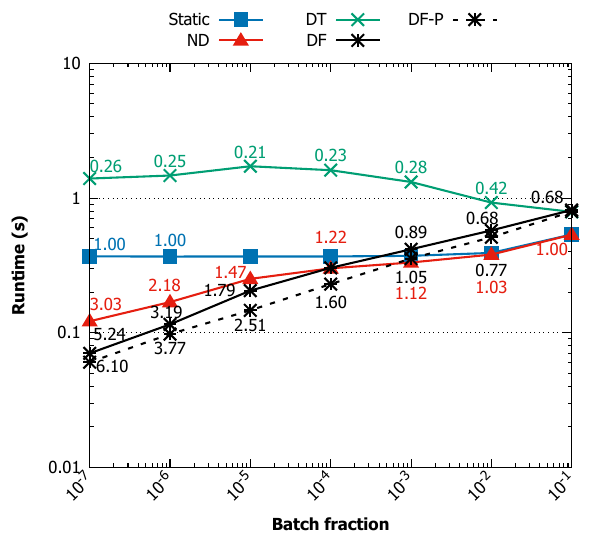}
  }
  \subfigure[Results on each graph]{
    \label{fig:8020-runtime--all}
    \includegraphics[width=0.58\linewidth]{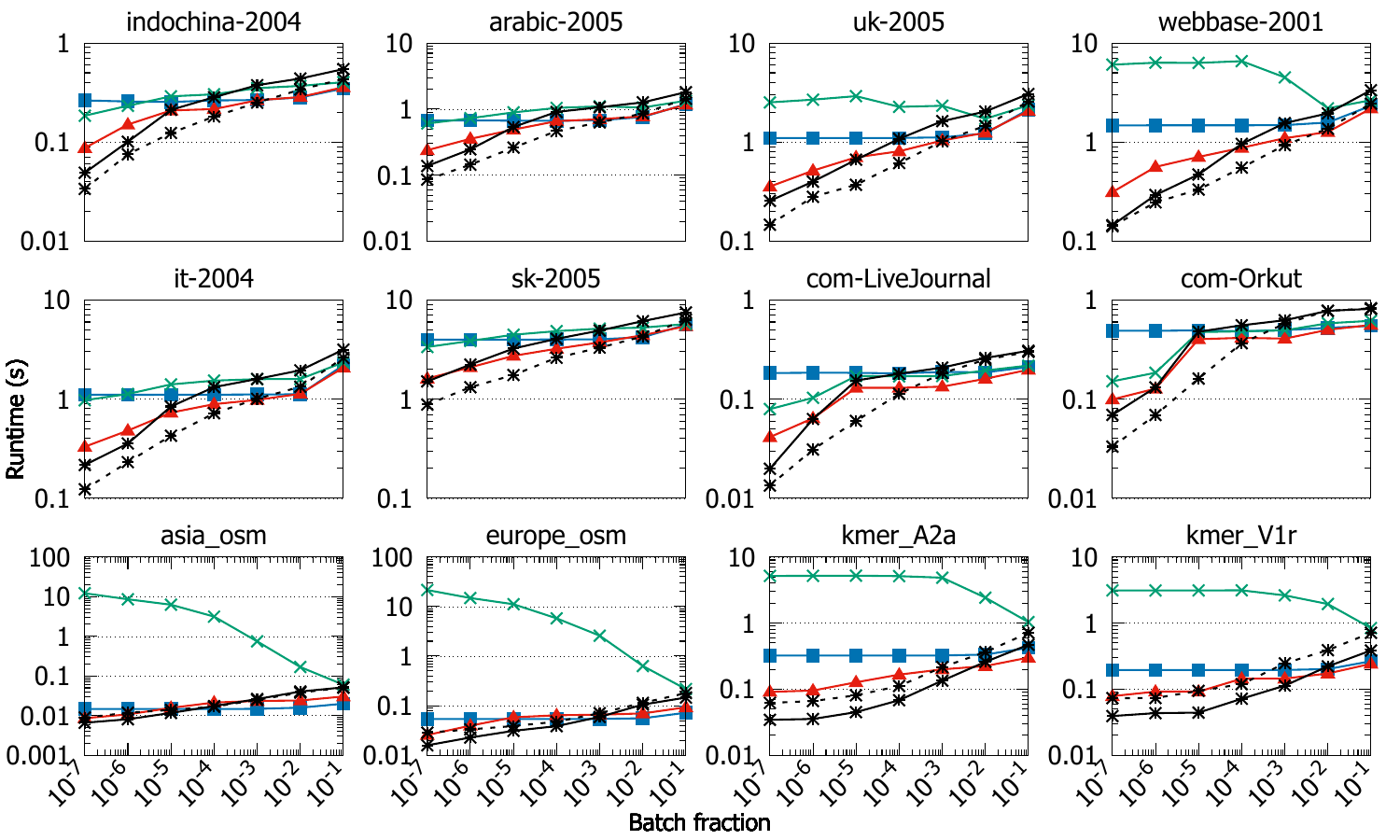}
  } \\[-1ex]
  \caption{Runtime (logarithmic scale) of GPU implementation for \textit{Static}, \textit{Naive-dynamic (ND)}, \textit{Dynamic Traversal (DT)}, \textit{Dynamic Frontier (DF)}, and \textit{Dynamic Frontier with Pruning (DF-P)} PageRank on large (static) graphs with generated random batch updates. Batch updates range in size from $10^{-7}|E|$ to $0.1|E|$ in multiples of $10$. These updates consist of $80\%$ edge insertions and $20\%$ edge deletions, mimicking realistic changes in a dynamic graph scenario. The right subfigure illustrates the runtime of each approach for individual graphs in the dataset, while the left subfigure presents overall runtimes (using geometric mean for consistent scaling across graphs). Additionally, the speedup of each approach relative to Static PageRank is labeled\ignore{on respective lines}.}
  \label{fig:8020-runtime}
\end{figure*}

%% file: src/fig-8020-error.tex
\begin{figure*}[hbtp]
  \centering
  \subfigure[Overall result]{
    \label{fig:8020-error--mean}
    \includegraphics[width=0.38\linewidth]{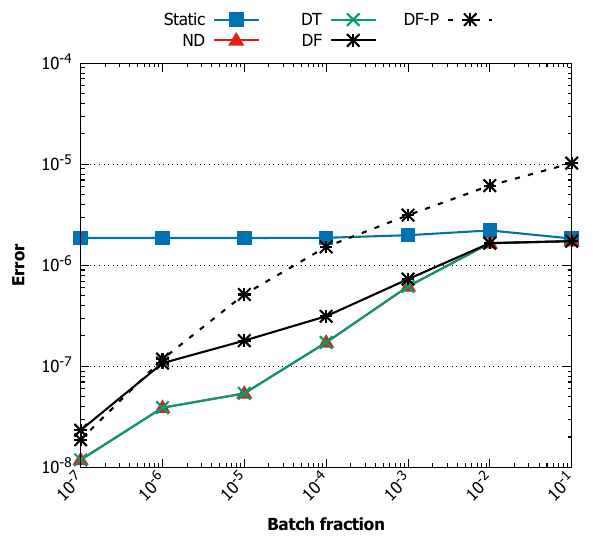}
  }
  \subfigure[Results on each graph]{
    \label{fig:8020-error--all}
    \includegraphics[width=0.58\linewidth]{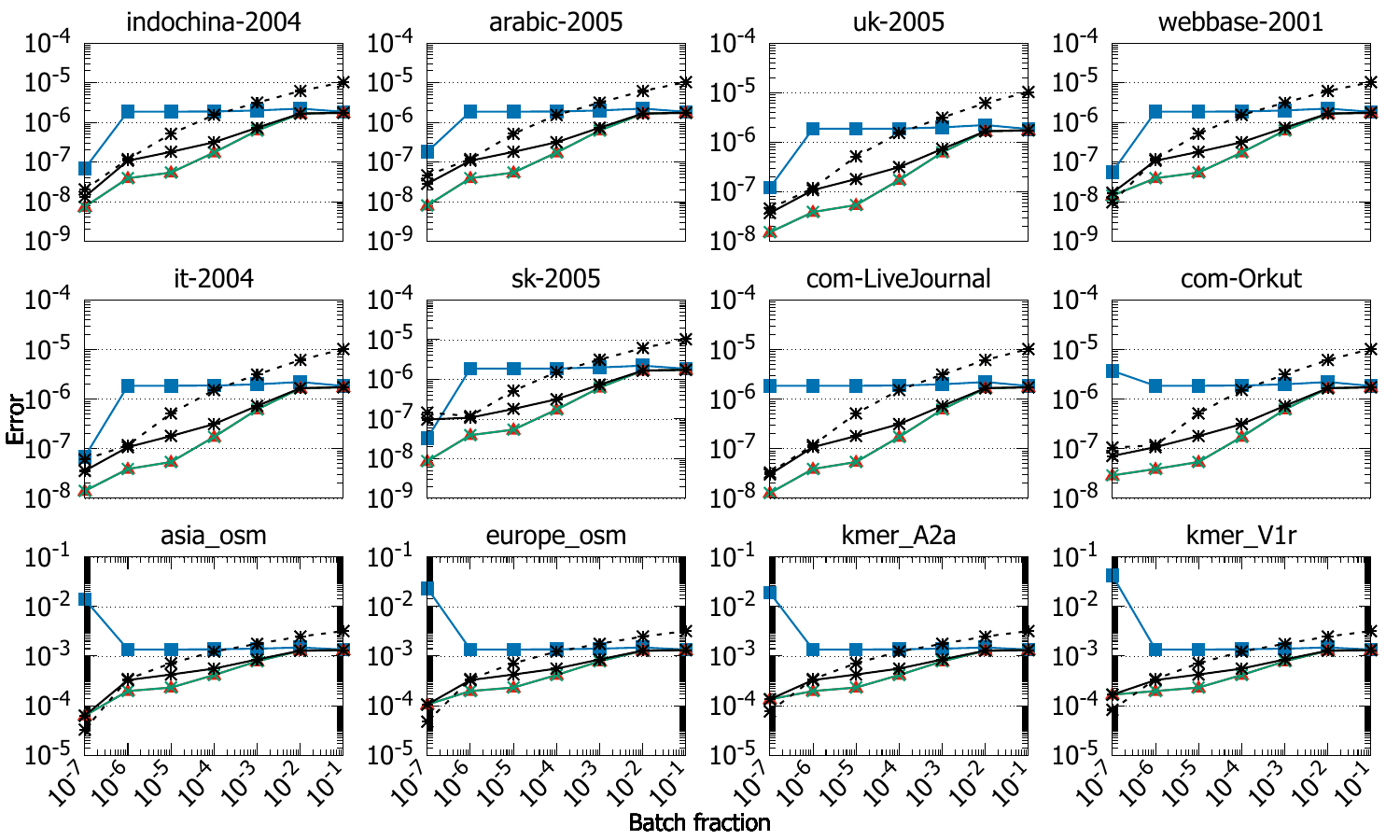}
  } \\[-1ex]
  \caption{Error comparison of our GPU implementation of \textit{Static}, \textit{Naive-dynamic (ND)}, \textit{Dynamic Traversal (DT)}, \textit{Dynamic Frontier (DF)}, and \textit{Dynamic Frontier with Pruning (DF-P)} PageRank on large (static) graphs with generated random batch updates, relative to a Reference Static PageRank (see Section \ref{sec:measurement}), using $L1$-norm. The size of batch updates range from $10^{-7} |E|$ to $0.1 |E|$ in multiples of $10$ (logarithmic scale), consisting of $80\%$ edge insertions and $20\%$ edge deletions to simulate realistic dynamic graph updates. The right subfigure depicts the error for each approach in relation to each graph, while the left subfigure showcases overall errors using geometric mean for consistent scaling across graphs.}
  \label{fig:8020-error}
\end{figure*}

%% file: src/fig-measure-affected.tex
\begin{figure}[!hbt]
  \centering
  \subfigure{
    \label{fig:measure-affected--batch}
    \includegraphics[width=0.98\linewidth]{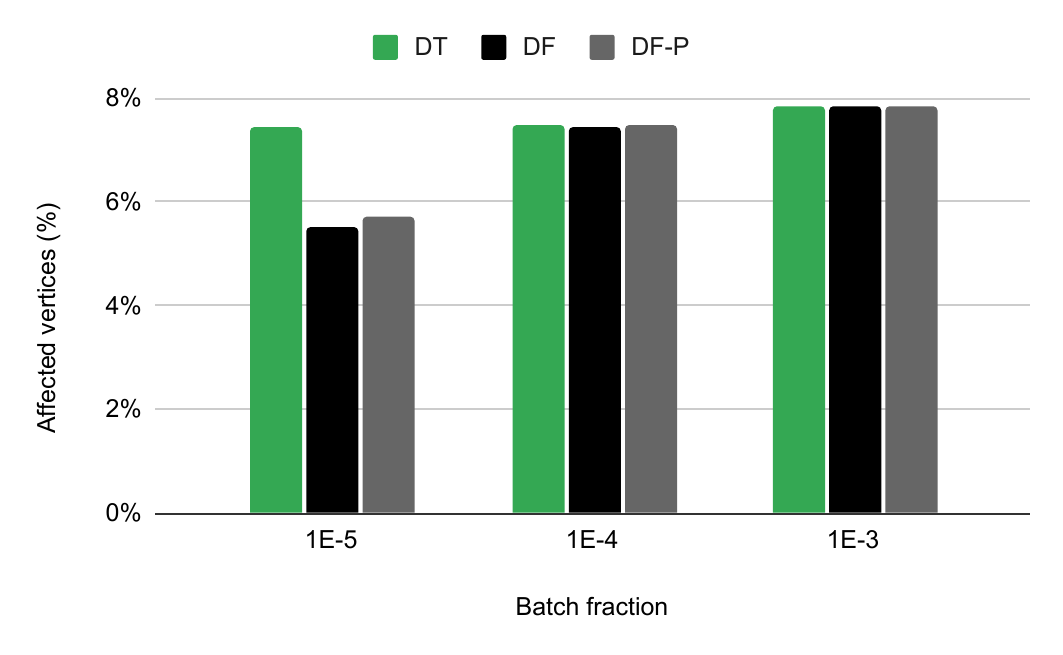}
  } \\[-2ex]
  \caption{Mean percentage of vertices marked as affected by \textit{Dynamic Traversal (DT)}, our improved \textit{Dynamic Frontier (DF)}, and \textit{Dynamic Frontier with Pruning (DF-P)} PageRank, on real-world graphs, with batch updates of size $10^{-5}|E_T|$ to $10^{-3}|E_T|$ (in multiples of $10$). DF and DF-P PageRank mark affected vertices incrementally --- thus, we count any vertex ever marked as affected. \su{TODO}}
  \label{fig:measure-affected}
\end{figure}

%% file: 06-conclusion.tex
In conclusion, in this report, we introduced one of the most efficient GPU implementations of Static PageRank, which involves recomputing PageRank scores from scratch. This implementation employs a synchronous pull-based approach for PageRank computation, devoid of atomics. It partitions and processes low and high in-degree vertices using two separate kernels. Furthermore, we presented our GPU implementation of Dynamic Frontier with Pruning (DF-P) PageRank, which incrementally expands (and contracts) the set of vertices likely to undergo rank changes. It is based on Static PageRank, and introduces additional partitioning between low and high out-degree vertices for the incremental expansion of the affected vertices set, implemented through two additional kernels. On a server equipped with an NVIDIA A100 GPU, we showed that our Static PageRank implementation outperforms Hornet and Gunrock's PageRank implementations by $31\times$ and $5.9\times$, respectively. Additionally, it achieves a speedup of $24\times$ compared to our multicore Static PageRank. Furthermore, DF-P PageRank demonstrates superior performance compared to Static PageRank, achieving a speedup of $2.1\times$ on real-world dynamic graphs and $3.1\times$ on large static graphs with random batch updates.

%% file: aa-appendix.tex
\section{Appendix}

\ignore{\subsection{Our ND PageRank implementation}}

\ignore{Algorithm \ref{alg:naive} outlines the psuedocode of our GPU-based Naive-dynamic PageRank. It takes as input the transpose of the current graph snapshot $G^{t'}$ and the previous rank vector $R^{t-1}$, and returns the updated rank vector $R$. We start by initializing the rank vectors $R$ and $R_{new}$ with the previous rank vector $R^{t-1}$ (line \ref{alg:frontier--initialize}). The rest of the algorithm is similar to Static PageRank (Section \ref{sec:static-impl}).}

\ignore{\input{src/alg-naive}}
\ignore{\input{src/alg-traversal}}

\subsection{Updating ranks of vertices in parallel}
\label{sec:update}

Algorithm \ref{alg:update} provides a psuedocode for updating the ranks of vertices in parallel. Here, the function \texttt{updateRanks()} takes as input the set of affected vertices $\delta_V$, affected neighbor vertices $\delta_N$, the previous and current rank vectors $R$ and $R_{new}$, respectively, the current input graph $G^t$, partitioned vertex IDs $P'$ (low in-degree vertices come first), and the number of vertices with low in-degree $N'_P$. It also requires parameters such as the damping factor $\alpha$, frontier tolerance $\tau_f$, and prune tolerance $\tau_p$.

The algorithm operates in two phases: a thread-per-vertex kernel (for low degree vertices) and a block-per-vertex kernel (for high degree vertices). In the \textit{thread-per-vertex kernel} (lines \ref{alg:update--thread-begin}-\ref{alg:update--thread-end}), we use each thread to process each low degree vertex in parallel, iterating over the partitioned vertex IDs ($P'$). For each vertex $v$, if it's not marked as affected, we skip it (line \ref{alg:update--affected}). Otherwise, we compute the new rank $r$ based on the incoming edges and ranks of neighboring vertices (lines \ref{alg:update--rank-begin}-\ref{alg:update--rank-end}). Depending on whether DF or DF-P PageRank is selected, we compute ranks using either Equation \ref{eq:pr} or \ref{eq:pr-prune}, respectively. Next, we compute the change in rank $\Delta r$ of the current vertex $v$ from its previous iteration (line \ref{alg:update--change}). If the relative change in rank of $v$, i.e., $\Delta r / \max(r, R[v])$, is within the prune tolerance $\tau_p$, we perform pruning by marking $v$ and no longer affected. However, if the relative change in rank of $v$ is above the frontier tolerance $\tau_f$, we mark the vertices whose neighbors must be incrementally marked as affected (the incremental marking of affected vertices is performed at a later point of time, using the \texttt{expandAffected()} function (Algorithm \ref{alg:affected}). Finally, we update the rank of vertex $v$ in the $R$ vector. In the \textit{block-per-vertex kernel} (lines \ref{alg:update--block-begin}-\ref{alg:update--block-end}), we use each thread block to process each high degree vertex in parallel, utilizing block-level parallelism. It involves similar operations as the thread-per-vertex kernel, but uses block-reduce operations and shared memory.

\input{src/alg-update}

\subsection{Parallel vertex partitioning by degree}
\label{sec:partition}

Algorithm \ref{alg:partition} outlines the psuedocode for parallel vertex partitioning by degree. It aims to split the vertices of the input graph $G(V, E)$ into two groups based on their degree: low-degree vertices and high-degree vertices. The algorithm provides partitioned vertex IDs $P$ with low-degree vertices being listed first, along with the number of vertices with low degree $N_P$, as its output.

In the function \texttt{partition()}, we start by initializing an empty buffer $B_k$ and the set of partitioned vertex IDs $P$ (line \ref{alg:partition--initialize}). We then proceed to populate $B_k$ with boolean values indicating whether each vertex has a degree less than or equal to a specified threshold $D_P$ (lines \ref{alg:partition--less-begin}-\ref{alg:partition--less-end}). Afterward, we perform an exclusive prefix sum operation on $B_k$ to determine the number of low-degree vertices $N_P$ (lines \ref{alg:partition--lscan-begin}-\ref{alg:partition--lscan-end}). In the subsequent loop, we assign low-degree vertices to the appropriate positions in the partitioned vertex IDs array $P$ (lines \ref{alg:partition--lpopulate-begin}-\ref{alg:partition--lpopulate-end}). We then repeat a similar process for high-degree vertices. We populate $B_k$ with boolean values indicating whether each vertex has a degree greater than $D_P$ (lines \ref{alg:partition--greater-begin}-\ref{alg:partition--greater-end}), perform another exclusive prefix sum operation on $B_k$ (line \ref{alg:partition--gscan}), and assign high-degree vertices to the appropriate positions in $P$ (lines \ref{alg:partition--gpopulate-begin}-\ref{alg:partition--gpopulate-end}). Finally, we return the partitioned vertex IDs $P$ along with the number of low-degree vertices $N_P$ (line \ref{alg:partition--return}).

\input{src/alg-partition}

\subsection{Parallel marking of affected vertices}
\label{sec:affected}

Algorithm \ref{alg:affected} presents the psuedocode for parallel marking of affected vertices, consisting of two functions: \texttt{initialAffected()} and \texttt{expandAffected()}.

The \texttt{initialAffected()} function performs an initial marking step of DF and DF-P PageRank. It takes as input the current graph snapshot $G^t$ and the sets of edge deletions $\Delta^{t-}$ and insertions $\Delta^{t+}$. Here, we first initialize two arrays, $\delta_V$ and $\delta_N$, which represent whether each vertex and its neighbors are affected, respectively (line \ref{alg:affected--iinitialize}). Next, for each edge deletion in $\Delta^{t-}$, we mark both the source and target vertices as affected (lines \ref{alg:affected--idel-begin}-\ref{alg:affected--idel-end}). Similarly, for each edge insertion in $\Delta^{t+}$, we mark the source vertex as affected (lines \ref{alg:affected--iins-begin}-\ref{alg:affected--iins-end}). Finally, we return $\delta_V$ and $\delta_N$ (line \ref{alg:affected--ireturn}).

The \texttt{expandAffected()} function propagates the affected status to neighboring vertices. It takes as input flags indicating whether each vertex is affected $\delta_V$ or its neighbors are affected $\delta_N$, the current graph snapshot $G^t$, partitioned vertex IDs $P$ with low degree vertices placed first, and the number of vertices with low degree $N_P$. This algorithm also operates in two phases: a thread-per-vertex kernel (for low degree vertices) and a block-per-vertex kernel (for high degree vertices). In the \textit{thread-per-vertex kernel} (lines \ref{alg:affected--ethread-begin}-\ref{alg:affected--ethread-end}), we use each thread to process each low degree vertex in parallel by iterating through the partitioned vertex IDs array $P$. For each vertex $u$ marked as affected in $\delta_N$, we invoke the \texttt{markNeighbors()} function to mark its neighbors as affected in $\delta_V$ (line \ref{alg:affected--etmark}). In the \textit{block-per-vertex kernel} (lines \ref{alg:affected--eblock-begin}-\ref{alg:affected--eblock-end}), we use each thread block to process each vertex in parallel, utilizing block-level parallelism. It involves similar operations as the thread-per-vertex kernel.

\input{src/alg-affected}

\subsection{Indirect Comparison with State-of-the-art PageRank Implementations (Static)}
\label{sec:static-comparison-indirect}

We now indirectly compare the performance of our GPU implementation of Static PageRank with other similar state-of-the-art implementations, listed in Table \ref{tab:compare-large}. Wang et al. \cite{wang2021grus} present Grus, a unified memory efficient high-performance graph processing framework for GPUs. They achieve a $1.2\times$ speedup over Gunrock \cite{wang2016gunrock} on the \textit{uk-2005} graph (refer to Table $4$ in their paper \cite{wang2021grus}). On the same graph, Wang et al. observe that Gunrock achieves a speedup of $4.6\times$ over Tigr \cite{nodehi2018tigr}. However, our implementation achieves an $8.6\times$ speedup over Gunrock on the same graph (Figure \ref{fig:compare--speedup}). Chen et al. \cite{chen2022atos} introduce Atos, a task-parallel GPU scheduler for graph analytics. They achieve a $3.2\times$ speedup over Gunrock on the \textit{indochina-2004} graph (refer to Table $1$ in their paper \cite{chen2022atos}). However, our implementation achieves a $24.4\times$ speedup over Gunrock on the same graph (Figure \ref{fig:compare--speedup}). In a subsequent work, Chen et al. \cite{chen2022scalable} extend their Atos dynamic scheduling framework to multi-node GPU systems supporting Partitioned Global Address Space (PGAS) style lightweight one-sided memory operations within and between nodes. Even with $4$ GPUs, they fail to surpass our speedup with respect to Gunrock on the \textit{indochina-2004} graph (refer to Table $4$ in their paper \cite{chen2022scalable}). On the same graph, Chen et al. observe that they achieve a speedup of $57.5\times$ over Galois with Gluon \cite{dathathri2018gluon} on a single GPU (see Table $5$ in their paper \cite{chen2022scalable}). Yang et al. \cite{yang2022graphblast} introduce GraphBLAST, a high-performance linear algebra-based graph framework for GPUs. On the \textit{indochina-2004} graph, they achieve a $2.2\times$/$1.2\times$ speedup over Gunrock (see Table $12$/$13$ in their paper \cite{yang2022graphblast}). Nonetheless, our implementation achieves a $24.4\times$ speedup over Gunrock on the same graph (Figure \ref{fig:compare--speedup}). Concessao et al. \cite{concessao2023meerkat} propose Meerkat, a library-based framework for dynamic graph algorithms that utilizes a GPU-tailored graph representation and exploits the warp-cooperative execution model. The PageRank implementation of Meerkat performs, on average, $1.7\times$ faster than Hornet \cite{busato2018hornet}. However, our Static PageRank outperforms Hornet by an average speedup of $31\times$ (Figure \ref{fig:compare--speedup}).

\ignore{\clearpage}

\input{src/fig-temporal-compare}
\input{src/fig-8020-runtime-compare}
\input{src/fig-8020-error-compare}
\input{src/fig-temporal-sx-mathoverflow}
\input{src/fig-temporal-sx-askubuntu}
\input{src/fig-temporal-sx-superuser}
\input{src/fig-temporal-wiki-talk-temporal}
\input{src/fig-temporal-sx-stackoverflow}

%% file: src/alg-naive.tex
\begin{algorithm}[!hbt]
\caption{Our GPU-based Naive-dynamic (ND) PageRank.}
\label{alg:naive}
\begin{algorithmic}[1]
\Require{$G^{t'}(V^{t'}, E^{t'})$: Transpose of current input graph}
\Require{$R^{t-1}$: Previous rank vector}
\Require{$R, R_{new}$: Rank vector in the previous, current iteration}
\Ensure{$P'$: Partitioned vertex IDs, low in-degree first }
\Ensure{$N'_P$: Number of vertices with low in-degree}
\Ensure{$\Delta R$: $L\infty$-norm between previous and current ranks}
\Ensure{$\tau$: Iteration tolerance}

\Statex

\Function{naiveDynamic}{$G^{t'}, R^{t-1}$}
  \State $\rhd$ Initialize ranks
  \State $R \gets R_{new} \gets R^{t-1}$ \label{alg:naive--initialize}
  \State $\rhd$ Partition vertex IDs by in-degree 
  \State $\{P', N'_P\} \gets partition(G^{t'})$ \label{alg:naive--partition} \Comment{Alg. \ref{alg:partition}}
  \State $\rhd$ Perform PageRank iterations
  \ForAll{$i \in [0 .. MAX\_ITERATIONS)$} \label{alg:naive--compute-begin}
    \State $updateRanks(\cdots, \cdots, R_{new}, R, G^{t'}, P', N'_P)$ \Comment{Alg. \ref{alg:update}} \label{alg:naive--update}
    \State $\Delta R \gets l_{\infty}NormDelta(R_{new}, R)$ \textbf{;} $swap(R_{new}, R)$ \label{alg:naive--error}
    \If{$\Delta R \leq \tau$} \textbf{break} \label{alg:naive--converged}
    \EndIf
  \EndFor \label{alg:naive--compute-end}
  \State \ReturnInline{$R$} \label{alg:naive--return}
\EndFunction
\end{algorithmic}
\end{algorithm}

%% file: src/alg-traversal.tex
\begin{algorithm}[!hbt]
\caption{Our GPU-based Dynamic Traversal (DT) PageRank.}
\label{alg:traversal}
\begin{algorithmic}[1]
\Require{$G^t(V^t, E^t), G^{t'}$: Current input graph, and its transpose}
\Require{$\Delta^{t-}, \Delta^{t+}$: Edge deletions and insertions (input)}
\Require{$R^{t-1}$: Previous rank vector}
\Require{$R, R_{new}$: Rank vector in the previous, current iteration}
\Ensure{$\delta_V, \delta_N$: Is a vertex, or neighbors of a vertex affected}
\Ensure{$P, P'$: Partitioned vertex IDs --- low out-, in-degree first }
\Ensure{$N_P, N'_P$: Number of vertices with low out-, in-degree}
\Ensure{$\Delta R$: $L\infty$-norm between previous and current ranks}
\Ensure{$\tau$: Iteration tolerance}

\Statex

\Function{dynamicTraversal}{$G^t, G^{t'}, \Delta^{t-}, \Delta^{t+}, R^{t-1}$}
  \State $\rhd$ Initialize ranks
  \State $R \gets R_{new} \gets R^{t-1}$ \label{alg:traversal--initialize}
  \State $\rhd$ Partition vertex IDs by out- and in-degree 
  \State $\{P, N_P\} \gets partition(G^t)$ \label{alg:traversal--partition-begin} \Comment{Alg. \ref{alg:partition}}
  \State $\{P', N'_P\} \gets partition(G^{t'})$ \label{alg:traversal--partition-end} \Comment{Alg. \ref{alg:partition}}
  \State $\rhd$ Mark all affected vertices \label{alg:traversal--mark-begin}
  \State $\{\delta_V, \delta_N\} \gets initialAffected(G^t, \Delta^{t-}, \Delta^{t+})$ \Comment{Alg. \ref{alg:affected}}
  \State $expandAffected(\delta_V, \delta_N, G^t, P, N_P)$ \label{alg:traversal--mark-end} \Comment{Alg. \ref{alg:affected}}
  \State $N_A \gets countOnes(\delta_V)$ \textbf{;} $\delta_N \gets \delta_V$
  \While{$true$}
    \State $\delta_F \gets \{\}$
    \State $expandAffected(\delta_V, \delta_N, G^t, P, N_P)$ \Comment{Alg. \ref{alg:affected}}
    \State $\delta_V \gets \delta_V$ \textbf{OR} $\delta_F$ \textbf{;} $swap(\delta_N, \delta_F)$
    \State $N_{A'} \gets countOnes(\delta_V)$
    \If{$N_{A'} = N_A$} \textbf{break}
    \EndIf
    \State $N_A \gets N_{A'}$
  \EndWhile
  \State $\rhd$ Perform PageRank iterations
  \ForAll{$i \in [0 .. MAX\_ITERATIONS)$} \label{alg:traversal--compute-begin}
    \State $\delta_N \gets \{\}$
    \State $updateRanks(\delta_V, \cdots, R_{new}, R, G^t, P', N'_P)$ \Comment{Alg. \ref{alg:update}} \label{alg:traversal--update}
    \State $\Delta R \gets l_{\infty}NormDelta(R_{new}, R)$ \textbf{;} $swap(R_{new}, R)$ \label{alg:traversal--error}
    \If{$\Delta R \leq \tau$} \textbf{break} \label{alg:traversal--converged}
    \EndIf
  \EndFor \label{alg:traversal--compute-end}
  \State \ReturnInline{$R$} \label{alg:traversal--return}
\EndFunction
\end{algorithmic}
\end{algorithm}

%% file: src/alg-update.tex
\begin{algorithm}[!hbt]
\caption{Updating ranks of vertices in parallel.}
\label{alg:update}
\begin{algorithmic}[1]
\Require{$G^t(V^t, E^t)$: Current input graph}
\Require{$R, R_{new}$: Rank vector in the previous, current iteration}
\Require{$\delta_V, \delta_N$: Is a vertex, or neighbors of a vertex affected}
\Require{$P'$: Partitioned vertex IDs --- low in-degree first }
\Require{$N'_P$: Number of vertices with low in-degree}
\Ensure{$\Delta r$: Change in rank of a vertex}
\Ensure{$\tau_f, \tau_p$: Frontier, prune tolerance}
\Ensure{$\alpha$: Damping factor}

\Statex

\Function{updateRanks}{$\delta_V, \delta_N, R_{new}, R, G^t, P', N'_P$}
  \State $\rhd$ A thread-per-vertex kernel
  \ForAll{$i \in [0, N'_P)$ \textbf{in parallel$_{thread/V}$}} \label{alg:update--thread-begin}
    \State $v \gets P'[i]$
    \If{\textbf{not} $\delta_V[v]$} \textbf{continue} \label{alg:update--affected}
    \EndIf
    \State $c \gets 0$ \textbf{;} $d \gets |G^t.out(v)|$ \label{alg:update--rank-begin}
    \State $C_0 \gets (1 - \alpha)/|V^t|$
    \ForAll{$u \in G^t.in(v)$}
      \State $c \gets c + R[u] / |G^t.out(u)|$
    \EndFor
    \If{\textbf{is \textit{DF-P}}}
      \State $r \gets 1/(1 - \alpha/d) * (C_0 + \alpha * (c - R[v]/d))$
    \Else
      \State $r \gets C_0 + \alpha * c$
    \EndIf \label{alg:update--rank-end}
    \State $\Delta r \gets |r - R[v]|$ \label{alg:update--change}
    \State $\rhd$ Prune $v$ if its relative rank change is small
    \If{\textbf{is \textit{DF-P} and} $\Delta r / \max(r, R[v]) \leq \tau_p$} \label{alg:update--prune-begin}
      \State $\delta_V[v] \gets 0$
    \EndIf \label{alg:update--prune-end}
    \State $\rhd$ Expand frontier if relative rank change is large
    \If{$\Delta r / \max(r, R[v]) > \tau_f$} \label{alg:update--expand-begin}
      \State $\delta_N[v] \gets 0$
    \EndIf \label{alg:update--expand-end}
    \State $\rhd$ Update rank of $v$
    \State $R[v] \gets r$ \label{alg:update--update}
  \EndFor \label{alg:update--thread-end}
  \State $\rhd$ Similarly with block-per-vertex kernel
  \State $\rhd$ (using block-reduce, requires shared memory)
  \ForAll{$i \in [N'_P, |V^t|)$ \textbf{in parallel$_{block/V}$}} \label{alg:update--block-begin}
    \State $\cdots$
  \EndFor \label{alg:update--block-end}
\EndFunction
\end{algorithmic}
\end{algorithm}

%% file: src/alg-partition.tex
\begin{algorithm}[!hbt]
\caption{Parallel vertex partitioning by degree.}
\label{alg:partition}
\begin{algorithmic}[1]
\Require{$G(V, E)$: Input graph}
\Ensure{$D_P$: Maximum degree of a low-degree vertex}
\Ensure{$B_k$: Temporary buffer used for partitioning}
\Ensure{$P$: Partitioned vertex IDs --- low degree vertices first }
\Ensure{$N_P$: Number of vertices with low degree}

\Statex

\Function{partition}{$G$}
  \State $P \gets B_k \gets \{\}$ \label{alg:partition--initialize}
  \State $\rhd$ Populate vertex IDs with degree $\leq D_P$
  \ForAll{$v \in V$ \textbf{in parallel}} \label{alg:partition--less-begin}
    \State $B_k[v] \gets |G.out(v)| \leq D_P$
  \EndFor \label{alg:partition--less-end}
  \State $B_k[|V|] \gets 0$ \label{alg:partition--lscan-begin}
  \State $exclusiveScan(B_k)$ \textbf{;} $N_P \gets B_k[|V|]$ \label{alg:partition--lscan-end}
  \ForAll{$v \in V$ \textbf{in parallel}} \label{alg:partition--lpopulate-begin}
    \If{$|G.out(v)| \leq D_P$} $P[B_k[v]] \gets v$
    \EndIf
  \EndFor \label{alg:partition--lpopulate-end}
  \State $\rhd$ Populate vertex IDs with degree $> D_P$
  \ForAll{$v \in V$ \textbf{in parallel}} \label{alg:partition--greater-begin}
    \State $B_k[v] \gets |G.out(v)| > D_P$
  \EndFor \label{alg:partition--greater-end}
  \State $exclusiveScan(B_k)$ \label{alg:partition--gscan}
  \ForAll{$v \in V$ \textbf{in parallel}} \label{alg:partition--gpopulate-begin}
    \If{$|G.out(v)| > D_P$} $P[N_P + B_k[v]] \gets v$
    \EndIf
  \EndFor \label{alg:partition--gpopulate-end}
  \State \ReturnInline{$\{P, N_P\}$} \label{alg:partition--return}
\EndFunction
\end{algorithmic}
\end{algorithm}

%% file: src/alg-affected.tex
\begin{algorithm}[!hbt]
\caption{Parallel marking of affected vertices.}
\label{alg:affected}
\begin{algorithmic}[1]
\Require{$G^t(V^t, E^t)$: Current input graph}
\Require{$\Delta^{t-}, \Delta^{t+}$: Edge deletions and insertions (input)}
\Require{$\delta_V, \delta_N$: Is a vertex, or neighbors of a vertex affected}
\Require{$P$: Partitioned vertex IDs (low-degree vertices first) }
\Require{$N_P$: Number of vertices with low degree}

\Statex

\Function{initialAffected}{$G^t, \Delta^{t-}, \Delta^{t+}$}
  \State $N_D \gets |\Delta^{t-}|$ \textbf{;} $N_I \gets |\Delta^{t+}|$ \label{alg:affected--iinitialize}
  \State $\rhd$ For edge deletions
  \ForAll{$i \in [0, N_D)$ \textbf{in parallel}} \label{alg:affected--idel-begin}
    \State $u \gets \Delta^{t-}[i].source$
    \State $v \gets \Delta^{t-}[i].target$
    \State $\delta_N[u] \gets 1$
    \State $\delta_V[v] \gets 1$
  \EndFor \label{alg:affected--idel-end}
  \State $\rhd$ For edge insertions
  \ForAll{$i \in [0, N_I)$ \textbf{in parallel}} \label{alg:affected--iins-begin}
    \State $u \gets \Delta^{t+}[i].source$
    \State $\delta_N[u] \gets 1$
  \EndFor \label{alg:affected--iins-end}
  \Return{$\{\delta_V, \delta_N\}$} \label{alg:affected--ireturn}
\EndFunction

\Statex

\Function{expandAffected}{$\delta_V, \delta_N, G^t, P, N_P$}
  \State $\rhd$ A thread-per-vertex kernel
  \ForAll{$i \in [0, N_P)$ \textbf{in parallel$_{thread/V}$}} \label{alg:affected--ethread-begin}
    \State $u \gets P[i]$
    \If{$\delta_N[u]$} $markNeighbors(\delta_V, G^t, u)$ \label{alg:affected--etmark}
    \EndIf
  \EndFor \label{alg:affected--ethread-end}
  \State $\rhd$ Similarly with block-per-vertex kernel
  \ForAll{$i \in [N_P, |V^t|)$ \textbf{in parallel$_{block/V}$}} \label{alg:affected--eblock-begin}
    \State $\cdots$
  \EndFor \label{alg:affected--eblock-end}
\EndFunction
\end{algorithmic}
\end{algorithm}

%% file: src/fig-temporal-compare.tex
\begin{figure*}[!hbt]
  \centering
  \subfigure[Overall Runtime \textbf{(GPU)}]{
    \label{fig:temporal-compare--runtime-overall}
    \includegraphics[width=0.48\linewidth]{out/temporal-summary-runtime-overall.pdf}
  }
  \subfigure[Overall Error in ranks obtained \textbf{(GPU)}]{
    \label{fig:temporal-compare--error-overall}
    \includegraphics[width=0.48\linewidth]{out/temporal-summary-error-overall.pdf}
  }
  \subfigure[Overall Runtime \textbf{(CPU)}]{
    \label{fig:temporal-compare--runtime-overall-cpu}
    \includegraphics[width=0.48\linewidth]{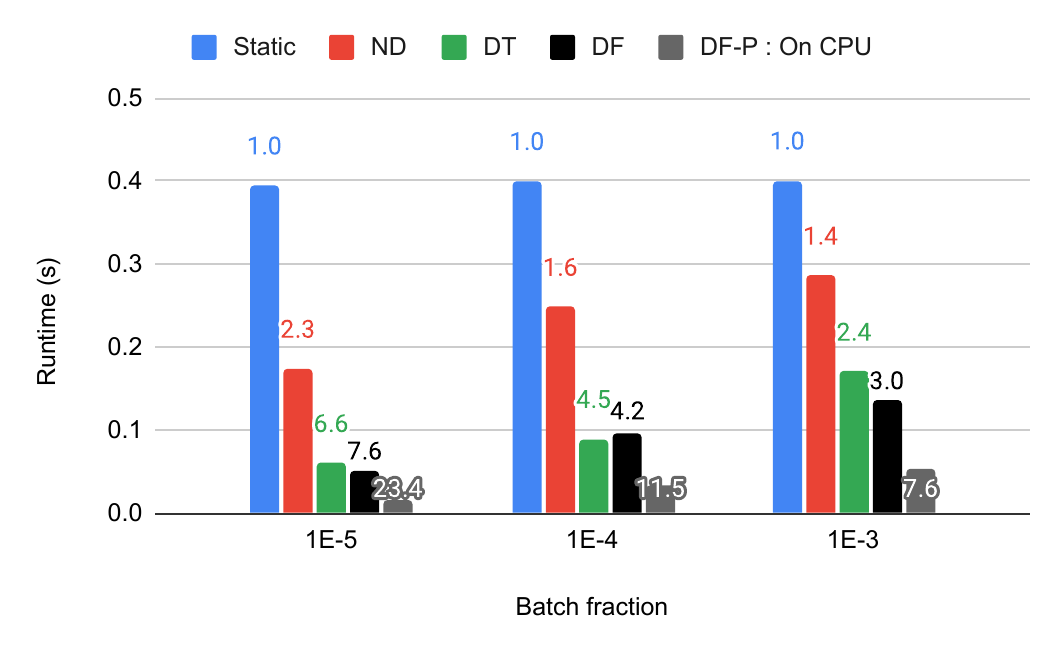}
  }
  \subfigure[Overall Error in ranks obtained \textbf{(CPU)}]{
    \label{fig:temporal-compare--error-overall-cpu}
    \includegraphics[width=0.48\linewidth]{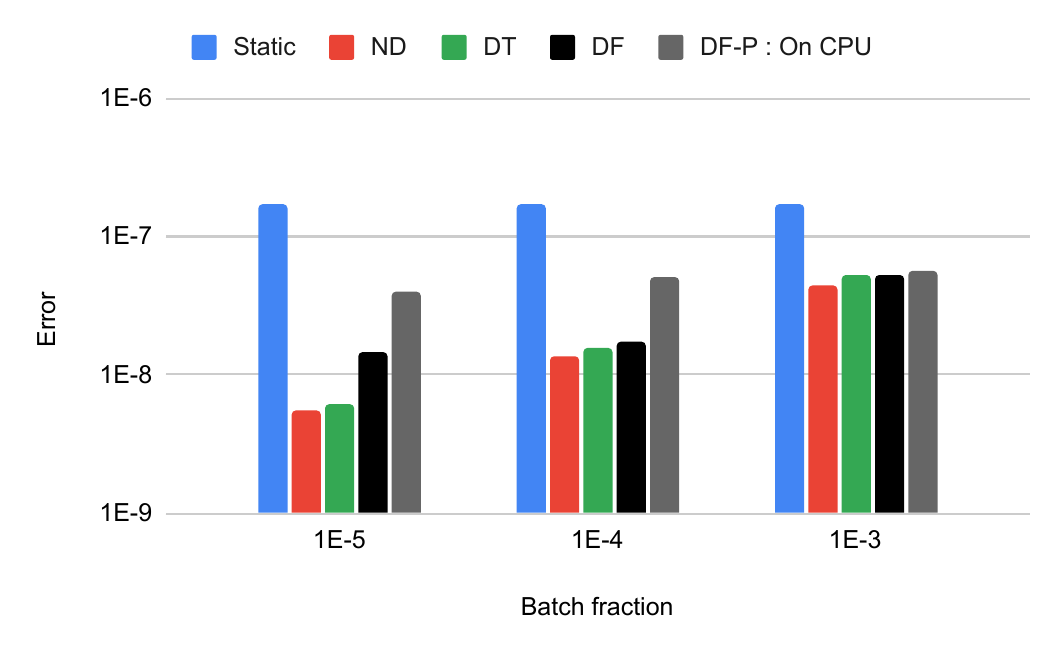}
  } \\[-2ex]
  \caption{Mean Runtime and Error in ranks obtained with our GPU / multicore CPU implementation \cite{sahu2024df} of \textit{Static}, \textit{Naive-dynamic (ND)}, \textit{Dynamic Traversal (DT)}, \textit{Dynamic Frontier (DF)}, and \textit{Dynamic Frontier with Pruning (DF-P)} PageRank on real-world dynamic graphs, with batch updates of size $10^{-5}|E_T|$ to $10^{-3}|E_T|$. Here, (a) and (b) show the overall runtime and error (relative to reference Static PageRank, see Section \ref{sec:measurement}) of the approaches on the GPU, while (c) and (d) show the overall runtime and error of the approaches on the CPU, across all temporal graphs. In (a) and (c), the speedup of each approach with respect to Static PageRank is labeled.}
  \label{fig:temporal-compare}
\end{figure*}

%% file: src/fig-8020-runtime-compare.tex
\begin{figure*}[hbtp]
  \centering
  \subfigure[Overall result \textbf{(GPU)}]{
    \label{fig:8020-runtime-compare--mean}
    \includegraphics[width=0.38\linewidth]{out/8020-runtime-mean.pdf}
  }
  \subfigure[Results on each graph \textbf{(GPU)}]{
    \label{fig:8020-runtime-compare--all}
    \includegraphics[width=0.58\linewidth]{out/8020-runtime-all.pdf}
  }
  \subfigure[Overall result \textbf{(CPU)}]{
    \label{fig:8020-runtime-compare--mean-cpu}
    \includegraphics[width=0.38\linewidth]{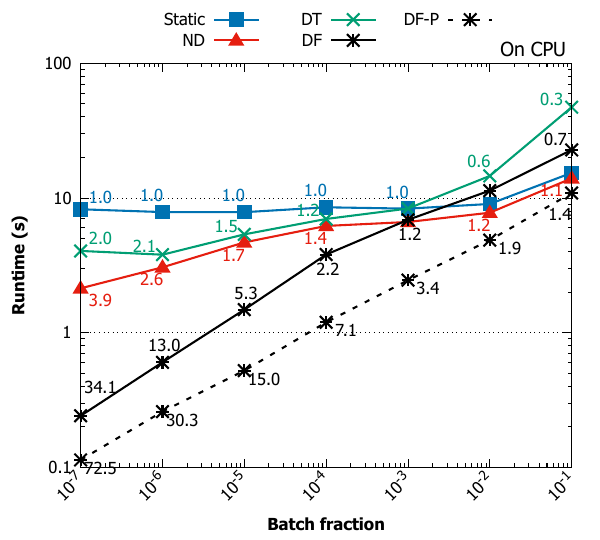}
  }
  \subfigure[Results on each graph \textbf{(CPU)}]{
    \label{fig:8020-runtime-compare--all-cpu}
    \includegraphics[width=0.58\linewidth]{out/8020-runtime-all.pdf}
  } \\[-1ex]
  \caption{Runtime (logarithmic scale) of our GPU implementation / multicore CPU implementation \cite{sahu2024df} of \textit{Static}, \textit{Naive-dynamic (ND)}, \textit{Dynamic Traversal (DT)}, \textit{Dynamic Frontier (DF)}, and \textit{Dynamic Frontier with Pruning (DF-P)} PageRank on large (static) graphs with generated random batch updates. Batch updates range in size from $10^{-7}|E|$ to $0.1|E|$ in multiples of $10$. These updates consist of $80\%$ edge insertions and $20\%$ edge deletions, mimicking realistic changes in a dynamic graph scenario. The right subfigures illustrate the runtime of each approach for individual graphs in the dataset, while the left subfigures present overall runtimes (using geometric mean for consistent scaling across graphs). Additionally, the speedup of each approach relative to Static PageRank is labeled on respective lines.}
  \label{fig:8020-runtime-compare}
\end{figure*}

%% file: src/fig-8020-error-compare.tex
\begin{figure*}[hbtp]
  \centering
  \subfigure[Overall result \textbf{(GPU)}]{
    \label{fig:8020-error-compare--mean}
    \includegraphics[width=0.38\linewidth]{out/8020-error-mean.pdf}
  }
  \subfigure[Results on each graph \textbf{(GPU)}]{
    \label{fig:8020-error-compare--all}
    \includegraphics[width=0.58\linewidth]{out/8020-error-all.pdf}
  } \\[-1ex]
  \subfigure[Overall result \textbf{(CPU)}]{
    \label{fig:8020-error-compare--mean-cpu}
    \includegraphics[width=0.38\linewidth]{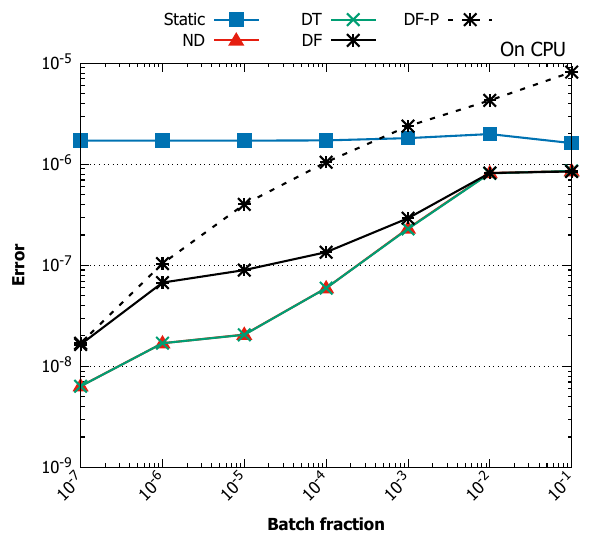}
  }
  \subfigure[Results on each graph \textbf{(CPU)}]{
    \label{fig:8020-error-compare--all-cpu}
    \includegraphics[width=0.58\linewidth]{out/8020-error-all.pdf}
  } \\[-1ex]
  \caption{Error comparison our GPU implementation / multicore CPU implementation \cite{sahu2024df} of \textit{Static}, \textit{Naive-dynamic (ND)}, \textit{Dynamic Traversal (DT)}, \textit{Dynamic Frontier (DF)}, and \textit{Dynamic Frontier with Pruning (DF-P)} PageRank on large (static) graphs with generated random batch updates, relative to a Reference Static PageRank (see Section \ref{sec:measurement}), using $L1$-norm. The size of batch updates range from $10^{-7} |E|$ to $0.1 |E|$ in multiples of $10$ (logarithmic scale), consisting of $80\%$ edge insertions and $20\%$ edge deletions to simulate realistic dynamic graph updates. The right subfigures depict the error for each approach in relation to each graph, while the left subfigures showcase overall errors using geometric mean for consistent scaling across graphs.}
  \label{fig:8020-error-compare}
\end{figure*}

%% file: src/fig-temporal-sx-mathoverflow.tex
\begin{figure*}[!hbt]
  \centering
  \subfigure[Runtime on consecutive batch updates of size $10^{-5}|E_T|$]{
    \label{fig:temporal-sx-mathoverflow--runtime5}
    \includegraphics[width=0.48\linewidth]{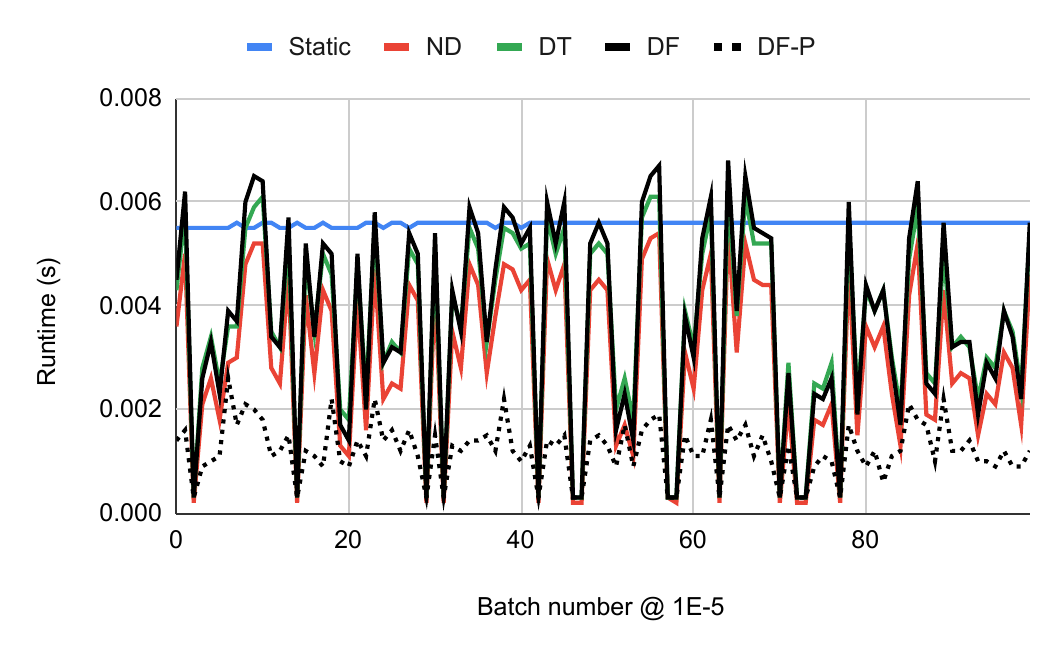}
  }
  \subfigure[Error in ranks obtained on consecutive batch updates of size $10^{-5}|E_T|$]{
    \label{fig:temporal-sx-mathoverflow--error5}
    \includegraphics[width=0.48\linewidth]{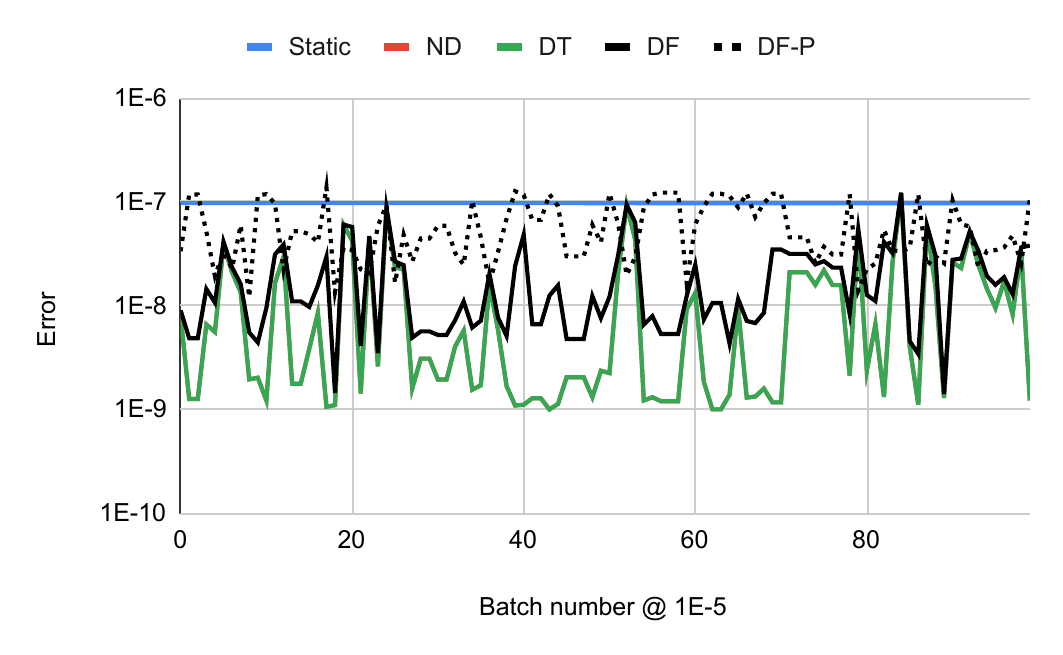}
  } \\[2ex]
  \subfigure[Runtime on consecutive batch updates of size $10^{-4}|E_T|$]{
    \label{fig:temporal-sx-mathoverflow--runtime4}
    \includegraphics[width=0.48\linewidth]{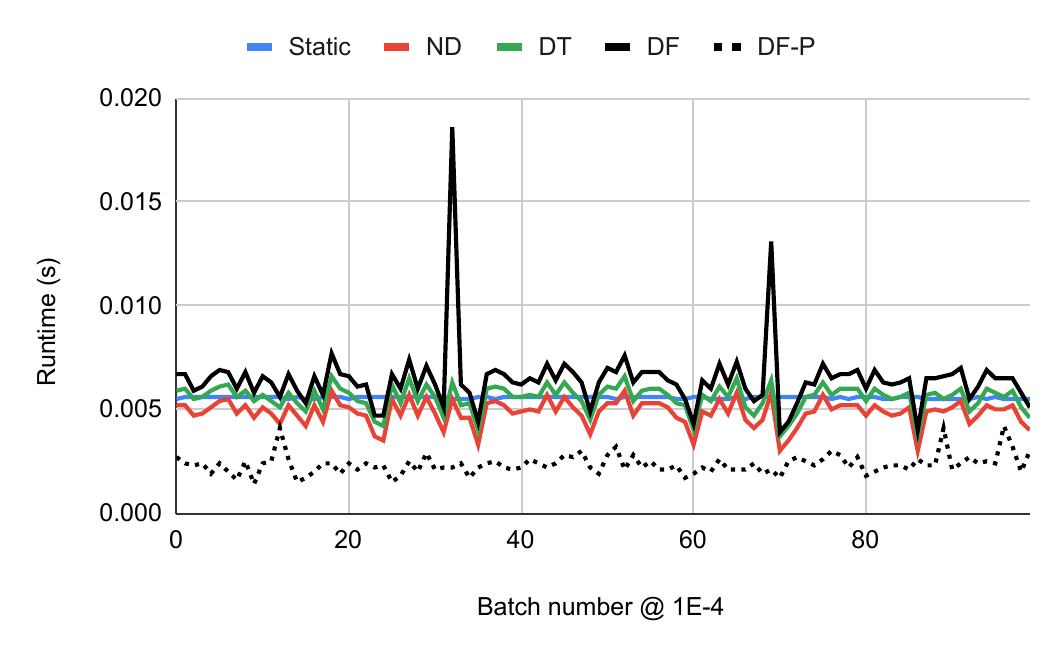}
  }
  \subfigure[Error in ranks obtained on consecutive batch updates of size $10^{-4}|E_T|$]{
    \label{fig:temporal-sx-mathoverflow--error4}
    \includegraphics[width=0.48\linewidth]{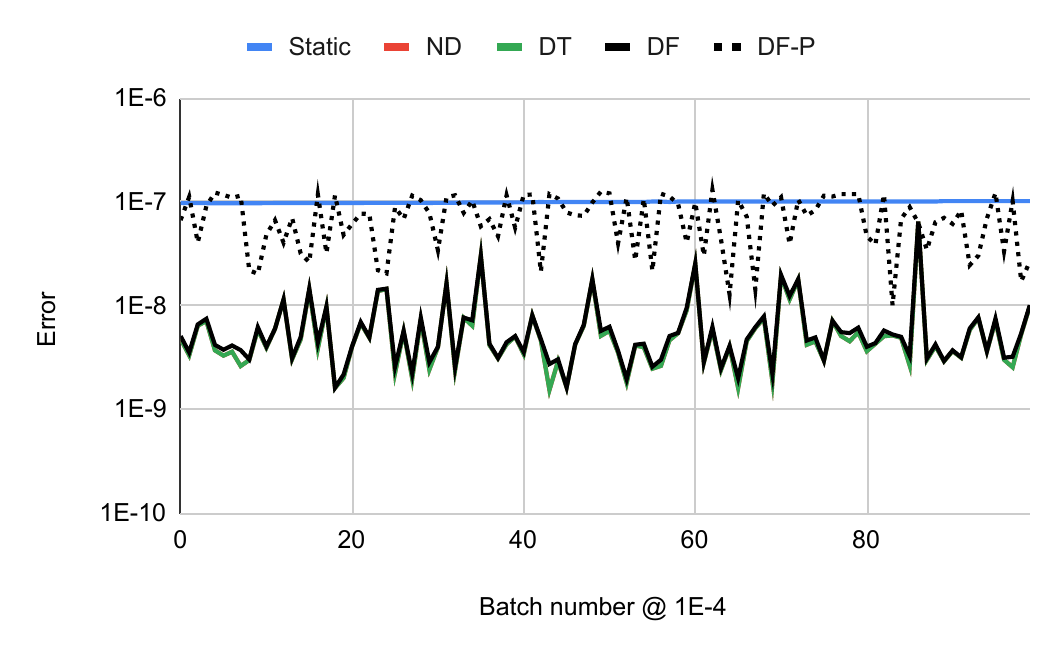}
  } \\[2ex]
  \subfigure[Runtime on consecutive batch updates of size $10^{-3}|E_T|$]{
    \label{fig:temporal-sx-mathoverflow--runtime3}
    \includegraphics[width=0.48\linewidth]{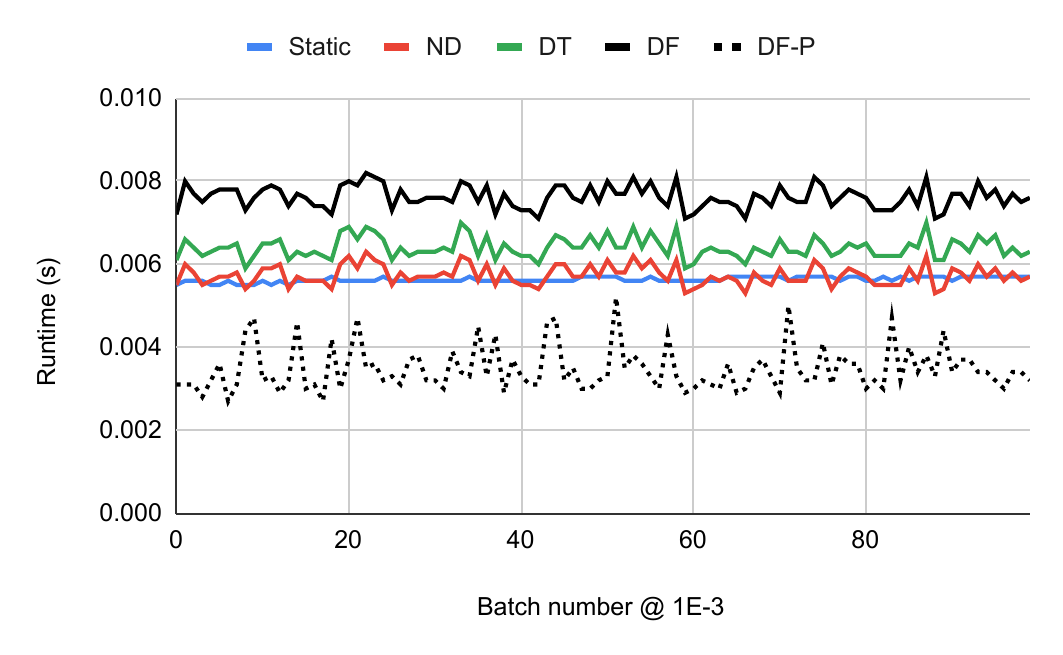}
  }
  \subfigure[Error in ranks obtained on consecutive batch updates of size $10^{-3}|E_T|$]{
    \label{fig:temporal-sx-mathoverflow--error3}
    \includegraphics[width=0.48\linewidth]{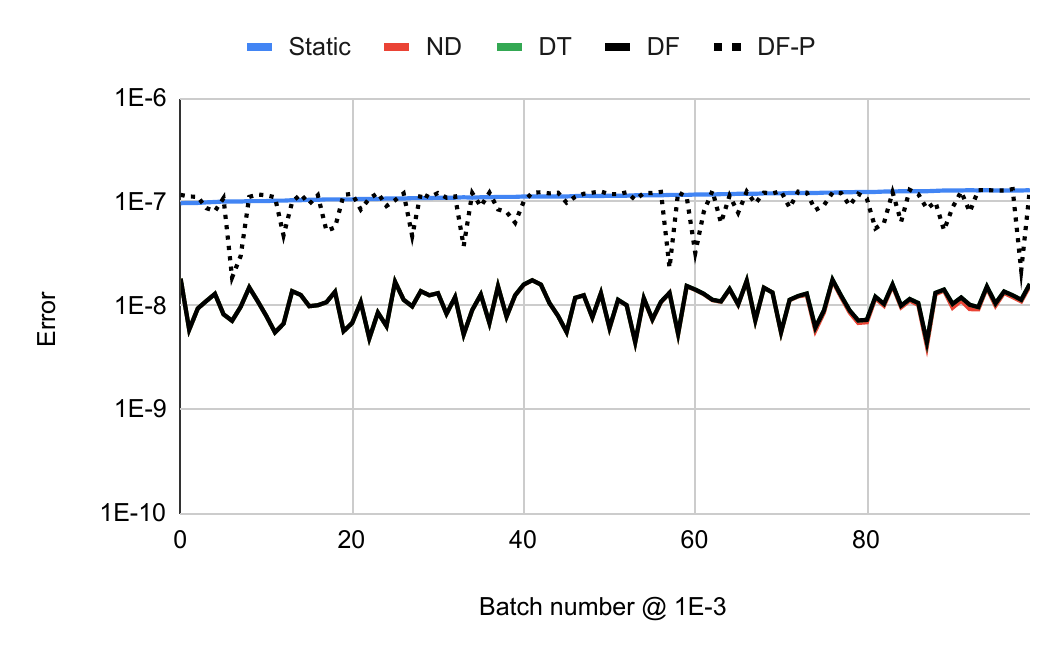}
  } \\[-2ex]
  \caption{Runtime and Error in ranks obtained with our GPU implementation of \textit{Static}, \textit{Naive-dynamic (ND)}, \textit{Dynamic Traversal (DT)}, \textit{Dynamic Frontier (DF)}, and \textit{Dynamic Frontier with Pruning (DF-P)} PageRank on the \textit{sx-mathoverflow} dynamic graph. The size of batch updates range from $10^{-5}|E_T|$ to $10^{-3}|E_T|$. The rank error with each approach is measured relative to ranks obtained with a reference Static PageRank run, as detailed in Section \ref{sec:measurement}.}
  \label{fig:temporal-sx-mathoverflow}
\end{figure*}

%% file: src/fig-temporal-sx-askubuntu.tex
\begin{figure*}[!hbt]
  \centering
  \subfigure[Runtime on consecutive batch updates of size $10^{-5}|E_T|$]{
    \label{fig:temporal-sx-askubuntu--runtime5}
    \includegraphics[width=0.48\linewidth]{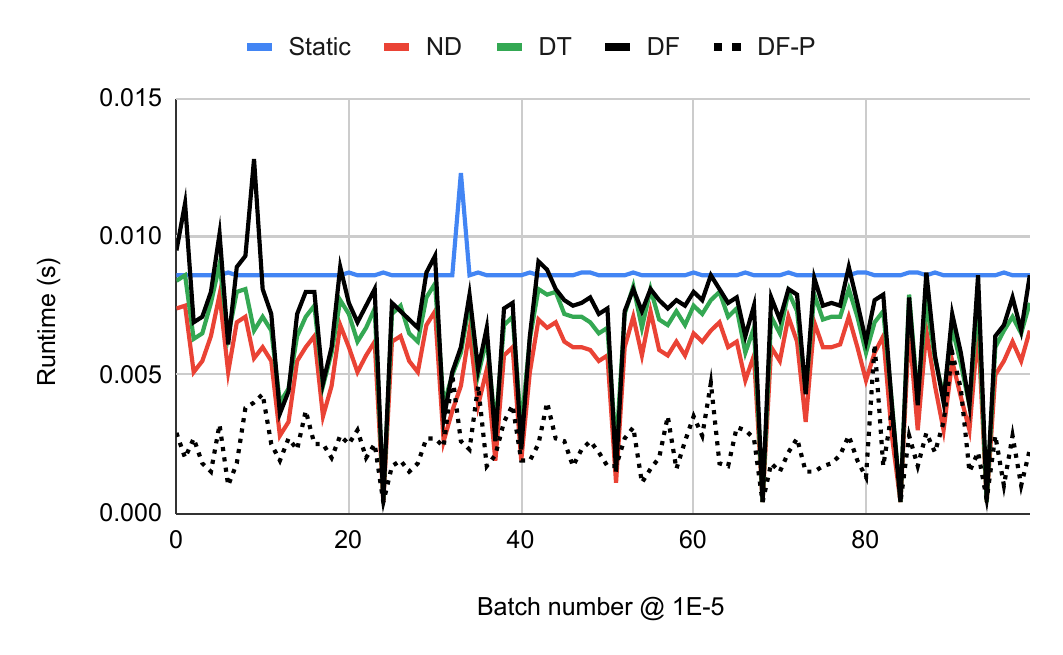}
  }
  \subfigure[Error in ranks obtained on consecutive batch updates of size $10^{-5}|E_T|$]{
    \label{fig:temporal-sx-askubuntu--error5}
    \includegraphics[width=0.48\linewidth]{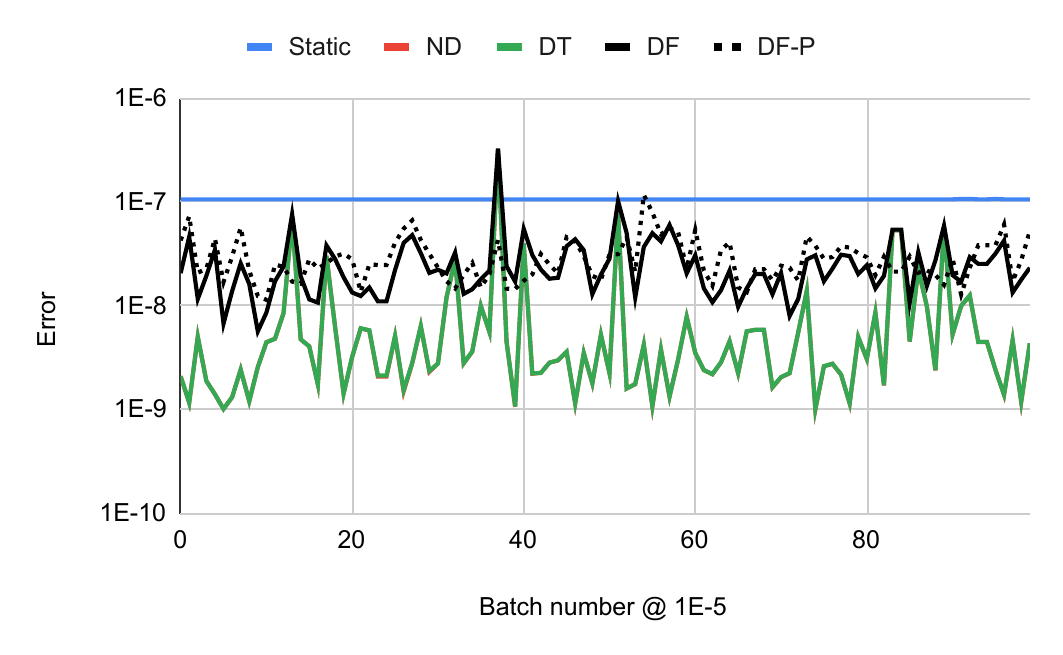}
  } \\[2ex]
  \subfigure[Runtime on consecutive batch updates of size $10^{-4}|E_T|$]{
    \label{fig:temporal-sx-askubuntu--runtime4}
    \includegraphics[width=0.48\linewidth]{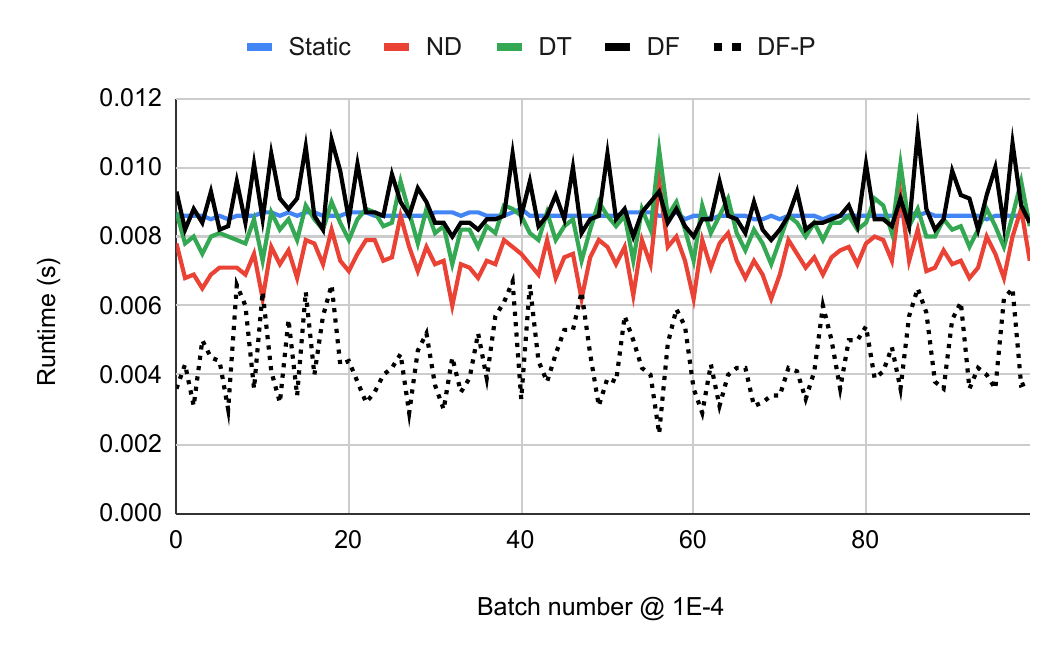}
  }
  \subfigure[Error in ranks obtained on consecutive batch updates of size $10^{-4}|E_T|$]{
    \label{fig:temporal-sx-askubuntu--error4}
    \includegraphics[width=0.48\linewidth]{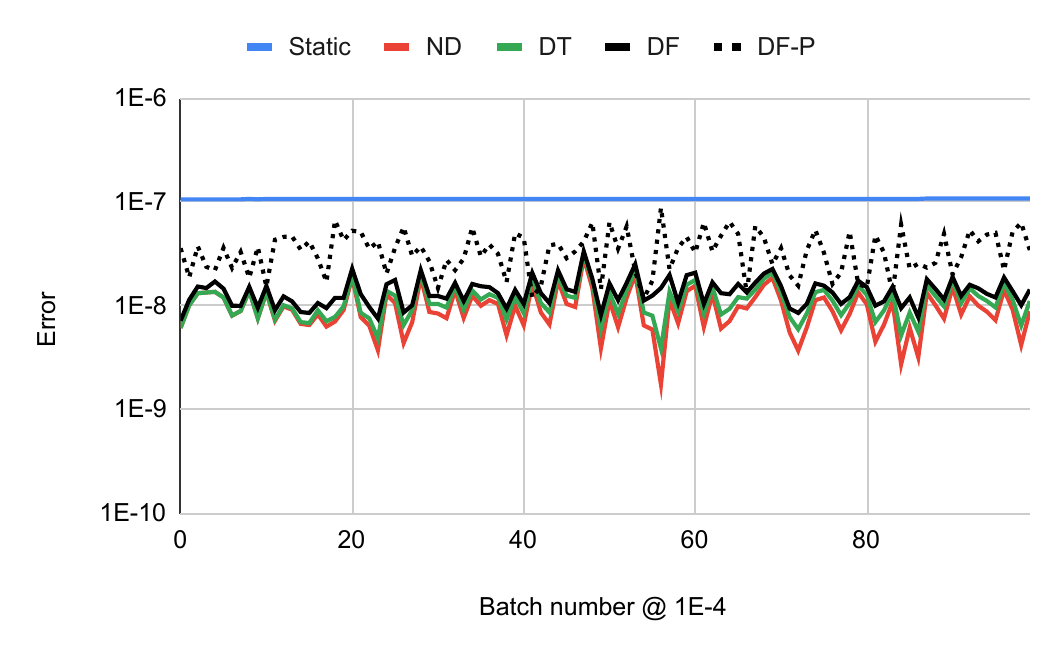}
  } \\[2ex]
  \subfigure[Runtime on consecutive batch updates of size $10^{-3}|E_T|$]{
    \label{fig:temporal-sx-askubuntu--runtime3}
    \includegraphics[width=0.48\linewidth]{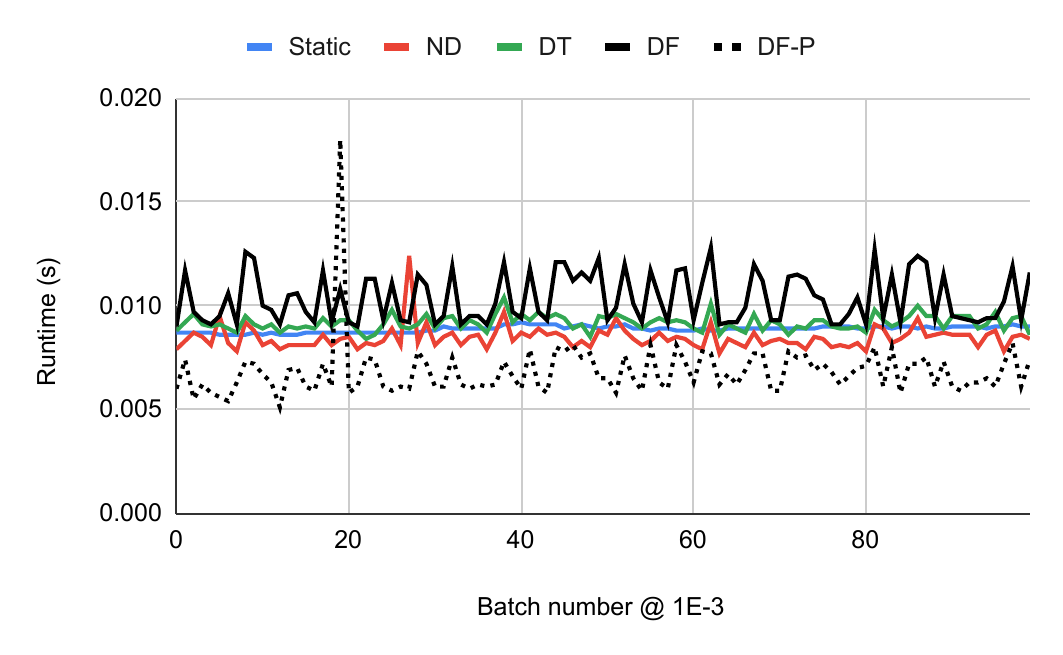}
  }
  \subfigure[Error in ranks obtained on consecutive batch updates of size $10^{-3}|E_T|$]{
    \label{fig:temporal-sx-askubuntu--error3}
    \includegraphics[width=0.48\linewidth]{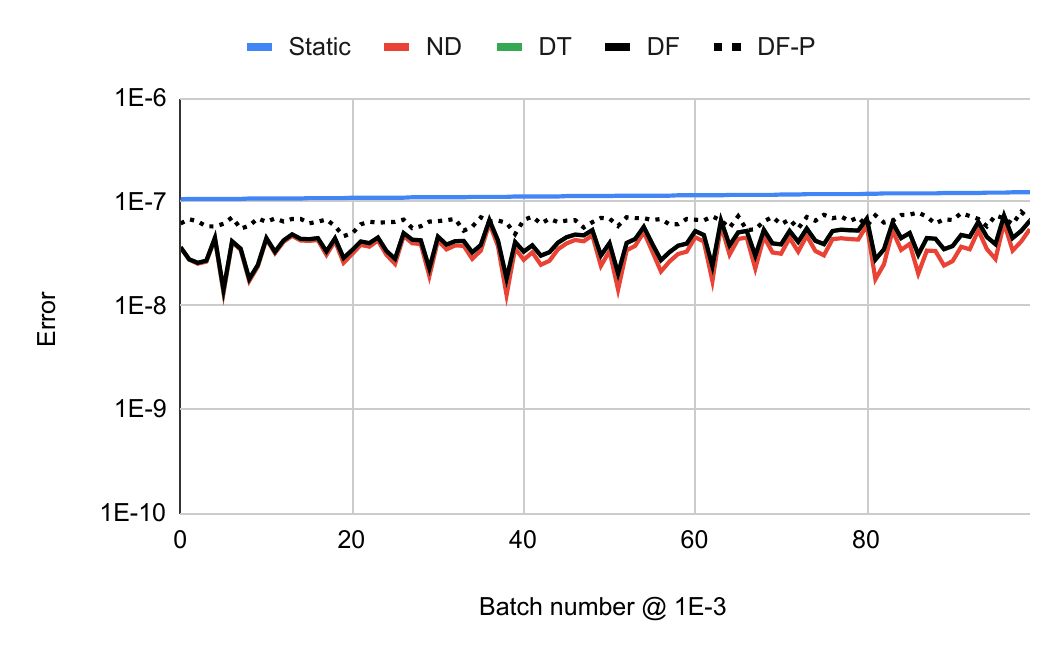}
  } \\[-2ex]
  \caption{Runtime and Error in ranks obtained with our GPU implementation of \textit{Static}, \textit{Naive-dynamic (ND)}, \textit{Dynamic Traversal (DT)}, \textit{Dynamic Frontier (DF)}, and \textit{Dynamic Frontier with Pruning (DF-P)} PageRank on the \textit{sx-askubuntu} dynamic graph. The size of batch updates range from $10^{-5}|E_T|$ to $10^{-3}|E_T|$. The rank error with each approach is measured relative to ranks obtained with a reference Static PageRank run, as detailed in Section \ref{sec:measurement}.}
  \label{fig:temporal-sx-askubuntu}
\end{figure*}

%% file: src/fig-temporal-sx-superuser.tex
\begin{figure*}[!hbt]
  \centering
  \subfigure[Runtime on consecutive batch updates of size $10^{-5}|E_T|$]{
    \label{fig:temporal-sx-superuser--runtime5}
    \includegraphics[width=0.48\linewidth]{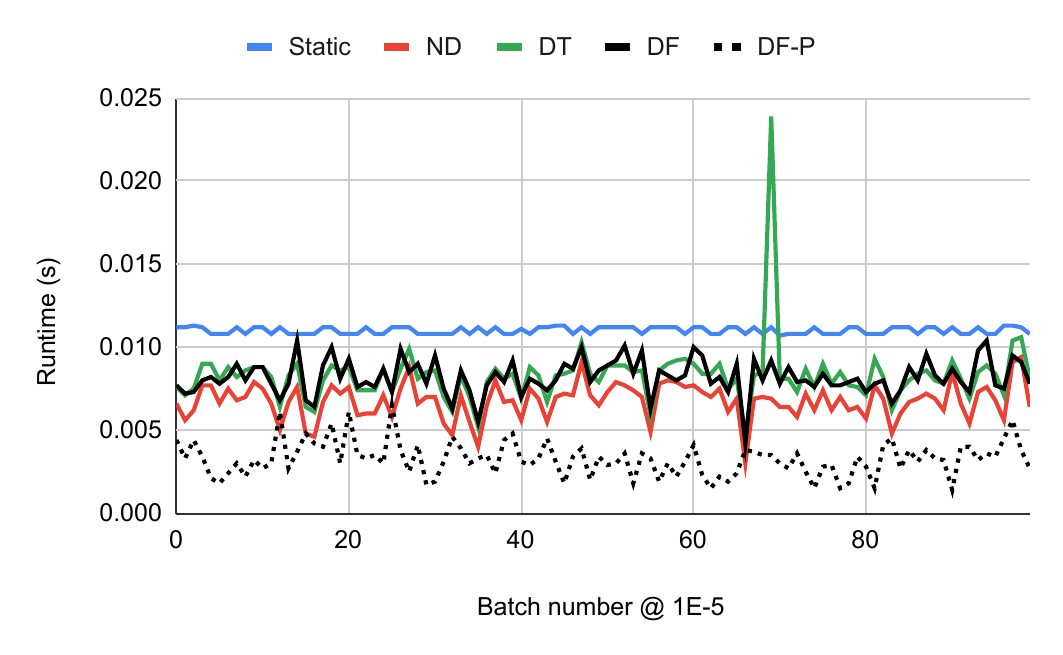}
  }
  \subfigure[Error in ranks obtained on consecutive batch updates of size $10^{-5}|E_T|$]{
    \label{fig:temporal-sx-superuser--error5}
    \includegraphics[width=0.48\linewidth]{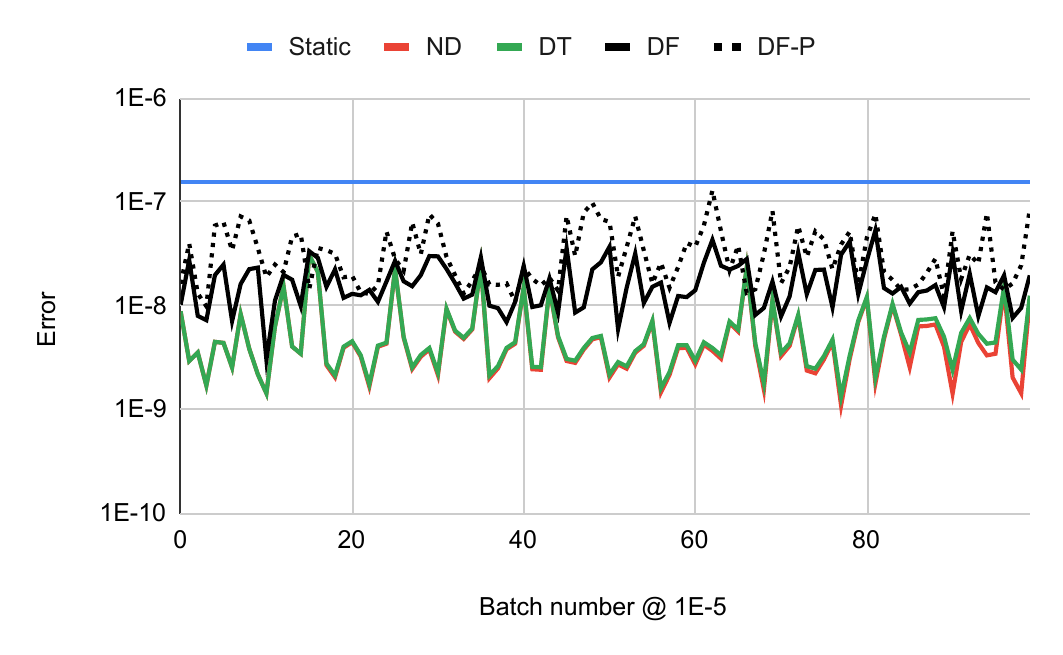}
  } \\[2ex]
  \subfigure[Runtime on consecutive batch updates of size $10^{-4}|E_T|$]{
    \label{fig:temporal-sx-superuser--runtime4}
    \includegraphics[width=0.48\linewidth]{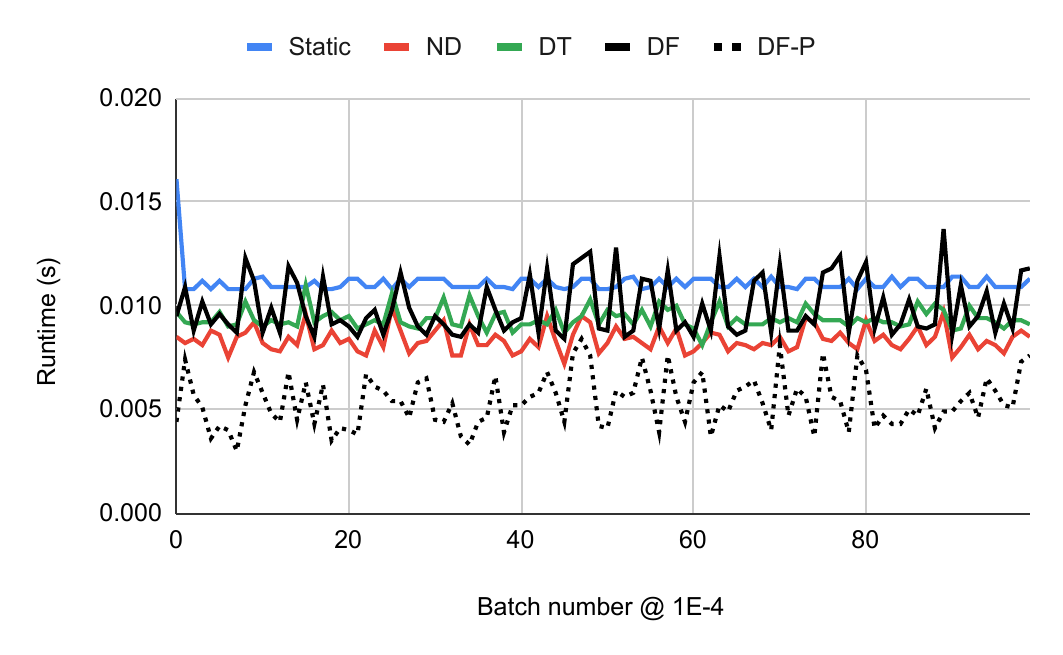}
  }
  \subfigure[Error in ranks obtained on consecutive batch updates of size $10^{-4}|E_T|$]{
    \label{fig:temporal-sx-superuser--error4}
    \includegraphics[width=0.48\linewidth]{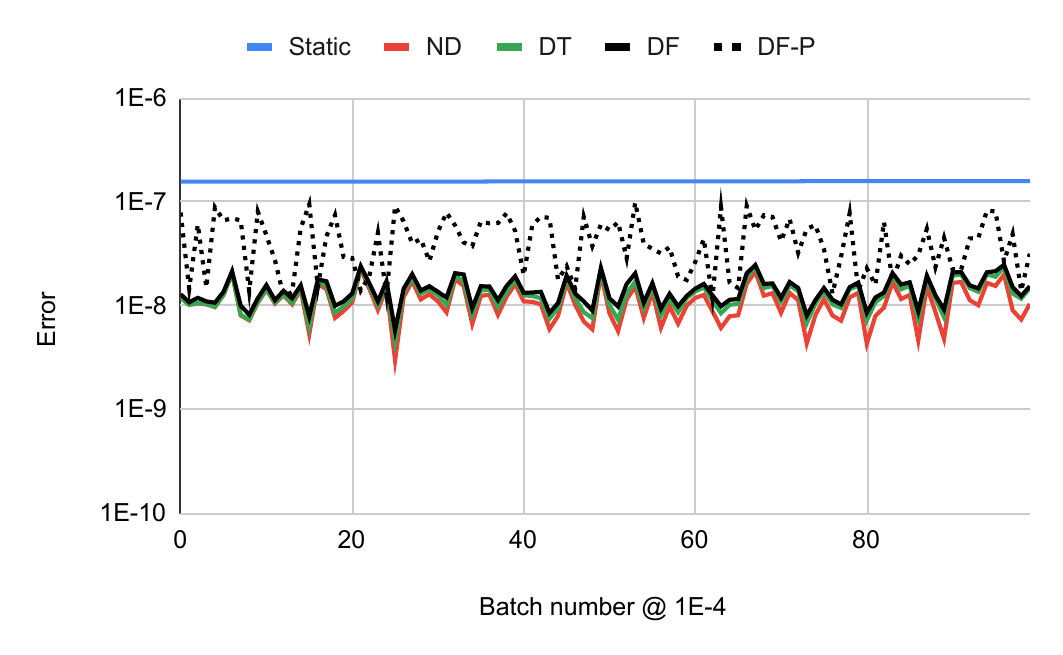}
  } \\[2ex]
  \subfigure[Runtime on consecutive batch updates of size $10^{-3}|E_T|$]{
    \label{fig:temporal-sx-superuser--runtime3}
    \includegraphics[width=0.48\linewidth]{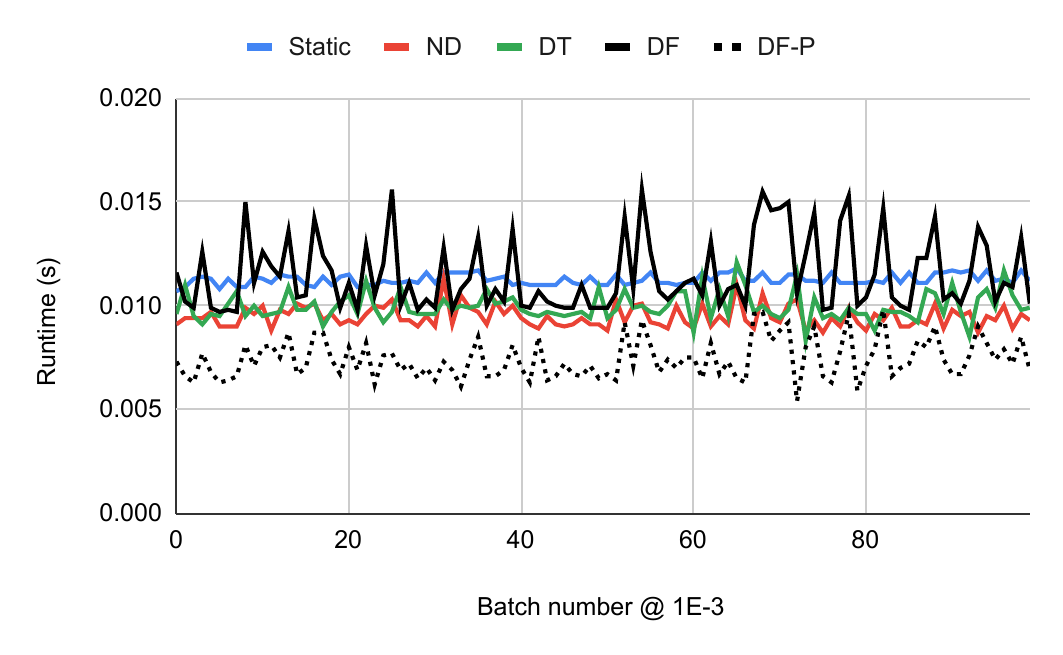}
  }
  \subfigure[Error in ranks obtained on consecutive batch updates of size $10^{-3}|E_T|$]{
    \label{fig:temporal-sx-superuser--error3}
    \includegraphics[width=0.48\linewidth]{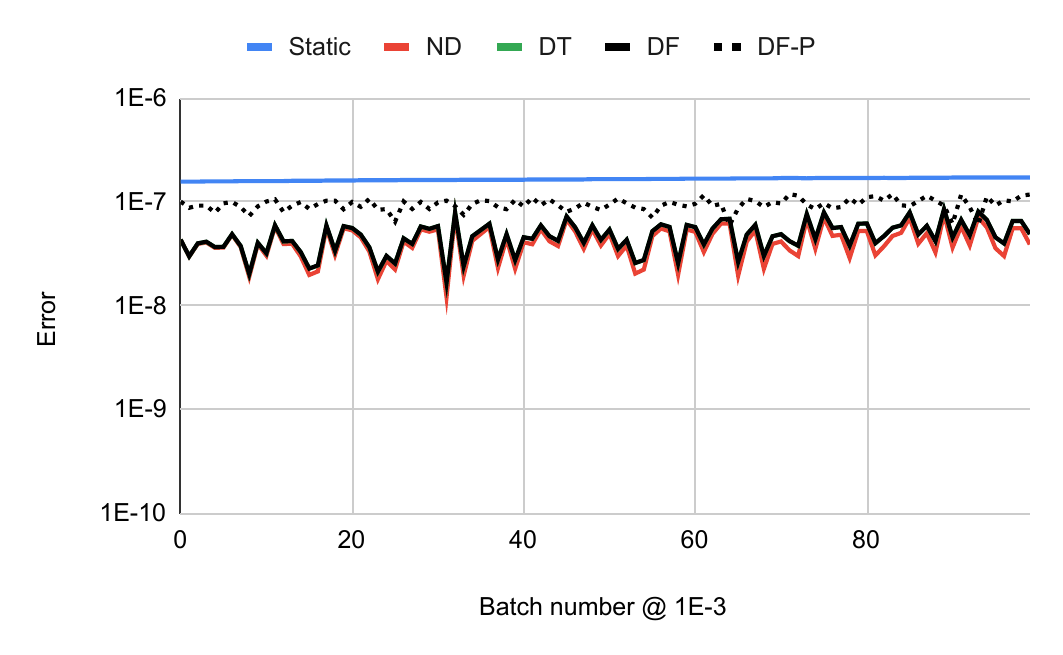}
  } \\[-2ex]
  \caption{Runtime and Error in ranks obtained with our GPU implementation of \textit{Static}, \textit{Naive-dynamic (ND)}, \textit{Dynamic Traversal (DT)}, \textit{Dynamic Frontier (DF)}, and \textit{Dynamic Frontier with Pruning (DF-P)} PageRank on the \textit{sx-superuser} dynamic graph. The size of batch updates range from $10^{-5}|E_T|$ to $10^{-3}|E_T|$. The rank error with each approach is measured relative to ranks obtained with a reference Static PageRank run, as detailed in Section \ref{sec:measurement}.}
  \label{fig:temporal-sx-superuser}
\end{figure*}

%% file: src/fig-temporal-wiki-talk-temporal.tex
\begin{figure*}[!hbt]
  \centering
  \subfigure[Runtime on consecutive batch updates of size $10^{-5}|E_T|$]{
    \label{fig:temporal-wiki-talk-temporal--runtime5}
    \includegraphics[width=0.48\linewidth]{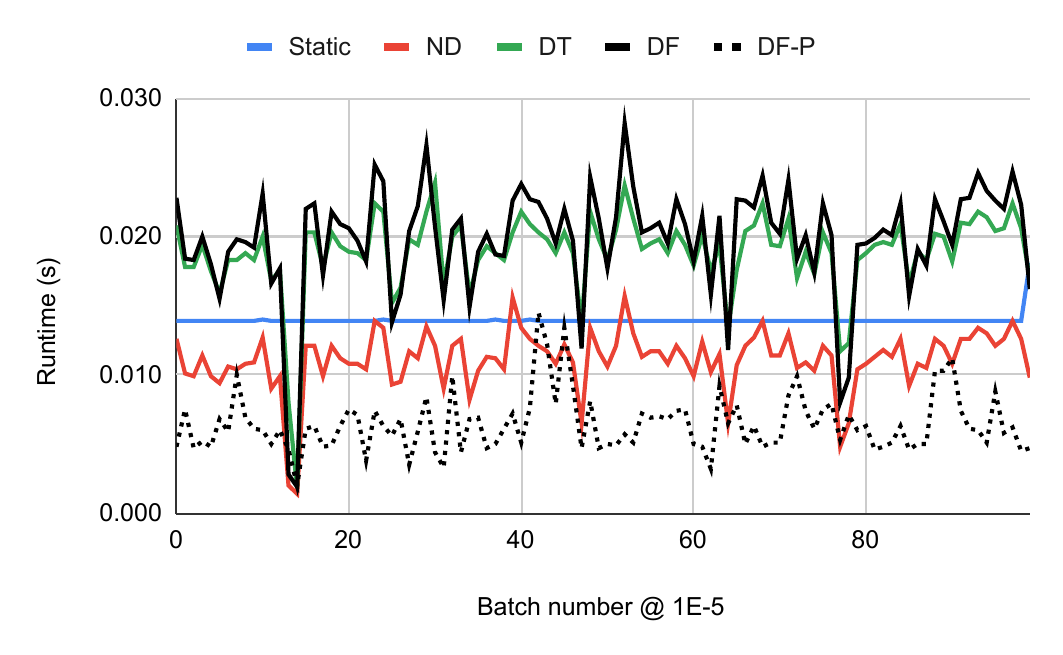}
  }
  \subfigure[Error in ranks obtained on consecutive batch updates of size $10^{-5}|E_T|$]{
    \label{fig:temporal-wiki-talk-temporal--error5}
    \includegraphics[width=0.48\linewidth]{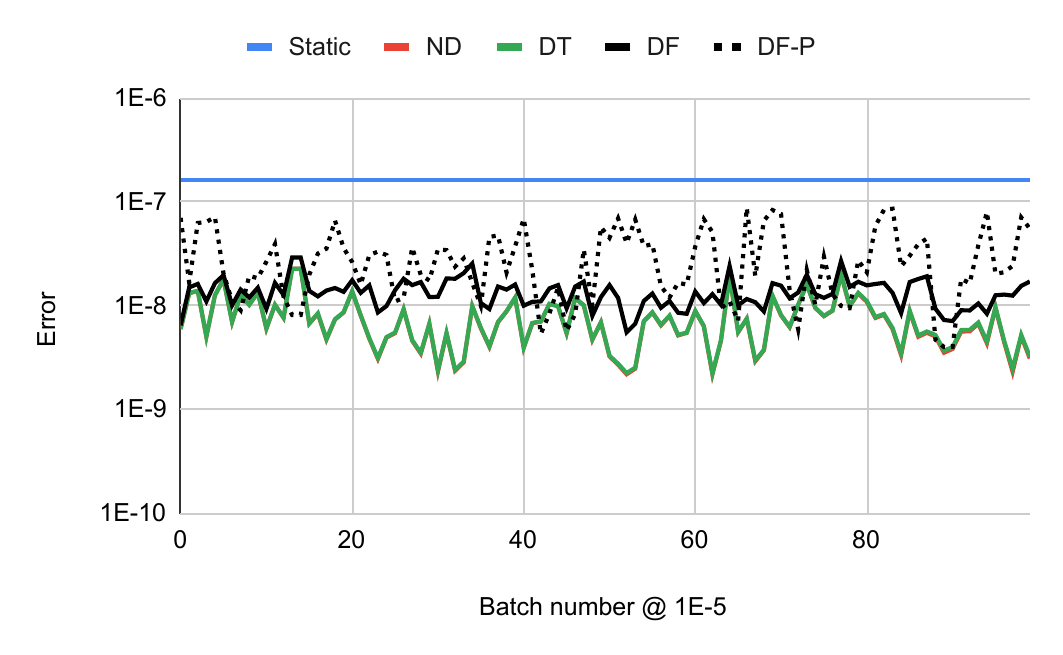}
  } \\[2ex]
  \subfigure[Runtime on consecutive batch updates of size $10^{-4}|E_T|$]{
    \label{fig:temporal-wiki-talk-temporal--runtime4}
    \includegraphics[width=0.48\linewidth]{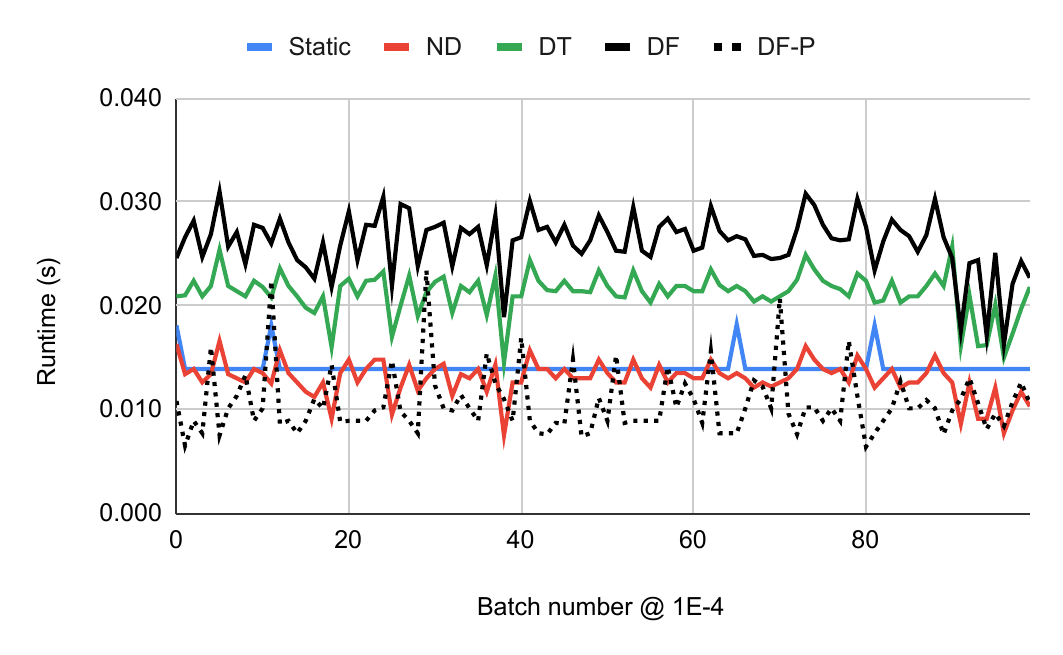}
  }
  \subfigure[Error in ranks obtained on consecutive batch updates of size $10^{-4}|E_T|$]{
    \label{fig:temporal-wiki-talk-temporal--error4}
    \includegraphics[width=0.48\linewidth]{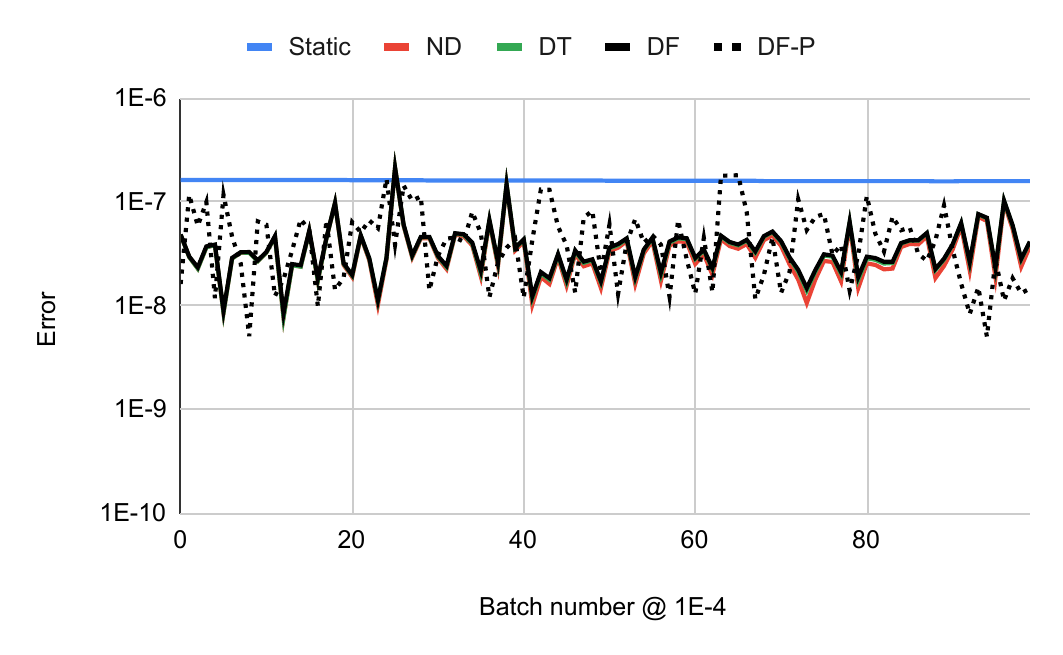}
  } \\[2ex]
  \subfigure[Runtime on consecutive batch updates of size $10^{-3}|E_T|$]{
    \label{fig:temporal-wiki-talk-temporal--runtime3}
    \includegraphics[width=0.48\linewidth]{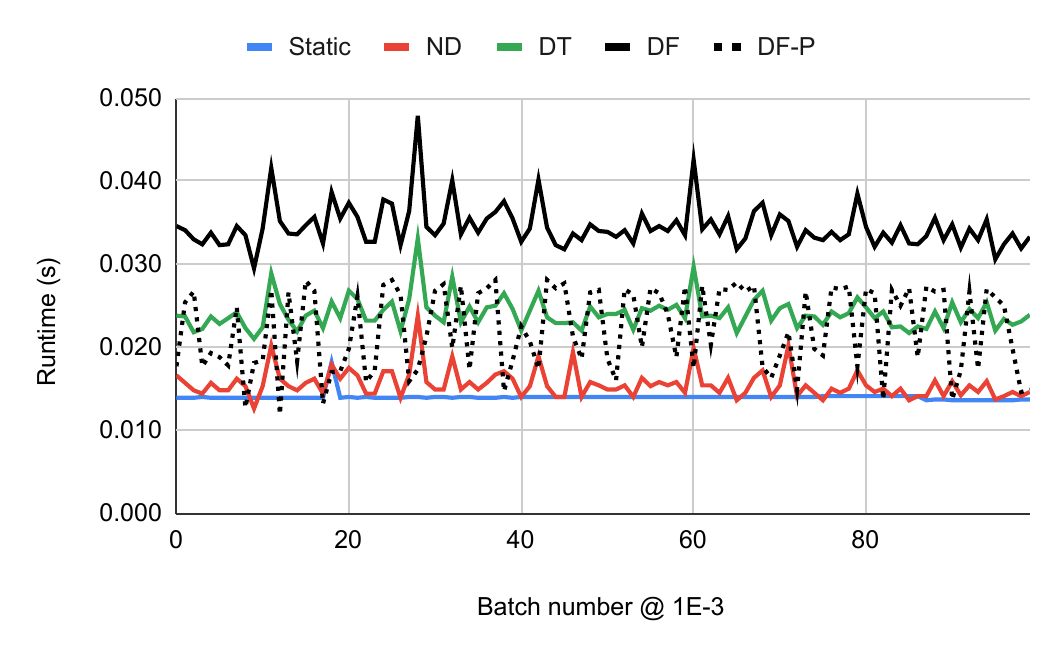}
  }
  \subfigure[Error in ranks obtained on consecutive batch updates of size $10^{-3}|E_T|$]{
    \label{fig:temporal-wiki-talk-temporal--error3}
    \includegraphics[width=0.48\linewidth]{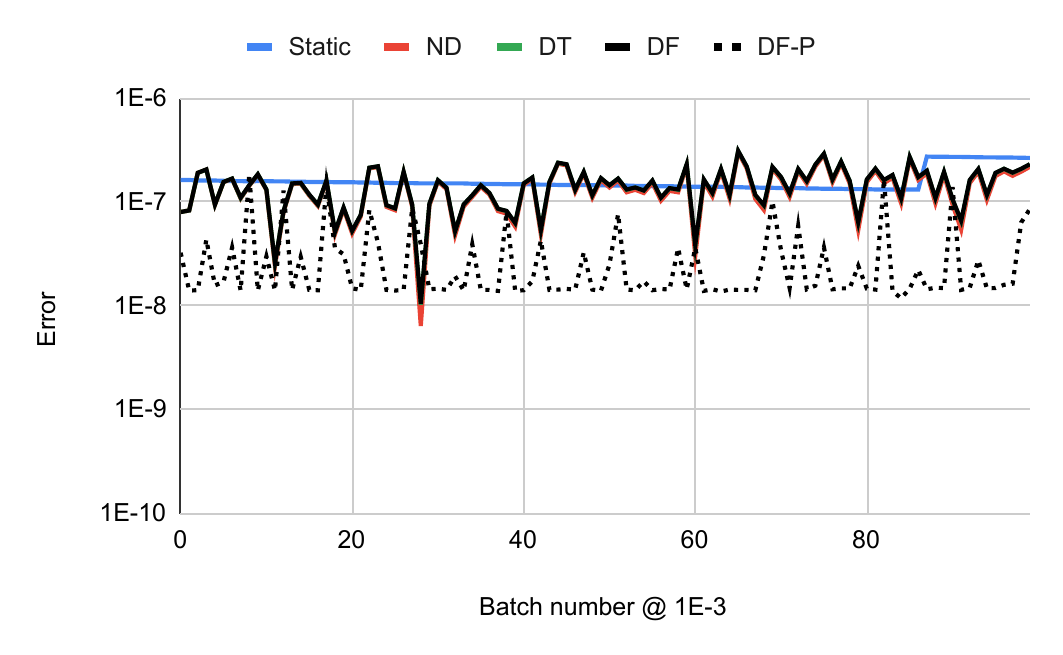}
  } \\[-2ex]
  \caption{Runtime and Error in ranks obtained with our GPU implementation of \textit{Static}, \textit{Naive-dynamic (ND)}, \textit{Dynamic Traversal (DT)}, \textit{Dynamic Frontier (DF)}, and \textit{Dynamic Frontier with Pruning (DF-P)} PageRank on the \textit{wiki-talk-temporal} dynamic graph. The size of batch updates range from $10^{-5}|E_T|$ to $10^{-3}|E_T|$. The rank error with each approach is measured relative to ranks obtained with a reference Static PageRank run, as detailed in Section \ref{sec:measurement}.}
  \label{fig:temporal-wiki-talk-temporal}
\end{figure*}

%% file: src/fig-temporal-sx-stackoverflow.tex
\begin{figure*}[!hbt]
  \centering
  \subfigure[Runtime on consecutive batch updates of size $10^{-5}|E_T|$]{
    \label{fig:temporal-sx-stackoverflow--runtime5}
    \includegraphics[width=0.48\linewidth]{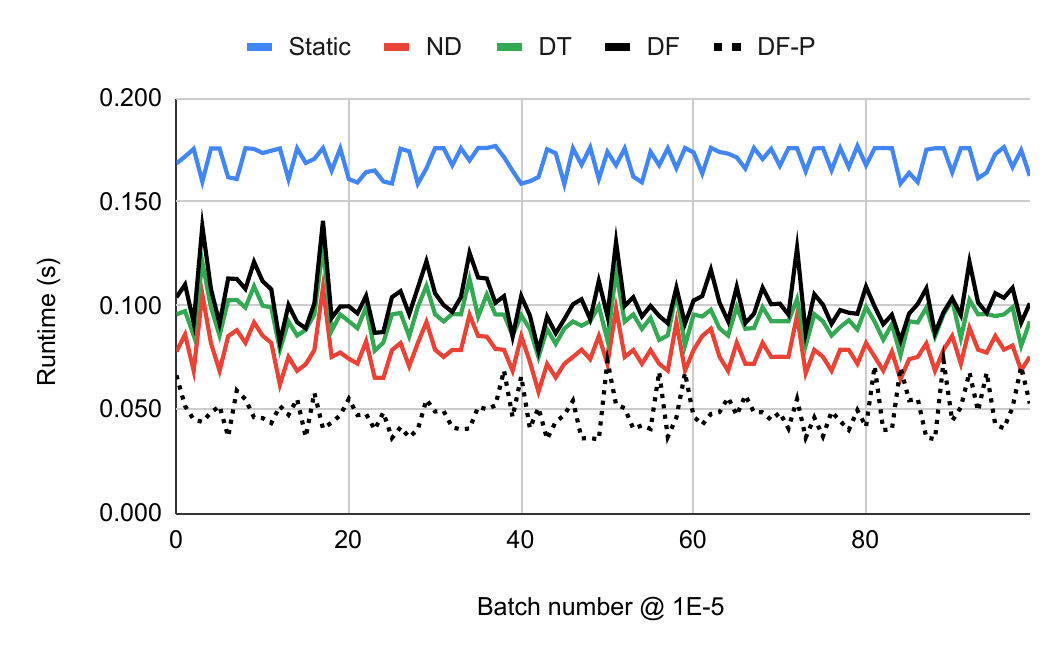}
  }
  \subfigure[Error in ranks obtained on consecutive batch updates of size $10^{-5}|E_T|$]{
    \label{fig:temporal-sx-stackoverflow--error5}
    \includegraphics[width=0.48\linewidth]{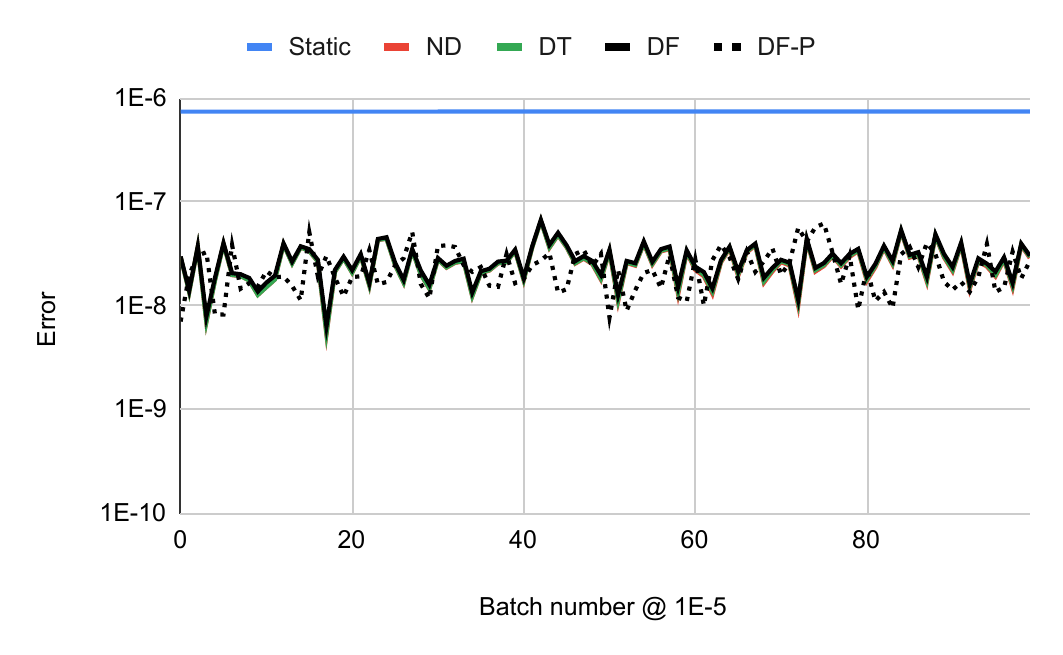}
  } \\[2ex]
  \subfigure[Runtime on consecutive batch updates of size $10^{-4}|E_T|$]{
    \label{fig:temporal-sx-stackoverflow--runtime4}
    \includegraphics[width=0.48\linewidth]{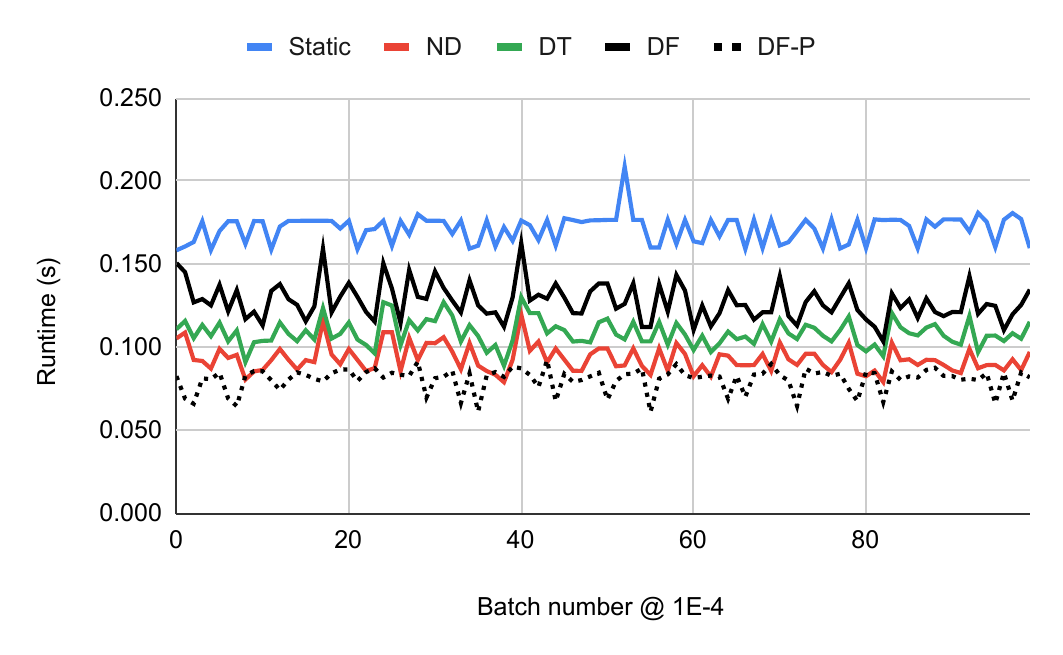}
  }
  \subfigure[Error in ranks obtained on consecutive batch updates of size $10^{-4}|E_T|$]{
    \label{fig:temporal-sx-stackoverflow--error4}
    \includegraphics[width=0.48\linewidth]{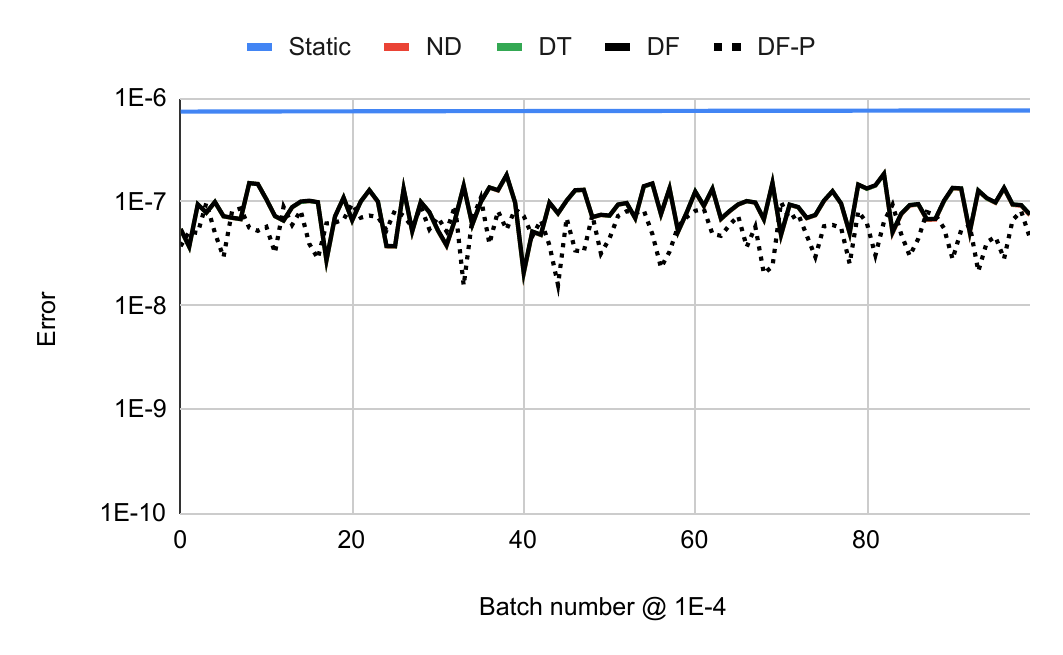}
  } \\[2ex]
  \subfigure[Runtime on consecutive batch updates of size $10^{-3}|E_T|$]{
    \label{fig:temporal-sx-stackoverflow--runtime3}
    \includegraphics[width=0.48\linewidth]{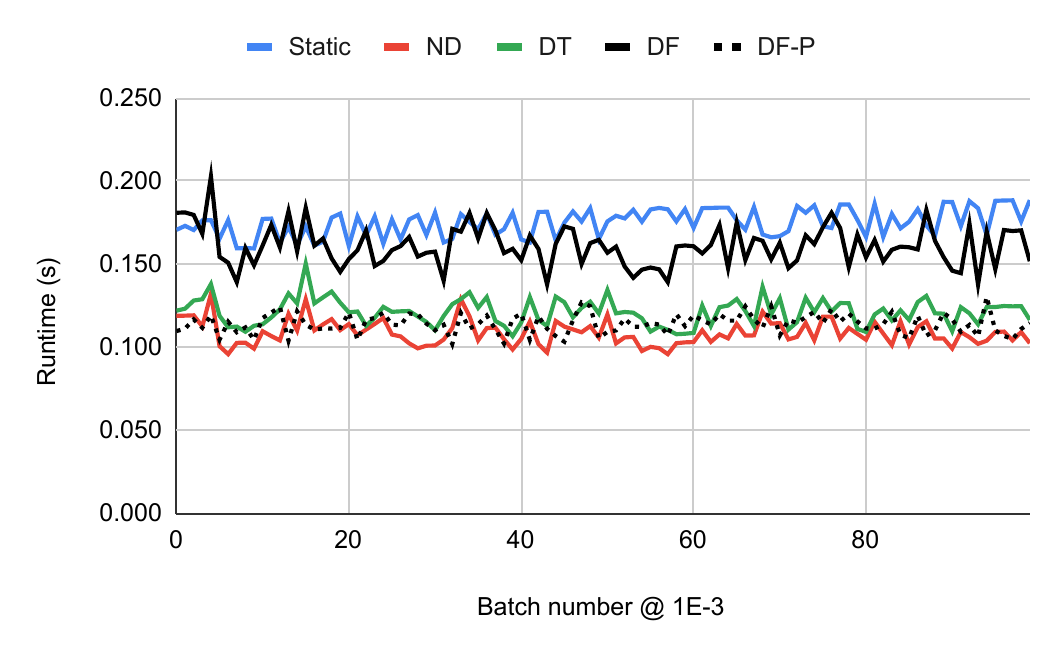}
  }
  \subfigure[Error in ranks obtained on consecutive batch updates of size $10^{-3}|E_T|$]{
    \label{fig:temporal-sx-stackoverflow--error3}
    \includegraphics[width=0.48\linewidth]{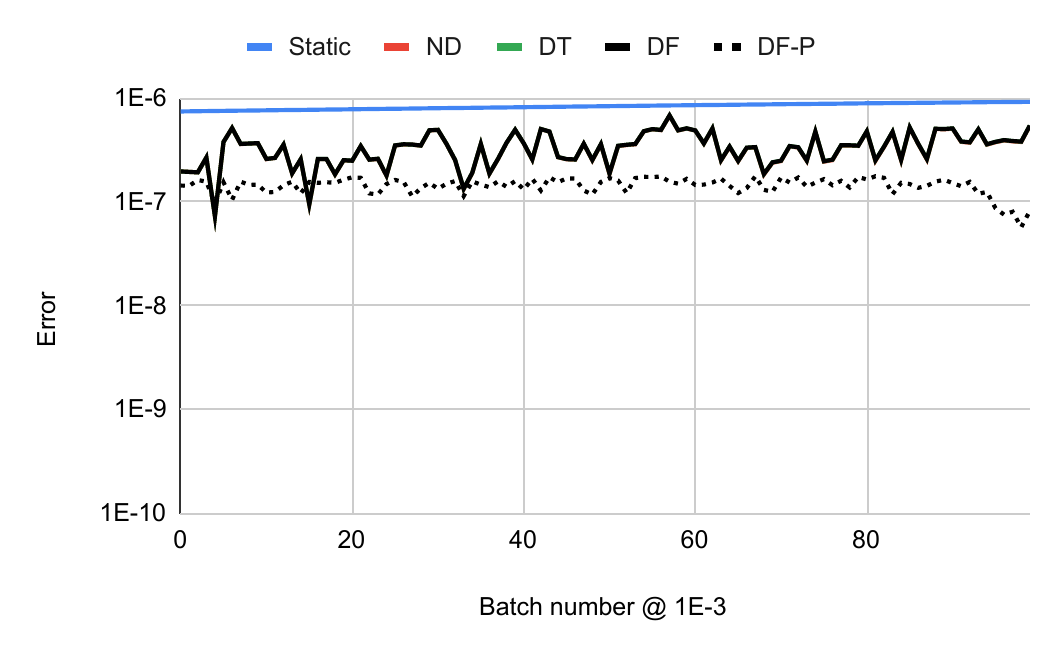}
  } \\[-2ex]
  \caption{Runtime and Error in ranks obtained with our GPU implementation of \textit{Static}, \textit{Naive-dynamic (ND)}, \textit{Dynamic Traversal (DT)}, \textit{Dynamic Frontier (DF)}, and \textit{Dynamic Frontier with Pruning (DF-P)} PageRank on the \textit{sx-stackoverflow} dynamic graph. The size of batch updates range from $10^{-5}|E_T|$ to $10^{-3}|E_T|$. The rank error with each approach is measured relative to ranks obtained with a reference Static PageRank run, as detailed in Section \ref{sec:measurement}.}
  \label{fig:temporal-sx-stackoverflow}
\end{figure*}